\documentclass[review,11pt,sort&compress]{elsarticle}

\usepackage{lineno,hyperref}

\journal{Progress in Nuclear and Particle Physics}

\usepackage{amsmath}
\usepackage{amssymb}
\usepackage[left=1in,right=1in,top=1in,bottom=1in]{geometry}
\usepackage[utf8]{inputenc}
\usepackage[usenames,dvipsnames]{color}
\usepackage{wrapfig}

\def\ba{\begin{eqnarray}}
\def\ea{\end{eqnarray}}
\def\be{\begin{eqnarray}}
\def\ee{\end{eqnarray}}
\newcommand{\intdP}{\int\!dP}

\begin{document}

\begin{frontmatter}

\title{Relativistic anisotropic hydrodynamics}

\author[a,b]{Mubarak Alqahtani}
\author[a]{Mohammad Nopoush}
\author[a]{and Michael Strickland}
\address[a]{Department of Physics, Kent State University, Kent, OH 44242 United States}
\address[b]{Imam Abdulrahman Bin Faisal University, Dammam 34212, Saudi Arabia}

\begin{abstract}

In this paper we review recent progress in relativistic anisotropic hydrodynamics.  We begin with a pedagogical introduction to the topic which takes into account the advances in our understanding of this topic since its inception.  We consider both conformal and non-conformal systems and demonstrate how one can implement a realistic equation of state using a quasiparticle approach.  We then consider the inclusion of non-spheroidal (non-ellipsoidal) corrections to leading-order anisotropic hydrodynamics and present the findings of the resulting second-order viscous anisotropic hydrodynamics framework.  We compare the results obtained in both the conformal and non-conformal cases with exact solutions to the Boltzmann equation and demonstrate that, in all known cases, anisotropic hydrodynamics best reproduces the exact solutions.  Based on this success, we then discuss the phenomenological application of anisotropic hydrodynamics.  Along these lines, we review techniques which can be used to convert a momentum-space anisotropic fluid into hadronic degrees of freedom by generalizing the original idea of Cooper-Frye freeze-out to momentum-space anisotropic systems.  And, finally, we present phenomenological results of 3+1d quasiparticle anisotropic hydrodynamic simulations and compare them to experimental data produced in 2.76 TeV Pb-Pb collisions at the LHC.  Our results indicate that anisotropic hydrodynamics provides a promising framework for describing the dynamics of the momentum-space anisotropic QGP created in heavy-ion collisions.

\end{abstract}

\begin{keyword}
quark-gluon plasma, hydrodynamics, anisotropic hydrodynamics, non-equilibrium dynamics
\end{keyword}

\end{frontmatter}

\setcounter{tocdepth}{2}
\tableofcontents


\section{Introduction}
\label{sec:intro}

The ongoing ultrarelativistic heavy-ion collision (URHIC) experiments at the Relativistic Heavy Ion Collider (RHIC) at Brookhaven National Laboratory (BNL) and the Large Hadron Collider (LHC) at the European Organization for Nuclear Research (CERN) aim to produce and study the properties of the quark-gluon plasma (QGP) using a variety of collision systems, e.g. AA, pA, dA, and pp over a wide range of center-of-mass energies.  One of the surprising findings of the experiments at RHIC was that the collective behavior of the soft hadrons $p_T \lesssim 2$ GeV was qualitatively well-described by ideal relativistic hydrodynamics \cite{Huovinen:2001cy,Hirano:2002ds,Kolb:2003dz}.  In these early days, due to the fact that ideal hydrodynamics implicitly relies on an assumption that the system is in perfect isotropic local thermal equilibrium, this led to the widespread supposition that the QGP created in URHICs underwent fast isotropization/thermalization on a timescale $\tau \sim 0.5$ fm/c (see e.g. \cite{Kolb:1999it,Heinz:2001ax,Heinz:2002un,Jacobs:2004qv,Muller:2005en,Strickland:2007fm}).

Despite subsequent relativistic hydrodynamics research \cite{Muronga:2001zk,Muronga:2003ta,Muronga:2004sf,Heinz:2005bw,Baier:2006um,Romatschke:2007mq,Baier:2007ix,Dusling:2007gi,Luzum:2008cw,Song:2008hj,Heinz:2009xj,El:2009vj,PeraltaRamos:2009kg,PeraltaRamos:2010je,Denicol:2010tr,Denicol:2010xn,Schenke:2010rr,Schenke:2011tv,Bozek:2011wa,Niemi:2011ix,Denicol:2011fa,Niemi:2012ry,Bozek:2012qs,Denicol:2012cn,Denicol:2012es,PeraltaRamos:2012xk,Jaiswal:2013npa,Jaiswal:2013vta,Calzetta:2014hra,Denicol:2014vaa,Denicol:2014mca,Jaiswal:2014isa}\footnote{These modern studies extended the seminal works of Mueller, Israel, and Stewart \cite{Muller:1967zza,Israel:1976tn,Israel:1979wp}}, which included the effect of dissipative corrections in the form of second-order viscous hydrodynamics (vHydro) and explicitly included pressure anisotropies in the form of the viscous stress tensor, it was hard to dislodge the idea that the QGP generated in URHICs was in perfect local isotropic equilibrium.  Simultaneously to these phenomenological developments, fundamental studies of the thermalization and isotropization of the QGP were undertaken in the context of quantum field theories in both the weak \cite{Heinz:1985vf,Mrowczynski:1988dz,Pokrovsky:1988bm,Mrowczynski:1993qm,Blaizot:2001nr,Romatschke:2003ms,Arnold:2003rq,Arnold:2004ih,Romatschke:2004jh,Arnold:2004ti,Mrowczynski:2004kv,Rebhan:2004ur,Rebhan:2005re,Romatschke:2005pm,Romatschke:2006nk,Romatschke:2006wg,Rebhan:2008uj,Fukushima:2011nq,Kurkela:2011ti,Kurkela:2011ub,Blaizot:2011xf,Attems:2012js,Kurkela:2012tq,Berges:2012iw,Blaizot:2013lga,Mrowczynski:2016etf} and strong coupling \cite{Chesler:2008hg,Grumiller:2008va,Chesler:2009cy,Albacete:2009ji,Wu:2011yd,Heller:2011ju,Chesler:2011ds,Heller:2012je,vanderSchee:2012qj,Romatschke:2013re,Casalderrey-Solana:2013aba,vanderSchee:2013pia} limits.  The conclusion of these studies was that, due to the extreme conditions in which the QGP is created, it is not possible to achieve isotropization on a sub-fm/c timescale.  This motivated the investigation of the impact of momentum-space anisotropies on QGP dynamics and signatures and, after many years, there is now a consensus in the theoretical community that the QGP possesses a high degree of momentum-space anisotropy at early times and near the dilute edges of the system~\cite{Ryblewski:2013jsa,Strickland:2014pga}.  In practice, one finds that, in these spacetime regions, the transverse pressure in the local rest frame greatly exceeds the longitudinal pressure and that, in the center of the fireball, it takes many fm/c for the system to become even approximately isotropic.  Faced with this, researchers began looking for ways to formulate hydrodynamics in a momentum-space anisotropic QGP.

A significant breakthrough occurred in this direction with two papers, one from Florkowski and Ryblewski \cite{Florkowski:2010cf} and the other from Martinez and Strickland \cite{Martinez:2010sc} in 2010.  In both papers, the authors considered a boost-invariant and transversally homogeneous system but in the first paper the authors postulated an equation governing entropy production in an anisotropic system, whereas in the second one the authors took moments of the Boltzmann equation in relaxation-time approximation.  These papers demonstrated that it was possible to formulate relativistic hydrodynamics using an intrinsically momentum-space anisotropic distribution in the local rest frame.  In addition, the Martinez and Strickland paper demonstrated that (a) the resulting dynamical equations could reproduce both the ideal hydrodynamics limit ($\eta/s =0$) and the free-streaming limit ($\eta/s=\infty$) and (b) that, in the case of weak momentum space isotropy ($P_L \approx P_T$), the formalism reproduced standard viscous hydrodynamics.  This framework has become known, generally, as {\em anisotropic hydrodynamics} (aHydro).  The original ideas presented in Refs.~\cite{Florkowski:2010cf,Martinez:2010sc} have since been generalized and extended to fully 3+1d systems with broken conformal symmetry.  These generalized frameworks have been applied successfully to heavy-ion collision phenomenology (see e.g. \cite{Ryblewski:2010ch,Florkowski:2011jg,Martinez:2012tu,Ryblewski:2012rr,Bazow:2013ifa,Tinti:2013vba,Nopoush:2014pfa,Tinti:2015xwa,Bazow:2015cha,Strickland:2015utc,Alqahtani:2015qja,Molnar:2016vvu,Molnar:2016gwq,Alqahtani:2016rth,Bluhm:2015raa,Bluhm:2015bzi,Alqahtani:2017jwl,Alqahtani:2017tnq}).  

In this review, rather than directly presenting the modern version of aHydro, we will attempt to make historical review, highlighting the important advances in stages in order to make the material more accessible to readers who are unfamiliar with the literature.  We begin with a pedagogical introduction to the topic which takes into account the advances in our understanding of this topic since its inception.  We consider both conformal and non-conformal systems and demonstrate how one can implement a realistic equation of state using a quasiparticle approach.  We then consider the inclusion of non-spheroidal (non-ellipsoidal) corrections to leading-order anisotropic hydrodynamics and present the findings of the resulting second-order viscous anisotropic hydrodynamics framework.  We compare the results obtained in both the conformal and non-conformal cases with exact solutions to the Boltzmann equation and demonstrate that, in all known cases, anisotropic hydrodynamics best reproduces the exact solutions.  Based on this success, we then discuss the phenomenological application of aHydro.  Along these lines, we review techniques which can be used to convert a momentum-space anisotropic fluid into hadronic degrees of freedom by applying the original idea of Cooper-Frye freeze-out to momentum-space anisotropic systems.  And, finally, we present phenomenological results of 3+1d quasiparticle anisotropic hydrodynamic simulations and compare them to experimental data produced in 2.76 TeV Pb-Pb collisions at the LHC.  Our results indicate that aHydro provides a promising framework for describing the dynamics of the momentum-space anisotropic QGP created in heavy-ion collisions.

\section*{Conventions and notation}

Unless otherwise indicated, the Minkowski metric tensor is taken to be ``mostly minus'', i.e. $g^{\mu\nu}={\rm diag}(+,-,-,-)$. The transverse projection operator $\Delta^{\mu\nu}\equiv g^{\mu\nu}{-}u^\mu u^\nu$ is used to project four-vectors and/or tensors into the space orthogonal to $u^\mu$. Parentheses and square brackets on indices denote symmetrization and anti-symmetrization, respectively, i.e. $A^{(\mu\nu)}\equiv\frac{1}{2}\left(A^{\mu\nu}{+}A^{\nu\mu}\right)$  and $A^{[\mu\nu]}\equiv\frac{1}{2}\left(A^{\mu\nu}{-}A^{\nu\mu}\right)$. Angle brackets on indices indicate projection with a four-index transverse projector, \mbox{$A^{\langle \mu \nu\rangle}\equiv\Delta^{\mu\nu}_{\alpha\beta}A^{\alpha\beta}$}, where $\Delta^{\mu\nu}_{\alpha\beta}\equiv\Delta^{(\mu}_\alpha\Delta^{\nu)}_\beta-\Delta^{\mu\nu}\Delta_{\alpha\beta}/3$ projects out the traceless and $u^\mu$-transverse components of a rank-two tensor.
 

\section{Historical foundations and pedagogical introduction}

In this section, we review the method presented originally in Ref.~\cite{Martinez:2010sc} in a concise and updated manner.  At the end, we will highlight the important findings.  We will attempt to present the material in a pedagogical manner so as to provide the reader a firm basis to build upon.  As in the original paper, we will make several simplifying assumptions:  (1) that the system is invariant under longitudinal boosts, (2) that the system is transversally homogenous, and (3) that the system is conformal (massless particles).  Hence, we consider here, a conformal 0+1d system.  In the course of this review, we will relax all of these assumptions.

\subsection{Moments of the Boltzmann equation}

The starting point for the derivation presented in Ref.~\cite{Martinez:2010sc} was the relativistic Boltzmann equation
\be
p^\mu \partial_\mu f(x,p) = - C[f] \, ,
\label{eq:boltzmann1}
\ee 
where $p^\mu$ and $x^\mu$ are the particle four-momentum and -position, respectively, $f$ is the one-particle distribution function, and $C[f]$ is the collisional kernel which includes both elastic and inelastic scatterings to all orders.  For this discussion, we will assume that the system is boost-invariant and transversally homogenous (0+1d) and that there is no chemical potential.  

To proceed, one can take moments of Eq.~(\ref{eq:boltzmann1}) using the integral operator
\be
\hat{\cal O}_n \, g = {\cal O}^{\mu_1 \mu_2 \cdots \mu_n}[g]  \equiv \int dP \, p^{\mu_1}p^{\mu_2} \cdots p^{\mu_n} \, g(p) \, ,
\ee
where
\be
\int dP \equiv N_{\rm dof} \int \frac{d^4p}{(2\pi)^4} \, 2\pi \delta(p^2 - m^2) \, 2 \theta(p_0) = N_{\rm dof} \int \frac{d^3{\bf p}}{(2\pi)^3} \frac{1}{E}  \, ,
\label{eq:invphase}
\ee
is the Lorentz-invariant momentum integration measure with $E = u_\mu p^\mu$ being the local rest frame energy and $N_{\rm dof}$ being the number of degrees of freedom (degeneracy).
Acting with $\hat{\cal O}_0$ on Eq.~(\ref{eq:boltzmann1}), one obtains the ``zeroth-moment'' of the Boltzmann equation
\be
\partial_\mu \left(\int dP \, p^\mu f \right) = -\int dP \, C[f] \,  .
\ee
Using the fact that the particle current is defined as
\be
J^\mu = \int dP \, p^\mu f \, ,
\ee
and defining a general moment of the collisional kernel via
\be
{\cal C}_r^{\mu_1 \mu_2 \cdots \mu_n} \equiv - \int dP \, (p \cdot u)^r p^{\mu_1}p^{\mu_2} \cdots p^{\mu_n} \, C[f] \, ,
\label{eq:calcdef}
\ee
we can write the zeroth moment compactly as
\be
\partial_\mu J^\mu = {\cal C}_0 \, .
\label{eq:0+1dneq}
\ee

Repeating this exercise using $\hat{\cal O}_1$, one obtains the first-moment of the Boltzmann equation
\be
\partial_\mu T^{\mu\nu} = {\cal C}^\nu_0 \, ,
\ee
where
\be
T^{\mu\nu} \equiv \int dP \, p^\mu p^\nu f \, ,
\ee
is the energy-momentum tensor.  For any valid microscopic model, the energy and momentum conserving delta function inherent in the collisional kernel ensures that ${\cal C}^\nu_0 = 0$.  If, instead, one works  with an effective collisional kernel such as the relaxation-time approximation (RTA) model~\cite{anderson1974relativistic}, the requirement that ${\cal C}^\nu_0 = 0$ becomes a constraint which must be enforced on any parameters appearing in the model's kernel.  We will return to this point later.  Hence one finds, in general, that the first moment of the Boltzmann equation results in the simple statement of energy and momentum conservation
\be
\partial_\mu T^{\mu\nu} = 0 \, .
\ee
In Ref.~\cite{Martinez:2010sc} the zeroth and first moments were used to obtain the necessary equations of motion, however, in order to bring the material up to date, we would like to also compare a different prescription in which one uses a linear combination of equations obtained from the second moment of the Boltzmann equation \cite{Tinti:2013vba}.  For this purpose, we note that, for a general moment, one obtains
\be
\partial_\mu I^{\mu\nu_1\nu_2\cdots\nu_n} = {\cal C}^{\nu_1\nu_2\cdots\nu_n}_0 \, ,
\label{eq:generaleom}
\ee
where $I^{\mu\nu_1\nu_2\cdots\nu_n} \equiv \int dP \, p^{\mu} p^{\nu_1} p^{\nu_2} \cdots p^{\nu_n} f$.

\subsection{Tensor basis}

At this point in the discussion it is helpful to write down the most general forms for the particle current $n^\mu$, energy-momentum tensor $T^{\mu\nu}$, and third-rank tensor $I^{\mu\nu\lambda}$.  For this purpose, we need only consider the symmetries of the system.  To begin, we introduce four 4-vectors which span spacetime in the local rest frame (LRF) \cite{Florkowski:2011jg,Martinez:2012tu}:
\ba
&&X^\mu_{0,{\rm LRF}} \equiv u^\mu_{\rm LRF} = (1,0,0,0) \, , \nonumber \\
&&X^\mu_{1,{\rm LRF}} \equiv X^\mu_{\rm LRF} = (0,1,0,0) \, , \nonumber \\
&&X^\mu_{2,{\rm LRF}} \equiv Y^\mu_{\rm LRF} = (0,0,1,0) \, , \nonumber \\
&&X^\mu_{3,{\rm LRF}} \equiv Z^\mu_{\rm LRF} = (0,0,0,1) \, .
\label{eq:rfbasis}
\ea
These four-vectors are orthonormal in all frames.  The vector $X^\mu_0$ is associated with the four-velocity of the LRF and is conventionally called $u^\mu$.  One can also identify $X^\mu_1 = X^\mu$, $X^\mu_2 = Y^\mu$, and $X^\mu_3 = Z^\mu$ as indicated above.\footnote{In the lab frame, the three space-like vectors $X^\mu_i$ can be written entirely in terms of $X^\mu_0=u^\mu$.  This is because $X^\mu_i$ can be obtained by a sequence of Lorentz transformations/rotations applied to the LRF expressions specified above \cite{Florkowski:2011jg,Martinez:2012tu}.}

The metric tensor can be expressed in terms of these four-vectors as
\be
g^{\mu \nu}= u^\mu u^\nu - \sum_i X^\mu_i X^\nu_i \, ,
\label{eq:gbasis}
\ee
where the sum extends over $i=1,2,3$.  In addition, the standard transverse projection operator, which is orthogonal to $X^\mu_0$, can be expressed in terms of the basis (\ref{eq:rfbasis})
\be
\Delta^{\mu \nu} = g^{\mu\nu} - u^\mu u^\nu = - \sum_i X^\mu_i X^\nu_i \, ,
\label{eq:transproj}
\ee
from which one finds $u_\mu \Delta^{\mu \nu} = u_\nu \Delta^{\mu \nu} = 0$.  The space-like components of the tensor basis are eigenvectors of this operator, i.e. $X_{i\mu} \Delta^{\mu \nu} = X^\nu_{i}$.

Using these basis vectors, we can expand any tensor.  For example, the number current, which is a rank-1 tensor, can be written in general as
\be
J^\mu = n u^\mu + \sum_i n_i X^\mu_i \, .
\label{eq:0+1dn}
\ee
Since the 0+1d distribution function is reflection symmetric in momentum-space around the $x$, $y$, and $z$ directions in the LRF, the space-like coefficients $n_i=0$, leaving in this case
\be
J^\mu = n u^\mu \, .
\ee

Likewise, since the the energy-momentum tensor is symmetric $T^{\mu\nu} = T^{\nu\mu}$, one has
\be
T^{\mu\nu} = t_{00} g^{\mu \nu}  + \sum_{i=1}^3 t_{ii} X^\mu_i X^\nu_i 
+ \sum_{\alpha,\beta=0 \atop \alpha>\beta}^3 t_{\alpha\beta} (X^\mu_\alpha X^\nu_\beta+X^\mu_\beta X^\nu_\alpha) \, ,
\ee
where the coefficients $t_{00}$, etc. are scalar fields.  Using the symmetries associated with the 0+1d case considered in this section, one can simplify this to \cite{Martinez:2012tu}
\be
T^{\mu\nu} = (\epsilon +P_T) u^\mu u^\nu  - P_T g^{\mu \nu}+ (P_L - P_T) Z^\mu Z^\nu \, ,
\ee
where $\epsilon$, $P_T$, and $P_L$ are the energy density, transverse pressure, and longitudinal pressure in the LRF.  Above, $L$ and $T$ correspond to the directions parallel and transverse to the anisotropy direction $\hat{\bf n} = \hat{\bf z}$, respectively.

Finally, we can repeat this exercise for the rank-three tensor $I^{\mu\nu\lambda}$
\ba
I &=& I_u \left[ u\otimes u \otimes u\right] 
\nonumber \\
&& \,+\, I_x \left[ u\otimes X \otimes X +X\otimes u \otimes X + X \otimes X \otimes u\right] 
\nonumber \\ 
&& \,+\,  I_y  \left[ u\otimes Y \otimes Y +Y\otimes u \otimes Y + Y\otimes Y \otimes u\right]
\nonumber \\
&& \,+\, I_z \left[ u\otimes Z \otimes Z + Z\otimes u \otimes Z + Z\otimes Z \otimes u\right] ,
 \label{eq:Theta}
\ea
where we have already made use of the 0+1d symmetries. All other possible combinations vanish by symmetry.  $I_u$ and $I_i$ can be found by appropriate projections, for example, $I_u$ can be obtained by taking $u_\mu u_\nu u_\lambda $ of $I^{\mu\nu\lambda}$ as follows
\be
I_u = u_\mu u_\nu u_\lambda I^{\mu\nu\lambda}=\int dP E^3 f \, .
\ee
We note that, for the 0+1d case, the corresponding basis vectors in the lab frame can be obtained using a longitudinal boost with the boost velocity equal to the spatial rapidity $\varsigma = {\rm arctanh}(z/t)$ since the system is boost-invariant.  This gives
\ba
u^\mu &=& (\cosh\varsigma,0,0,\sinh\varsigma) \, , \nonumber \\
X^\mu &=& (0,1,0,0) \, , \nonumber \\
Y^\mu &=& (0,0,1,0) \, , \nonumber \\
Z^\mu &=& (\sinh\varsigma,0,0,\cosh\varsigma) \, .
\label{eq:0+1dbasis}
\ea
We will make use of these forms in the forthcoming discussion.

\subsection{The conformal 0+1d aHydro distribution and bulk variables}

To proceed one must specify a form for the distribution function which appears in the various integrals above.  In anisotropic hydrodynamics one allows the one-particle distribution function in the LRF to be intrinsically momentum-space anisotropic.  For this introductory presentation, we are considering a 0+1d system, in which case it suffices to introduce a single momentum-space anisotropy parameter $\xi$, as follows,
\be
f(x,p) = f_{\rm RS}(x,p) = f_{\rm eq}\left( \frac{\sqrt{{\bf p}^2 + \xi(x) p_z^2}}{\Lambda(x)},\frac{\mu(x)}{\Lambda(x)} \right) ,
\label{eq:rsform}
\ee
where $f_{\rm RS}$ indicates the Romatschke-Strickland form \cite{Romatschke:2003ms,Romatschke:2004jh}, $f_{\rm eq}(\hat{E}) = 1/[\exp(\hat{E})+a]$ is an equilibrium distribution function with $a=0,1,$ and $-1$ corresponding to classical, Fermi-Dirac, and Bose-Einstein statistics, respectively, $\xi(x)$ is the local anisotropy parameter, $\Lambda(x)$ is the local scale parameter which reduces to the temperature in the isotropic limit, $\xi(x) \rightarrow 0$, and $\mu(x)$ is the local chemical potential.  Note that all momenta appearing above are specified in the LRF of the system.  We will formulate this in an explicitly Lorentz-invariant manner in a forthcoming section, but for now we will continue with this form.  In what follows, we will additionally assume zero chemical potential, $\mu=0$.

Using this form, it is possible to evaluate the necessary moments of the distribution function analytically.  For example, in the conformal case considered in this section, the number density in the LRF can be factorized into a function that depends solely on $\xi$ and another function which depends solely on the scale $\Lambda$
\ba
n(\xi,\Lambda) &\equiv& \int \frac{d^3 p}{(2\pi)^3 } f_{\rm eq}\left(\sqrt{{\bf p}^2 + \xi(x) p_z^2}/\Lambda(x)\right) \nonumber \\
&=& \frac{1}{\sqrt{1+\xi}} \int \frac{d^3 p}{(2\pi)^3 } f_{\rm eq}\left(|{\bf p}|/\Lambda(x)\right) \nonumber \\
&=& \frac{1}{\sqrt{1+\xi}} n_{\rm eq}(\Lambda) \, ,
\label{eq:0+1dna}
\ea
where, in order to evaluate the integral, we have made a change of variables to $\bar{p}_z \equiv \sqrt{1+\xi}\,p_z$, then relabeled $\bar{p}_z \rightarrow p_z$, and recognized that the remaining integral is nothing but the isotropic number density evaluated at the momentum scale $\Lambda$.  Similarly, the components of the energy-momentum tensor can be evaluated analytically, e.g. the LRF energy density can be factorized
\ba
\epsilon &=& \int dP E^2 f_{\rm eq}\left(\sqrt{{\bf p}^2 + \xi(x) p_z^2}/\Lambda(x)\right) \nonumber \\
&=& \int \frac{d^3 p}{(2\pi)^3 } \sqrt{p_x^2 + p_y^2 + p_z^2} \, f_{\rm eq}\left(\sqrt{{\bf p}^2 + \xi(x) p_z^2}/\Lambda(x)\right)  \nonumber \\
&=& \frac{1}{\sqrt{1+\xi}} \int \frac{d^3 p}{(2\pi)^3 } |{\bf p}| \sqrt{\sin^2\theta + \frac{\cos^2\theta}{1+\xi}} \, f_{\rm eq}\left(|{\bf p}|/\Lambda(x)\right)  \nonumber \\
&=& \underbrace{\left( \frac{1}{2\sqrt{1+\xi}} \int d(\cos\theta)  \sqrt{\sin^2\theta + \frac{\cos^2\theta}{1+\xi}} \right)}_{\equiv {\cal R}(\xi)} \;
\underbrace{\int \frac{d^3 p}{(2\pi)^3 } |{\bf p}| \, f_{\rm eq}\left(|{\bf p}|/\Lambda(x)\right)}_{= \epsilon_{\rm eq}(\Lambda)} \, .
\ea
Performing the angular integration and repeating this exercise for the transverse and longitudinal pressures given by $T^{xx}=T^{yy}$ and $T^{zz}$, respectively, one obtains
\ba
\epsilon &=& {\cal R}(\xi) \epsilon_{\rm eq}(\Lambda) \, , \nonumber \\
P_T &=& {\cal R}_T(\xi) P_{\rm eq}(\Lambda) \, , \nonumber \\
P_L &=& {\cal R}_L(\xi) P_{\rm eq}(\Lambda) \, , 
\ea
with
\ba
{\cal R}(\xi) &=& \frac{1}{2}\left[\frac{1}{1+\xi}
+\frac{\arctan\sqrt{\xi}}{\sqrt{\xi}} \right] ,
\nonumber
\\
{\cal R}_{T}(\xi) &=& \frac{3}{2 \xi} 
\left[ \frac{1+(\xi^2-1){\cal R}(\xi)}{\xi + 1}\right] ,
\nonumber
\\
{\cal R}_{L}(\xi) &=& \frac{3}{\xi} 
\left[ \frac{(\xi+1){\cal R}(\xi)-1}{\xi+1}\right] ,
\label{eq:Rfuncs}
\ea
which satisfy $3{\cal R} = 2 {\cal R}_T + {\cal R}_L$.  This follows from the fact that the conformal energy-momentum tensor is traceless, ${T^\mu}_\mu = 0$.  

Turning, finally to the rank--three tensor, using the 0+1d aHydro distribution function, one finds
\ba
I_u &=& {\cal S}_u(\xi) I_{\rm eq}(\Lambda) \, , \nonumber \\
I_x = I_y & = & {\cal S}_T(\xi) I_{\rm eq}(\Lambda) \, , \nonumber \\
I_z &=& {\cal S}_L(\xi) I_{\rm eq}(\Lambda) \, ,
\ea
with $I_{\rm eq}(\Lambda) =  \frac{1}{3} \int dP E^3 f_{\rm eq}$ and
\ba
{\cal S}_u(\xi) &=& \frac{3+2\xi}{(1+\xi)^{3/2}} \, , \nonumber \\
{\cal S}_T(\xi) &=& \frac{1}{\sqrt{1+\xi}} \, , \nonumber \\
{\cal S}_L(\xi) &=& \frac{1}{(1+\xi)^{3/2}} \, ,
\ea
which satisfy $2 {\cal S}_T + {\cal S}_L = {\cal S}_u$.

\subsection{The 0+1d equations of motion}

We will now put together the pieces presented in the previous subsections in order to obtain the aHydro equations of motion.  Starting with the zeroth-moment we can use Eqs.~(\ref{eq:0+1dneq}) and (\ref{eq:0+1dn}) to obtain
\be
D_u n + n \theta_u = {\cal C}_0 \, ,
\ee
where $D_u = u^\mu \partial_\mu$ is the co-moving derivative and $\theta_u = \partial_\mu u^\mu$ is the expansion scalar.  Using the 0+1d basis vector (\ref{eq:0+1dbasis}) and transforming to Milne coordinates using $\tau = t \cosh(\varsigma)$ and $z = t \sinh(\varsigma)$, one obtains
\be
\partial_\tau n + \frac{n}{\tau} = {\cal C}_0 \, .
\label{eq:neq}
\ee
In order to reach the final form we will need to specify the collisional kernel.  Following Ref.~\cite{Martinez:2010sc} we will assume that the collisional kernel is given by the RTA form
\be
C[f] = \frac{p \cdot u}{\tau_{\rm eq}} \left[ f - f_{\rm eq}(T) \right] ,
\label{eq:rtadef}
\ee
where $\tau_{\rm eq} = 5\bar\eta/T$ is the relaxation time \cite{Denicol:2010xn,Denicol:2011fa}, which must be inversely proportional to the local (effective) temperature T in the conformal case and $\bar\eta=\eta/s$ is the shear viscosity to entropy density ratio.

As mentioned previously, in order to conserve energy and momentum it is necessary that the first moment of the collisional kernel vanish, i.e. $\int dP \, p^\mu C[f] = 0$.  This is trivially satisfied for $\mu=1,2,3$ in RTA due to the symmetries of $f$ and $f_{\rm eq}$ and for $\nu=0$ it results in the so-called Landau matching condition
\be
\epsilon(\xi,\Lambda) = \epsilon_{\rm eq}(T) \, .
\ee
Using this and Eq.~(\ref{eq:Rfuncs}), in the 0+1d conformal case, one obtains
\be
T = {\cal R}^{1/4}(\xi) \Lambda \, .
\label{eq:conformallandaumatching}
\ee
The temperature determined in this manner will be called the {\em effective temperature}.  In the end it is a stand-in for the local energy density which is well-defined both in and out of equilibrium.

Evaluating ${\cal C}_0$ using Eq.~(\ref{eq:0+1dna}), one obtains
\be
{\cal C}_0 = \frac{n_{\rm eq}}{\tau_{\rm eq}} \left( \frac{1}{\sqrt{1+\xi}} - R^{3/4}(\xi) \right) .
\ee
Again using Eq.~(\ref{eq:0+1dna}), we can expand the left-hand-side of Eq.~(\ref{eq:neq}) in terms of derivatives of $\xi$ and $\Lambda$.  Doing so and simplifying the result gives
\be
 \frac{1}{1+\xi} \partial_\tau \xi - \frac{6}{\Lambda} \partial_\tau \Lambda - \frac{2}{\tau} =  \frac{2}{\tau_{\rm eq}} \left( 1- R^{3/4}(\xi)  \sqrt{1+\xi}  \right) .
\label{eq:zerothmoment}
\ee

Performing a similar manipulation on the first moment equation, starting from the simplified form of energy conservation in 0+1d, i.e.
\be
\label{eq:energydens}
\frac{\partial {\epsilon (\tau)}}{\partial \tau}=-\frac{\epsilon (\tau)+P_L(\tau)}{\tau} \, ,
\ee
one obtains
\be
\frac{{\cal R}'(\xi)}{{\cal R}(\xi)} \partial_\tau \xi + \frac{4}{\Lambda} \partial_\tau \Lambda = 
\frac{1}{\tau} \left[ \frac{1}{\xi(1+\xi){\cal R}(\xi)} - \frac{1}{\xi} - 1 \right] .
\label{eq:firstmoment}
\ee
Note that the three equations related to momentum conservation are, once again, automatically satisfied if the distribution function is reflection symmetric in momentum space.

Finally, turning to the second moment, the $zz$ projection of Eq.~(\ref{eq:generaleom}) with $n=2$ gives
\be
(\log {\cal S}_L)' \partial_\tau \xi + 5 \partial_\tau \!\log \Lambda + \frac{3}{\tau} = \frac{1}{\tau_{\rm eq}} \left[ \frac{{\cal R}^{5/4}}{{\cal S}_L} -1 \right] ,
\label{eq:Izzeq}
\ee
and the $xx$ and $yy$ projections both give
\be
(\log {\cal S}_T)' \partial_\tau \xi + 5 \partial_\tau \!\log \Lambda + \frac{1}{\tau} = \frac{1}{\tau_{\rm eq}} \left[ \frac{{\cal R}^{5/4}}{{\cal S}_T} -1 \right] .
\label{eq:Ixxeq}
\ee
Combining the $zz$ projection minus one-third of the sum of the $xx$, $yy$, and $zz$ projections gives
\be
\frac{1}{1+\xi} \partial_\tau\xi - \frac{2}{\tau} + \frac{{\cal R}^{5/4}(\xi)}{\tau_{\rm eq}} \xi \sqrt{1+\xi} = 0\, .
\label{eq:2ndmomf}
\ee

Summarizing, from the zeroth, first, and second moments of the Boltzmann equation we obtain three equations of motion given by Eqs.~(\ref{eq:zerothmoment}), (\ref{eq:firstmoment}), and (\ref{eq:2ndmomf}).  Of course, we can continue in this manner ad-infinitum to the third moment, etc., however, in practice we only need two equations of motion to evolve $\xi$ and $\Lambda$.  Since energy-momentum conservation is sacrosanct, it must be included in the set.  In addition, since higher moments are sensitive to high-momentum behavior of the distribution and we are looking for equations that describe the long wavelength dynamics, we are guided naturally to consider the lowest possible momentum-moments.  Therefore, based on the equations presented thus far, that leaves two possibilities:  (a) zeroth+first and (b) first+second.  In order to decide which option to use in practice, ones needs to consider the near equilibrium (small anisotropy) limit.  One can show that the option (a) does not reproduce the correct near-equilibrium equations if one uses $\tau_{\rm eq} = 5\bar\eta/T$; however, as we will demonstrate in the next subsection, option (b) automatically reproduces the correct near-equilibrium limit \cite{Tinti:2013vba}.\footnote{In Ref.~\cite{Martinez:2010sc} the authors used option (a), but fixed it by hand by adjusting the relaxation time by a factor two in order to match Israel-Stewart theory in the near-equilibrium limit.}

Note that above we painted an agnostic view concerning whether one should use the zeroth moment of the Boltzmann equation as one of the equations of motion for 0+1d conformal aHydro and concluded that the combination of the first and second moments was more appropriate because the near-equilibrium limit was correctly reproduced.  While this is an accurate statement, we now understand that the zeroth moment should only be used if one is trying to describe the evolution of a system at finite chemical potential.  At finite chemical potential, one must evolve $\mu$, $\Lambda$, and $\xi$ using the zeroth, first, and second moments \cite{Molnar:2016gwq}.  In what follows, we will continue to work at zero chemical potential, in which case it makes sense to proceed with the first and second moments of the Boltzmann equation.\footnote{For a discussion of the inclusion of a chemical potential and the role of the zeroth moment, we refer the reader to Ref.~\cite{Molnar:2016gwq}.}

\subsection{Relation to second-order viscous hydrodynamics in the small anisotropy limit}
\label{sec:smallaniso}

In order to make the connection to standard second-order vHydro, one can rewrite Eq.~(\ref{eq:2ndmomf}) in terms of the single shear stress tensor component $\pi \equiv {\pi^\varsigma}_\varsigma$ necessary for a conformal 0+1d system, i.e. $P_T = P_{\rm eq}(T) + \pi/2$ and $P_L = P_{\rm eq}(T) - \pi$.  To start, we note that the energy conservation equation (\ref{eq:energydens}) can be expressed in terms of $\pi$ as
\be
\tau \partial_\tau  \! \log \epsilon  = -\frac{4}{3} + \frac{\pi}{\epsilon} \, .
\label{eq:firstmom}
\ee 

\begin{figure}[t!]
\centerline{
\includegraphics[width=.475\linewidth]{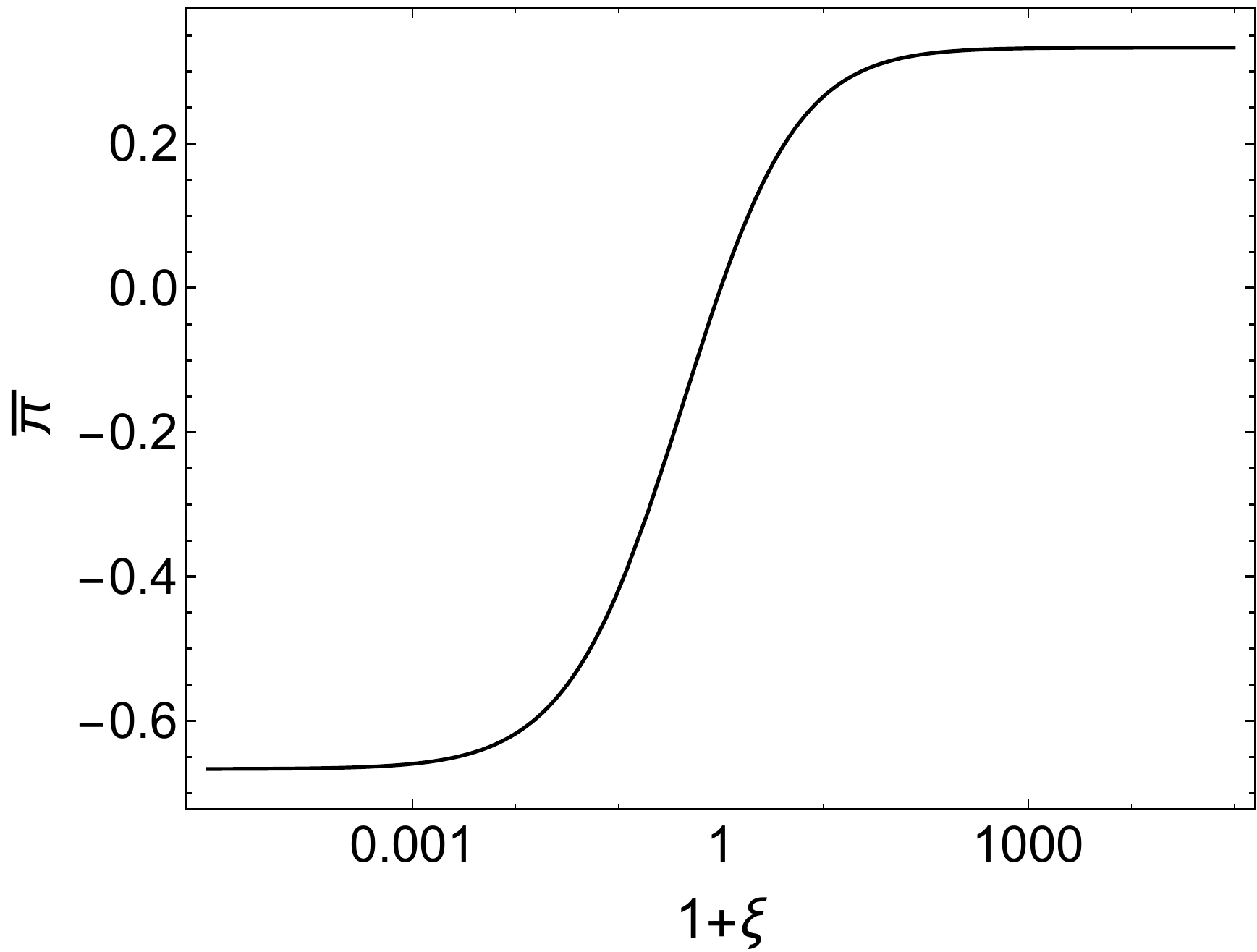}
\hspace{2mm}
\includegraphics[width=.475\linewidth]{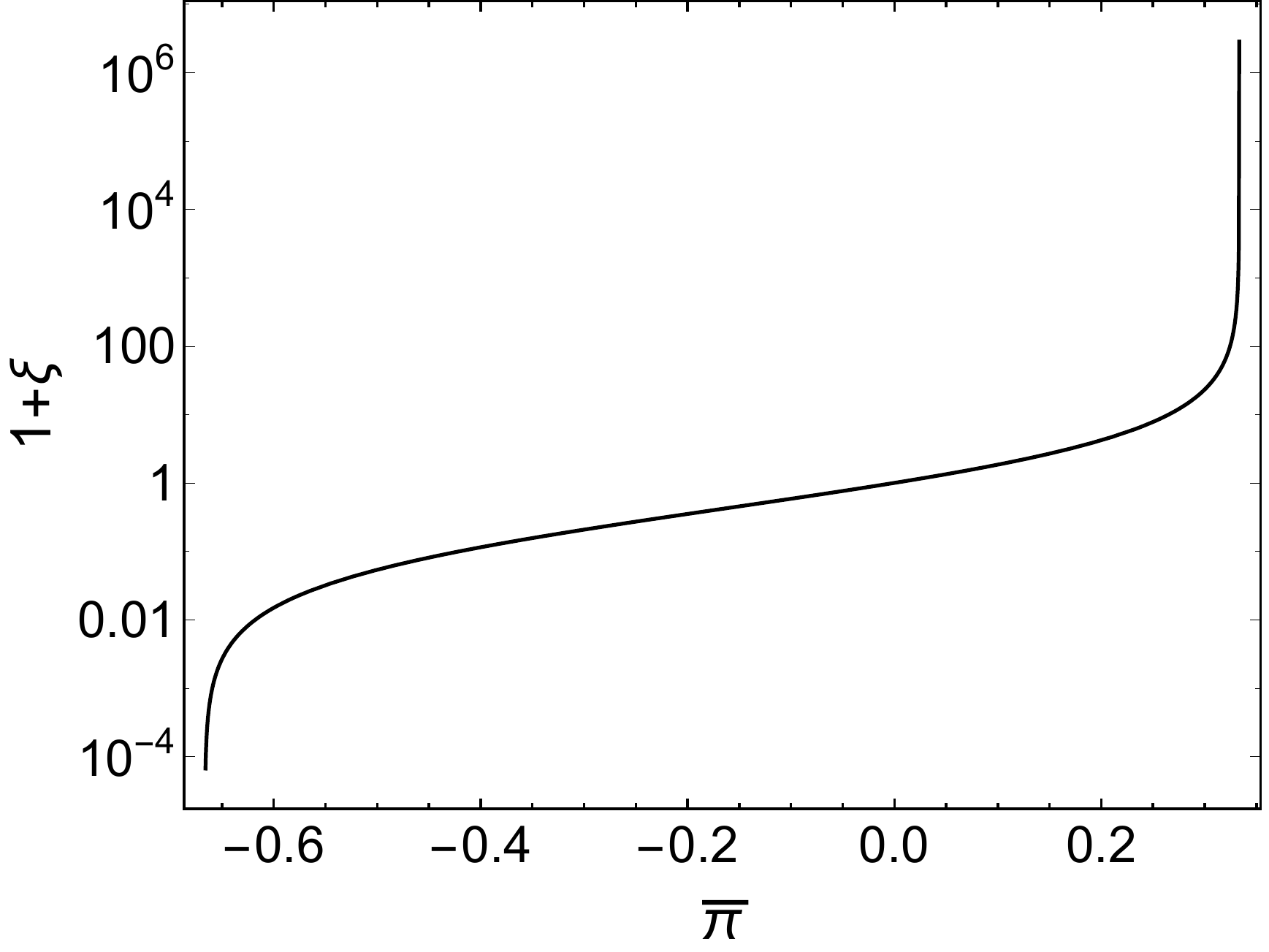}
}
\caption{The left panel shows $\overline{\pi}$ as a function of $\xi$ found using Eq.~(\ref{eq:pixirel}).  The right panel shows $\xi$ as a function of $\overline{\pi}$ found using numerical inversion of Eq.~(\ref{eq:pixirel}).}  
\label{fig:pibar}
\end{figure}

To relate $\pi$ and $\xi$ one can use $\pi = P_{\rm eq} - P_L$, to obtain
\be
\overline\pi(\xi) \equiv \frac{\pi}{\epsilon} = \frac{1}{3} \left[ 1 - \frac{{\cal R}_L(\xi)}{\cal R(\xi)} \right]  .
\label{eq:pixirel}
\ee 
In the left panel of Fig.\ \ref{fig:pibar} we plot $\overline{\pi}$ as a function of $\xi$ determined via Eq.~(\ref{eq:pixirel}) and, in the right panel, we plot $\xi$ as a function of $\overline{\pi}$ determined via numerical inversion of Eq.~(\ref{eq:pixirel}).  Importantly, one observes that in aHydro $\overline\pi$ is bounded, $-2/3 < \overline{\pi} < 1/3$.  This is related to the positivity of the longitudinal and transverse pressures which naturally emerges in this framework. Note that, at leading order in aHydro all pressures are positive, however, if one includes the next-to-leading-order corrections in $\delta\tilde{f}$ then it is possible to also describe negative longitudinal pressures (see Sec.~\ref{sec:vaHydro} for more information).  That said, using a purely particle-based transport model at very early times does not capture the essence of the physics necessary and, instead, one should consider generalizing aHydro to include background chromofields which are evolved self-consistently with the distribution function.  In this way, the chromofield pressure contribution would result in a negative longitudinal pressure in a manner which is closer to the spirit of the colored-glass-condensate framework.

One can easily show that $\overline\pi$ is related to the shear inverse Reynolds number via \cite{Denicol:2012cn} 
\be
R_\pi^{-1} \equiv \frac{\sqrt{\pi^{\mu\nu} \pi_{\mu\nu}}}{P_{\rm eq}} = 3 \sqrt{\frac{3}{2}} |\overline\pi| \, ,
\label{eq:reynoldsnumber}
\ee
from which one can see that the shear inverse Reynolds number measures the relative magnitude of the shear viscous correction to the energy-momentum tensor compared to the isotropic pressure.  We also note that, as a consequence of Eq.~(\ref{eq:reynoldsnumber}), a series in $\overline\pi$ can be loosely understood as a series in $R_\pi^{-1}$.

Using Eq.~(\ref{eq:pixirel}), one can show that
\be
\frac{\partial_\tau\pi}{\epsilon} = \overline\pi^\prime \partial_\tau\xi + \overline\pi \partial_\tau \!\log\epsilon \, ,
\ee
which upon using Eqs.~(\ref{eq:pixirel}) and (\ref{eq:firstmom}) gives
\be
\partial_\tau\xi = \frac{1}{\overline\pi'} \left[ \frac{\partial_\tau\pi}{\epsilon} + \frac{\pi}{\epsilon\tau} \left( \frac{4}{3} - \frac{\pi}{\epsilon}  \right)  \right] ,
\label{eq:xidot2}
\ee
where $\overline\pi' \equiv d\overline\pi/d\xi$.

Plugging (\ref{eq:xidot2}) into (\ref{eq:2ndmomf}), one obtains
\be
\frac{\partial_\tau\pi}{\epsilon} + \frac{\pi}{\epsilon\tau} \left( \frac{4}{3} - \frac{\pi}{\epsilon}  \right) - \left[ \frac{2(1+\xi)}{\tau} - \frac{{\cal H}(\xi)}{\tau_{\rm eq}} \right]\overline\pi'(\xi) = 0\, ,
\label{eq:2ndmomf3}
\ee
with
\be
{\cal H}(\xi) \equiv \xi (1+\xi)^{3/2}{\cal R}^{5/4}(\xi) \, ,
\ee
and the understanding that $\xi = \xi(\overline\pi)$ where $\xi(\overline\pi)$ is the inverse function of $\overline\pi(\xi)$ (see right panel of Fig.~\ref{fig:pibar}).  Written in this form, we can see explicitly that the aHydro second-moment equation sums an infinite number of terms in the expansion in the inverse Reynolds number (\ref{eq:reynoldsnumber}).  This follows because the quantity in square brackets in Eq.~(\ref{eq:2ndmomf3}) is a function that contains all orders in $\xi$ and, hence, $\overline\Pi$.  As we will see subsequently, this is extremely important because the exact attractor possesses a large Reynolds number in the limit $\tau T \rightarrow 0$.   Before proceeding to this, in the next subsection we expand this equation in powers of $\xi$ through second order (near-equilibrium limit) and compare it to standard vHydro.

\subsubsection*{Small-$\xi$ expansion}

In order make the connection to standard vHydro, one can expand Eq.~(\ref{eq:2ndmomf3}) in a Taylor-series in $\xi$ around $\xi=0$. In order to accomplish this we need the $\xi$-expansions of the non-linear functions appearing. To ${\cal O}(\xi^2)$, one finds
\ba
\overline\pi &=& \frac{8}{45} \xi  \left[1 - \frac{13}{21} \xi + {\cal O}(\xi^2) \right] , \nonumber \\
\overline\pi^\prime  &=& \frac{8}{45} \left[1 - \frac{26}{21} \xi + \frac{131}{105} \xi^2 + {\cal O}(\xi^3) \right] , \nonumber \\
(1+\xi)\overline\pi^\prime  &=& \frac{8}{45} \left[1 - \frac{5}{21} \xi + \frac{1}{105} \xi^2 + {\cal O}(\xi^3) \right] , \nonumber \\
{\cal H} &=& \xi + \frac{2}{3} \xi^2 + {\cal O}(\xi^3) \, .
\ea
With this, we can invert the relationship between $\overline\pi$ and $\xi$, obtaining
\be
\xi = \frac{45}{8} \overline\pi \left[ 1 + \frac{195}{56} \overline\pi + {\cal O}(\pi^2)  \right] , 
\ee
which results in
\ba
\overline\pi^\prime  &=& \frac{8}{45} - \frac{26}{21} \overline\pi + \frac{1061}{392} \overline\pi^2 + {\cal O}(\overline\pi^3) \, , \nonumber \\
(1+\xi)\overline\pi^\prime  &=& \frac{8}{45} - \frac{5}{21} \overline\pi - \frac{38}{49} \overline\pi^2 + {\cal O}(\overline\pi^3) \, , \nonumber \\
{\cal H} &=& \frac{45}{8} \overline\pi \left[ 1 + \frac{405}{56} \overline\pi + {\cal O}(\overline\pi^3) \right] ,  \nonumber \\
{\cal H} \overline\pi^\prime &=&  \overline\pi + \frac{15}{56}  \overline\pi^2 + {\cal O}(\overline\pi^3) \, .
\ea

Using this expansion in Eq.~(\ref{eq:2ndmomf3}) and keeping terms through linear order in $\pi$ gives
\be
\partial_\tau\pi - \frac{4 \eta}{3 \tau_\pi \tau} + \frac{38}{21} \frac{\pi}{\tau} = - \frac{\pi}{\tau_\pi} \, , %
\label{eq:2ndmomf4}
\ee
where we have made use of the fact that for a conformal (massless) system one can eliminate the energy density by expressing it in terms of the transport coefficients
\be
\epsilon = \frac{15}{4} \frac{\eta}{\tau_{\rm eq}} \, ,
\ee
and relabeled $\tau_{\rm eq} \rightarrow \tau_\pi$ in order to express the equations in standard second order hydrodynamics form.  Equation~(\ref{eq:2ndmomf4}) agrees with previously obtained RTA second-order vHydro results \cite{Denicol:2010xn,Denicol:2012cn,Denicol:2014loa,Jaiswal:2013vta,Jaiswal:2013npa}, demonstrating that, in the limit of small momentum-space anistropy, aHydro automatically reproduces the correct second-order vHydro equations if one uses the first and second moments of the Boltzmann equation to generate the dynamical equations.\footnote{The reader may be wondering how the agreement between aHydro and vHydro is maintained if one goes beyond leading-order to include the $\delta \tilde{f}$ corrections.  In practice, one finds that equations of motion are affected in such a way that the near-equilibrium vHydro limit is unaffected, however, agreement with known exact solutions of the Boltzmann equation is improved when including the residual anisotropy tensor corrections.  For more information concerning next-to-leading order aHydro, the vHydro limit, and comparisons to available exact solutions see Secs.~\ref{sec:vaHydro} and \ref{sec:exactsolutions}.}

\subsection{Ideal and free streaming limits}

In the previous subsection we proved that, using the first and second moments of the Boltzmann equation, aHydro reduces to vHydro (vHydro) in the small anisotropy limit.  In RTA, the small anisotropy limit is appropriate when the relaxation time of the system $\tau_{\rm eq} = 5 \bar\eta/T$ is very small, which occurs in the limit that $\bar\eta \rightarrow 0$ (or $T \rightarrow \infty$).  From the previous subsection it is easy to see that when $\bar\eta \rightarrow 0$ and hence $\tau_{\rm eq} = \tau_{\pi} \rightarrow 0$, the aHydro equations reproduce the equation of motion of ideal hydrodynamics \cite{Martinez:2010sc}.  Importantly, however, the equations also contain the free streaming (FS) limit.  To see this, we consider the opposite limit, namely the limit $\bar\eta \rightarrow \infty$, which corresponds to $\tau_{\rm eq} \rightarrow \infty$.  

Taking the $\tau_{\rm eq} \rightarrow \infty$ limit in Eq.~(\ref{eq:2ndmomf}) one obtains
\ba
&& \frac{1}{1+\xi} \partial_\tau\xi = \frac{2}{\tau}\, , 
\label{eq:fsxi}
\ea
which has a solution of the form
\be
\xi_{\rm FS} = (1 + \xi_0) \left(\frac{\tau}{\tau_0} \right)^2 -1 \, .
\ee
Using (\ref{eq:fsxi}), the energy conservation equation (\ref{eq:firstmoment}) then becomes
\be
\partial_\tau \Lambda = 0 \, ,
\ee
which tells us that in the FS limit $\Lambda_{\rm FS} = \Lambda_0$.  These solutions for $\xi$ and $\Lambda$ correspond precisely with the analytic result for the case of 0+1d free streaming \cite{Baym:1984np,Mauricio:2007vz,Martinez:2009mf,Martinez:2009ry}.  The fact that aHydro can reproduce the ideal limit, the free streaming limit, and second-order vHydro in the limit of small $\eta/s$ makes it a unique approach to dissipative dynamics.  Typically, one must rely on arguments based on near-equilibrium limits to obtain fluid dynamical equations, however, aHydro shows that it may also be possible to describe certain classes of far-from-equilibrium dynamics using an optimized fluid-dynamical approach.  Note, however, demonstrating that aHydro reproduces the free streaming limit has only been proven for Bjorken and Gubser flows.  It would be interesting to see if a general proof could be constructed, but at this point in time, we are not aware of such a proof.

\begin{figure}[t]
\centerline{
\includegraphics[width=.495\linewidth]{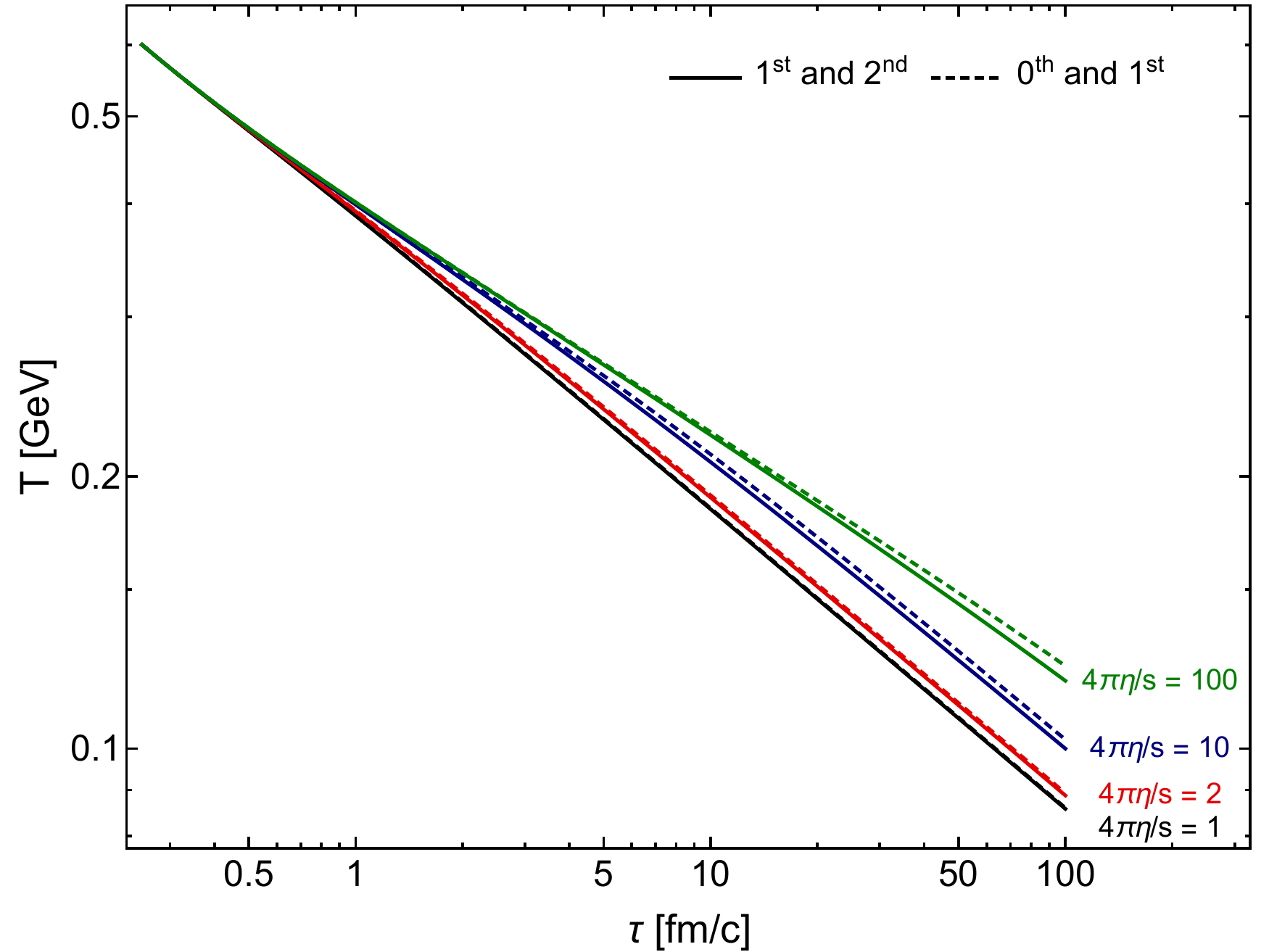}
\hspace{2mm}
\includegraphics[width=.495\linewidth]{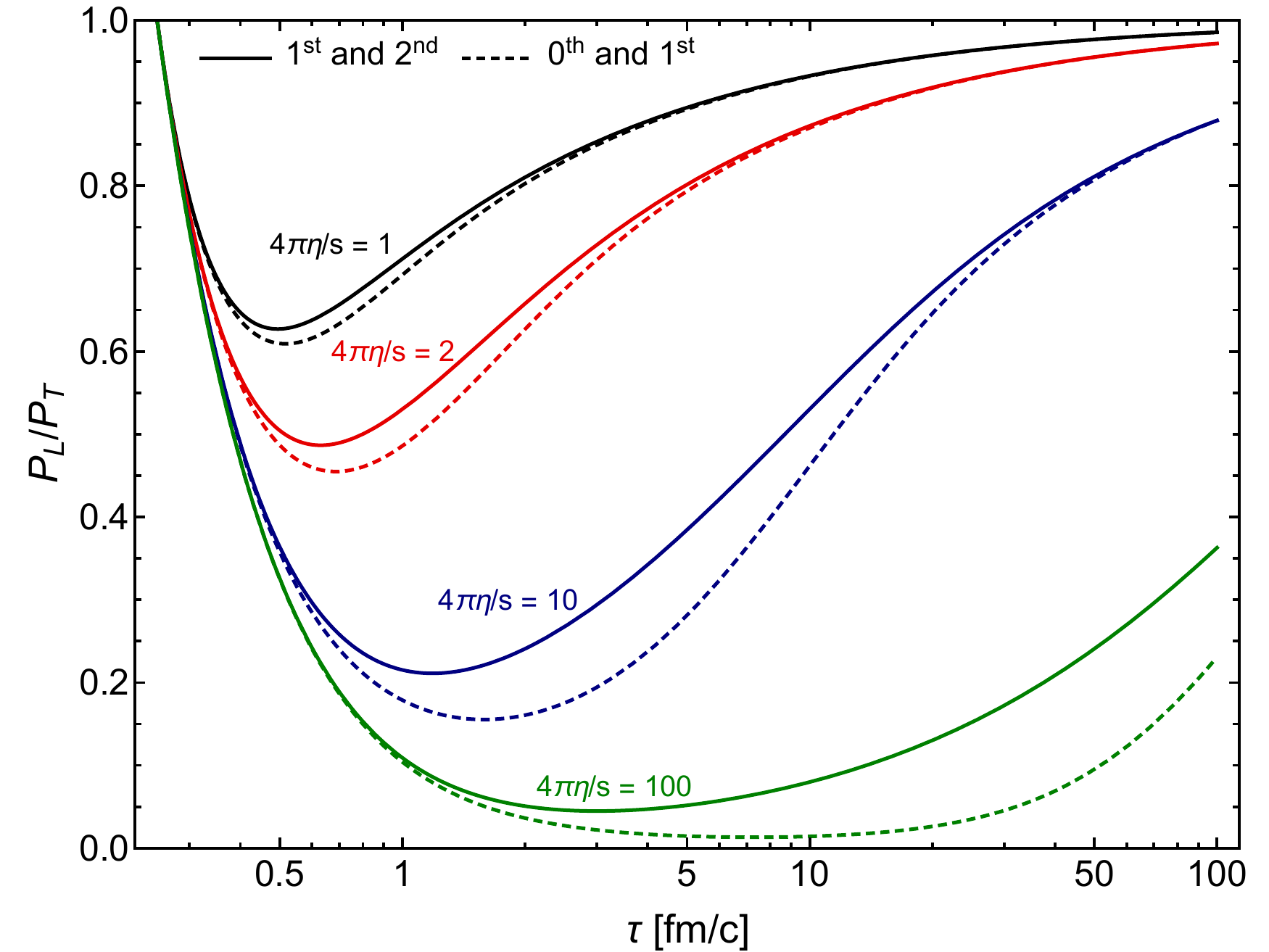}
}
\caption{(Color online) The effective temperature $T$ (left) and the 0+1d pressure anisotropy  $P_L/P_T$ (right) versus proper time predicted by two versions of aHydro:  (i) Using the first and second moments (solid lines) \cite{Tinti:2013vba} and (ii) using the the zeroth and first moment prescription of Ref.~\cite{Martinez:2010sc} (dashed lines).  The initial conditions for both schemes were $T_0 = 600$ MeV, $\xi_0 = 0$, $\tau_0 = 0.25$ fm/c.   }  
\label{fig:01compare}
\end{figure}

\begin{figure}[t]
\centerline{
\includegraphics[width=.495\linewidth]{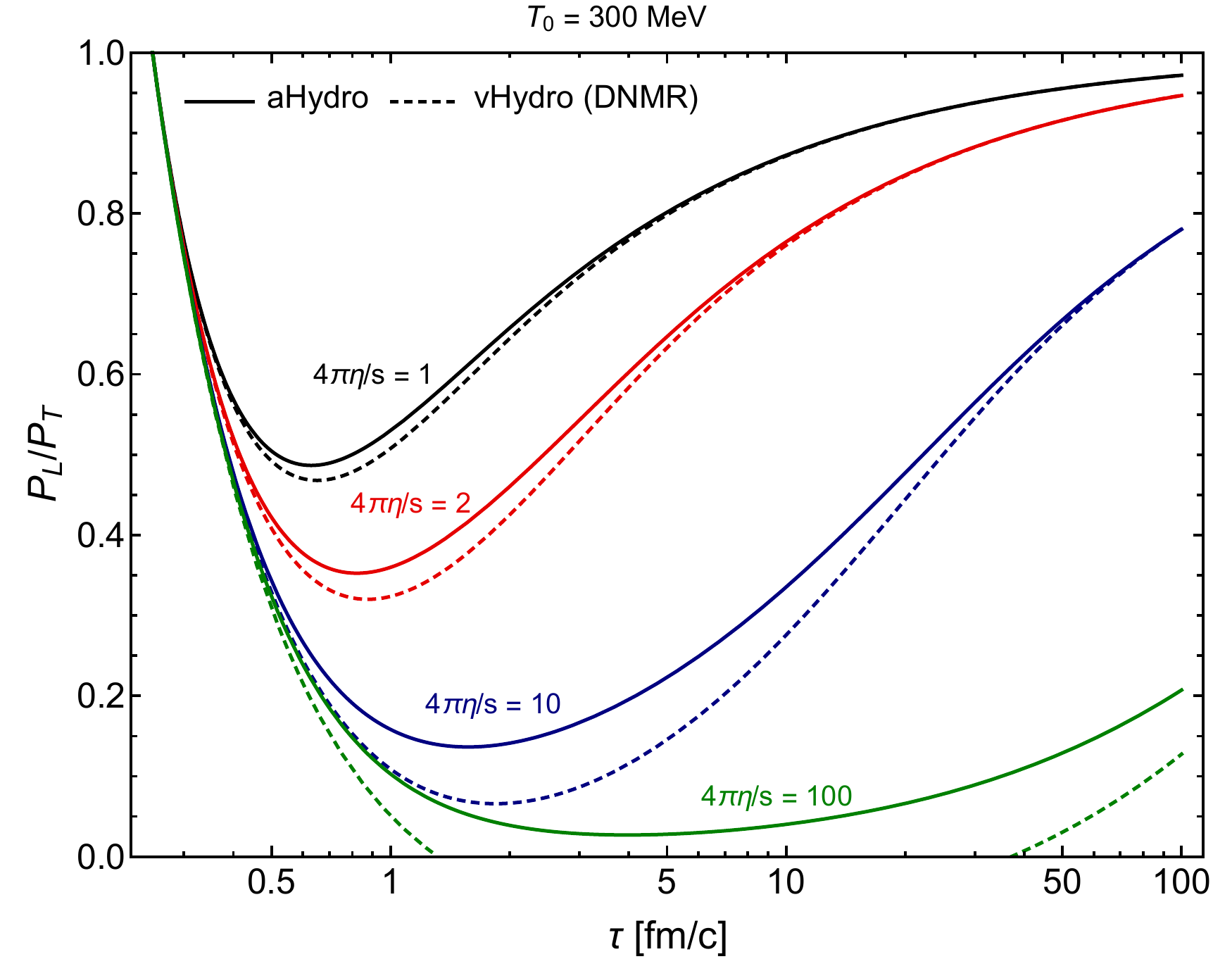}
\hspace{2mm}
\includegraphics[width=.495\linewidth]{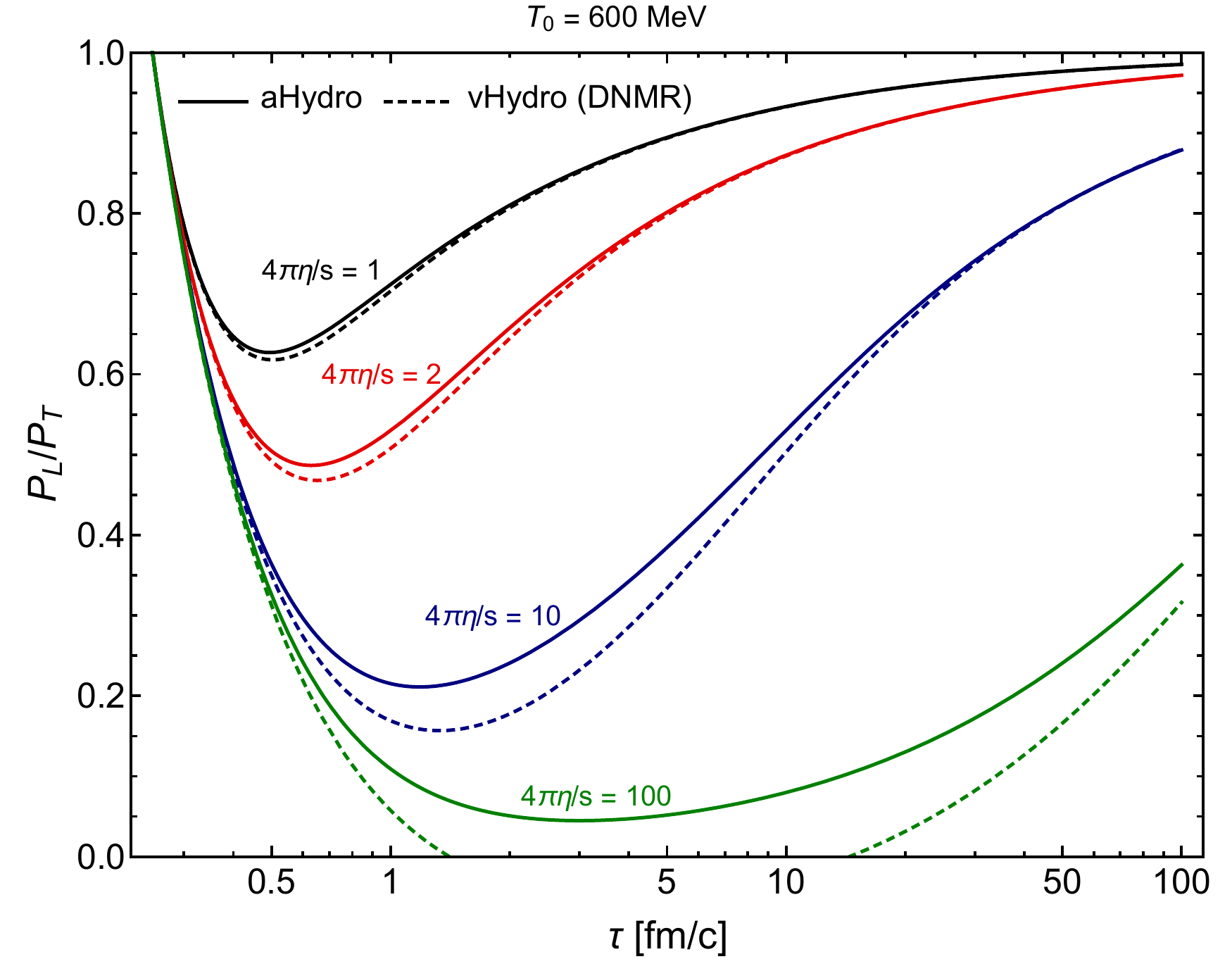}
}
\caption{(Color online) A comparison of the 0+1d pressure anisotropy  $P_L/P_T$ versus proper time predicted by aHydro and DNMR vHydro for $T_0 = 600$ MeV (left) and $T_0 = 300$ MeV (right) with $\tau_0 = $ 0.25 fm/c.  In both panels we took the system to be initially isotropic corresponding to $\xi_0 = 0$ and $\pi_0=0$ for aHydro and vHydro, respectively.  The four sets of solid and dashed lines correspond to $4\pi\eta/s \in \{1,2,10,100\}$.
}  
\label{fig:01compare2}
\end{figure}

\subsection{aHydro numerical solution and comparison to vHydro}

Next, we turn to the numerical solution of the 0+1d conformal aHydro and vHydro equations.  In Fig.~\ref{fig:01compare} we plot the effective temperature and pressure anisotropy ($P_L/P_T$) as a function of proper time ($\tau$).  For these figures we used an initial condition of $T_0 = 600$ MeV, $\xi_0 = 0$, $\tau_0 = 0.25$ fm/c.  The solid lines were generated using Eqs.~(\ref{eq:firstmoment}) and (\ref{eq:2ndmomf}) with $\tau_{\rm eq} = 5 \bar\eta/T$ and the effective temperature given by Eq.~(\ref{eq:conformallandaumatching}).  The dashed lines were generated using Eqs.~(\ref{eq:zerothmoment}) and (\ref{eq:firstmoment}) with $\tau_{\rm eq} = 5 \bar\eta/(2T)$ using the original prescription in Ref.~\cite{Martinez:2010sc}.  As can be seen from the left panel the temperature evolution predicted by both aHydro schemes is quite similar, however, the pressure anisotropy predicted by the two schemes is quantitatively different, particularly at large values of $\eta/s$.  That being said, comparing the qualitative aspects we find similar predictions from both schemes.  Since the scheme which uses 1st and 2nd moments automatically reproduces the near-equilibrium limit it is natural to use this scheme if one uses the method of moments of the Boltzmann equation.\footnote{Once again we remind the reader that we now understand that it is appropriate to use the zeroth moment only if one is considering finite chemical potential in which it can be used to obtain an equation of motion for the chemical potential.}

In Fig.~\ref{fig:01compare2}, we compare the 0+1d pressure anisotropy  $P_L/P_T$ as a function of proper time predicted by the aHydro 1st and 2nd moment scheme and Denicol-Niemi-Molnar-Rischke (DNMR) vHydro \cite{Denicol:2012cn} for $T_0 = 300$ MeV (left) and $T_0 = 600$ MeV (right) with $\tau_0 = $ 0.25 fm/c.  For DNMR vHydro we numerically solved Eqs.~(\ref{eq:firstmom}) and (\ref{eq:2ndmomf4}).  DNMR vHydro is a complete second-order treatment which is based on kinetic theory.  In both panels of Fig.~\ref{fig:01compare2} we took the system to be initially isotropic in momentum space, corresponding to $\xi_0 = 0$ and $\pi_0=0$ for aHydro and vHydro, respectively.  As one can see from this figure, aHydro and vHydro agree qualitatively concerning the magnitude of the pressure anisotropy.  Comparison of the left and right panels demonstrates that lower initial temperatures result in larger deviations from isotropy.  This is to be expected since the relaxation time scales inversely with the local effective temperature.  Finally, we note that there are sizable quantitative differences between aHydro and vHydro evolutions.  Importantly, we note that in both panels, the vHydro pressure ratio is observed to go negative for large values of $\eta/s$.  This is indicative of a complete breakdown of vHydro which is not surprising when $\eta/s$ is large since traditional vHydro are based on a linearization around local isotropic equilibrium\footnote{Traditional viscous hydrodynamics is a linearization since a dual expansion and truncation in Knudsen number and inverse Reynolds number is performed.  This procedure is expected to work only very close to equilibrium.}  and the corrections are proportional to this ratio.  We note, however, that at even lower temperatures one sees this breakdown for even small values of $\eta/s$.  Finally, we note that apart from this apparent violation of positivity of the longitudinal pressure, it's hard to say whether aHydro or vHydro provide a more quantitatively reliable description of the system's evolution.  In a forthcoming section, we will compare the various schemes developed with exact solutions of the Boltzmann equation which are available in some simple situations.  From these comparisons we will learn that the aHydro framework provides the most quantitatively reliable method.

\subsection{Summary}

In this section, we presented the basic ingredients of the aHydro formalism in the case of a conformal system which is transversally homogeneous and boost invariant (0+1d).  We demonstrated that aHydro reproduces the ideal hydrodynamics and free streaming limits.   In addition, we demonstrated that it reproduces the correct equations of second-order vHydro in the limit of small anisotropy.  This sets the stage for (a) extending the formalism to 3+1d by relaxing all of the symmetries assumed in the prior subsection and (b) applying the formalism to non-conformal (massive) gases.  In the next section, we will do this at leading order in the aHydro expansion.


\section{3+1d leading-order aHydro for non-conformal systems}
\label{sec:lonoconformal}

In this section we will present 3+1d aHydro for a non-conformal QGP at leading-order in the aHydro expansion.  By ``leading-order" we simply mean that we take into account possible momentum-space anisotropies by allowing for a generalized ellipsoidal form for the one-particle distribution function and ignore any possible deviations from the assumed form.  In aHydro, one assumes that the full one-particle distribution function is given by a leading-order term of generalized Romatschke-Strickland form \cite{Romatschke:2003ms,Romatschke:2004jh}
plus a correction term which accounts for deviations from the generalized ellipsoidal form
\be
f(x,p) = f_{\rm eq}\!\left(\frac{1}{\lambda}\sqrt{p_\mu \Xi^{\mu\nu} p_\nu}, \frac{\mu}{\lambda} \right) + \delta \tilde{f} \, ,
\label{eq:genf}
\ee
where $\lambda$ is an energy scale which becomes the temperature in the isotropic equilibrium limit and $\mu$ is the chemical potential.\footnote{We have called the scale $\lambda$ here to emphasize that it not necessarily equal to the momentum scale $\Lambda$ associated with the canonical Romatschke-Strickland form used in the previous section.  The explicit relation between the two scales can be found in Sec.~IIC of Ref.~\cite{Nopoush:2014pfa}.} The anisotropy tensor has the form $\Xi^{\mu\nu} \equiv u^\mu u^\nu + \xi^{\mu\nu} - \Delta^{\mu\nu} \Phi$ where $\xi^{\mu \nu}$ is a symmetric traceless tensor obeying $ u_\mu \xi ^{\mu \nu} = 0$ and $ {\xi^\mu}_\mu = 0 $, $\Phi$ is the bulk degree of freedom, and $\Delta^{\mu\nu}$ is the transverse projector defined in the conventions and notation block in the beginning of the review \cite{ Martinez:2012tu, Nopoush:2014pfa}. Using the ellipsoidal form (\ref{eq:genf}) and the tracelessness of $\xi^{\mu\nu}$, we are left with six independent parameters out of the seven original parameters $\Phi$, $\Xi_{ii}$, and $\Xi_{ij}=\Xi_{ji}$. Combining these six with the three independent parameters which describe $u^\mu$ and the one for the momentum scale $\lambda$, we arrive at ten degrees of freedom, which suffice to describe the dynamics of the ten independent components of the energy-momentum tensor.  In thermal equilibrium, the distribution function $f_{\rm eq}(x)$ can be identified as Fermi-Dirac, Bose-Einstein, or Maxwellian distribution. Unless otherwise indicated, we will take the Boltzmann form.  Finally, in this section, following the leading-order aHydro model, we ignore the non-ellipsoidal and non-exponential deviations accounted for by $\delta\tilde{f}$.  We will return to this issue in Sec.~\ref{sec:vaHydro} where we will discuss second-order aHydro (dubbed vaHydro in the literature) in which one uses orthonormal polynomial expansions of $\delta{\tilde f}$ similar to standard vHydro to compute these corrections systematically.

In what follows in this section, we will consider a 3+1d system consisting of particles with a temperature independent mass following Ref.~\cite{Nopoush:2014pfa}.  The introduction of the mass scale will allow us to study bulk viscous corrections in the context of aHydro.  For simplicity, we will assume vanishing chemical potential herein.

\subsection{Basis vectors}

As mentioned previously, the lab frame basis vectors for a general 3+1d system can be obtained by a set of Lorentz transformations applied to the LRF basis vectors \cite{Martinez:2012tu, Ryblewski:2010ch}. The set of Lorentz transformations correspond to a longitudinal boost by $\vartheta$ along the beam line, a rotation by \mbox{$\varphi\equiv \tan^{-1}(u_y/u_x)$} around the beam line, and a transverse boost $\theta_\perp$, which together yield
\ba
u^\mu &\equiv& (u_0 \cosh\vartheta,u_x,u_y,u_0 \sinh\vartheta) \, , \nonumber\\
X^\mu &\equiv& \Big(u_\perp\cosh\vartheta,\frac{u_0 u_x}{u_\perp},\frac{u_0 u_y}{u_\perp},u_\perp\sinh\vartheta\Big) , \nonumber \\ 
Y^\mu &\equiv& \Big(0,-\frac{u_y}{u_\perp},\frac{u_x}{u_\perp},0\Big)  , \nonumber \\
Z^\mu &\equiv& (\sinh\vartheta,0,0,\cosh\vartheta ) \, ,
\label{eq:4vectors}
\ea
where $u_\perp\equiv \sqrt{u_x^2+u_y^2}=\sqrt{u_0^2-1} = \sinh\theta_\perp$ and $u_0 = \cosh\theta_\perp$ is the Lorentz factor associated with the transverse boost.  Note that these basis vectors reduce to the 0+1d form specified in Eq.~(\ref{eq:0+1dbasis}) under the assumption of transverse homogeneity ($u_y \rightarrow 0$ followed by $u_x \rightarrow 0$) and boost invariance ($\vartheta \rightarrow \varsigma$).

\subsection{Diagonal ellipsoidal form}

In order to simplify the formalism, in this section we will make an additional assumption, namely that the anisotropy tensor $\Xi^{\mu\nu}$ is diagonal in the local rest frame, i.e.~$\Xi^{\mu\nu}_{\rm LRF} = {\rm diag}(0,\xi_x,\xi_y,\xi_z)$.  This assumption is exact for central collisions if smooth Glauber-like initial conditions are used, however, would be violated if event-by-event fluctuations are included.  In general, we must also consider the effect of off-diagonal anisotropies.  For non-central collisions, results from standard vHydro simulations find that the off-diagonal components of the shear are small \cite{Song:2009gc}, which suggests that one can treat the off-diagonal anisotropies perturbatively.  In forthcoming sections (Secs.~\ref{sec:anisomatching} and \ref{sec:vaHydro}, respectively), we will discuss how to relax this assumption in the context of leading- and second-order aHydro.

Due to the tracelessness of the $\xi^{\mu\nu}$ tensor, one has $\xi_x + \xi_y + \xi_z = 0$.  Expanding the argument of the square root appearing in Eq.~(\ref{eq:genf}) in the LRF gives
\be
f(x,p) = f_{\rm eq}\!\left(\frac{1}{\lambda} \sqrt{p_\mu \Xi^{\mu\nu} p_\nu} \right) 
=  f_{\rm eq}\!\left(\frac{1}{\lambda}\sqrt{\sum_i \frac{p_i^2}{\alpha_i^2} + m^2}\right)  \, ,
\label{eq:fform}
\ee
where $i\in \{x,y,z\}$ and we have introduced some more convenient anisotropy parameters
\be
\alpha_i \equiv (1 + \xi_i + \Phi)^{-1/2} \, .
\label{eq:alphadef}
\ee  
This set of three anisotropy parameters $\alpha_i$ replace the two independent components of $\boldsymbol\xi$ and the degree of freedom $\Phi$.  Using Eq.~(\ref{eq:alphadef}) and ${\xi^\mu}_\mu=0$, one has
\be
\Phi = \frac{1}{3} \sum_i \alpha_i^{-2} - 1 \, .
\ee

\subsection{Dynamical equations}

Assuming a diagonal anisotropy tensor, the energy-momentum tensor can be expressed as
\be
T^{\mu\nu}=\epsilon u^\mu u^\nu+P_x X^\mu X^\nu+P_y Y^\mu Y^\nu+P_z Z^\mu Z^\nu \, .
\label{eq:T-expan}
\ee
The associated energy density and pressures can be expressed as
\ba
\epsilon  &=& {\cal H}_{3}({\boldsymbol\alpha},\hat{m}) \, \lambda^4 \, ,\nonumber \\
P_i &=& {\cal H}_{3i}({\boldsymbol\alpha},\hat{m}) \, \lambda^4 \, ,
\label{eq:E_P_difenitions}
\ea
with $ i \in \{x,y,z\}$ and the ${\cal H}$-functions are
\ba 
{\cal H}_3({\boldsymbol\alpha},\hat{m}) &\equiv&\tilde{N} \alpha \int d^3\hat{p} \, R(\hat{\bf p},\boldsymbol\alpha)  \, f_{\rm eq}\!\left(\!\sqrt{\hat{p}^2 + \hat{m}^2}\right) , \\
{\cal H}_{3i}({\boldsymbol\alpha},\hat{m}) &\equiv& \tilde{N} \alpha \, \alpha_i^2 \int d^3\hat{p} \, R_i(\hat{\bf p},\boldsymbol\alpha) \, f_{\rm eq}\!\left(\!\sqrt{\hat{p}^2 + \hat{m}^2}\right) ,
\label{eq:hfuncs3}
\ea
where $\boldsymbol\alpha = (\alpha_x,\alpha_y,\alpha_z)$, $\hat{p}=p/\lambda$, $\hat{m}=m/\lambda$, $i \in \{x,y,z\}$, $\alpha \equiv \prod_i \alpha_i$, and $\tilde{N}\equiv N_{\rm dof}/(2\pi)^3$.  The $R$ and $R_i$ functions appearing above are
\ba 
R(\hat{\bf p},\boldsymbol\alpha) &\equiv& \sqrt{\alpha_x^2 \,  \hat{p}_x^2+\alpha_y^2 \,  \hat{p}_y^2+\alpha_z^2 \,   \hat{p}_z^2+\hat{m}^2} \, , \\
R_i(\hat{\bf p},\boldsymbol\alpha) &\equiv& \frac{\hat{p}_i^2}{R(\hat{\bf p},\boldsymbol\alpha)} \, .
\ea
More details concerning the ${\cal H}$-functions and their efficient evaluation can be found Refs.~\cite{Nopoush:2014pfa,Alqahtani:2015qja,Alqahtani:2016rth,Alqahtani:2017tnq}.  Note that, in the isotropic limit, $\boldsymbol\alpha \rightarrow {\bf 1}$, and assuming Boltzmann statistics, one has $\lambda \rightarrow T$ and
\ba
{\cal H}_3 &\rightarrow& {\cal H}_{3,{\rm eq}}(1,\hat{m}_{\rm eq}) = 4 \pi \tilde{N} \hat{m}_{\rm eq}^2
 \Big[ 3 K_{2}\left( \hat{m}_{\rm eq} \right) + \hat{m}_{\rm eq} K_{1} \left( \hat{m}_{\rm eq} \right) \Big] , \\
{\cal H}_{3i} &\rightarrow& {\cal H}_{3i,{\rm eq}}(1,\hat{m}_{\rm eq}) = 4 \pi \tilde{N} \hat{m}_{\rm eq}^2 K_2\left( \hat{m}_{\rm eq}\right) ,
\ea
where $\hat{m}_{\rm eq} =  m/T$.

\subsubsection{First moment}

The first moment of Boltzmann equation results in four equations
\ba
D_u\epsilon +\epsilon \theta_u+ P_x u_\mu D_xX^\mu+ P_y u_\mu D_yY^\mu +P_z u_\mu D_zZ^\mu &=&0\, , \nonumber\\
D_x P_x+P_x\theta_x -\epsilon X_\mu D_uu^\mu -P_y X_\mu D_yY^\mu - P_z X_\mu D_zZ^\mu &=& 0\,, \nonumber\\
D_y P_y+P_y \theta_y-\epsilon Y_\mu D_uu^\mu -P_x Y_\mu D_xX^\mu - P_z Y_\mu D_zZ^\mu  &=& 0\,, \nonumber\\
D_z P_z+P_z \theta_z-\epsilon Z_\mu D_uu^\mu- P_x Z_\mu D_xX^\mu - P_y Z_\mu D_yY^\mu &=& 0 \, ,
\label{eq:1stmoment}
\ea
where $D_u \equiv u^\mu \partial_\mu$, $D_x \equiv X^\mu \partial_\mu$, $D_y \equiv Y^\mu \partial_\mu$, and $D_z \equiv Z^\mu \partial_\mu$.  Likewise, the expansion scalars are $\theta_u = \partial_\mu u^\mu$, $\theta_x = \partial_\mu X^\mu$, $\theta_y = \partial_\mu Y^\mu$, and $\theta_z = \partial_\mu Z^\mu$.  Explicit expressions for these derivative operators and expansion scalars can be found in Refs.~\cite{Nopoush:2014pfa,Alqahtani:2015qja,Alqahtani:2016rth,Alqahtani:2017tnq}.

\subsubsection{Second moment}

Using the diagonal ellipsoidal form, the tensor-basis coefficient functions for the second moment (\ref{eq:Theta}) are~\cite{Nopoush:2014pfa}
\ba
I_i(\boldsymbol\alpha,\lambda,m)  &=& \alpha \, \alpha_i^2 \, I_{\rm eq}(\lambda,m) \, , \nonumber \\ 
I_{\rm eq}(\lambda,m) &=&  4 \pi {\tilde N} \lambda^5 \hat{m}^3 K_3(\hat{m}) \, ,
\ea
where $\alpha = \prod_i \alpha_i$ and $K_3$ is a modified Bessel function of the second kind.

Using the diagonal form of the anisotropy tensor, there are seven independent microscopic variables, $\alpha_x$, $\alpha_y$, $\alpha_z$, $u_x$, $u_y$, $\vartheta$, and $\lambda$ and, additionally, one must use the Landau-matching condition to determine the effective temperature $T$.  Therefore, we need only eight equations in order to evolve the system.   Since there are, in general, ten independent tensor components of the second moment of Boltzmann equation, we need to select an appropriate reduced set of components to evolve.  For this purpose, we take the three diagonal projections of the equation of motion of the third moment, i.e. $X_\mu X_\nu \partial_\alpha I^{\alpha \mu \nu}$, $Y_\mu Y_\nu \partial_\alpha I^{\alpha \mu \nu}$, and $Z_\mu Z_\nu \partial_\alpha I^{\alpha \mu \nu}$ giving~\cite{Nopoush:2014pfa}
\be
D_u I_i + I_i (\theta_u + 2 u_\mu D_i X_i^\mu)
= \frac{1}{\tau_{\rm eq}} \Big[ I_{\rm eq}(T,m) - I_i \Big] ,
\label{eq:2ndmoment} 
\ee
where we have specialized to the case of the RTA collisional kernel and $i \in \{x,y,z\}$.

The effective temperature $T$ appearing above can be determined using the requirement of energy-momentum conservation, which results in the following matching condition between the non-equilibrium kinetic energy density and the equilibrium energy density
\be
{\cal H}_3({\boldsymbol\alpha},\hat{m}) \lambda^4 = {\cal H}_{3,\rm eq}(1,\hat{m}_{\rm eq}) T^4.
 \label{eq:matching}
\ee
In the end, for diagonal ellipsoidal 3+1d aHydro evolution, we have eight equations resulting from the first  (\ref{eq:1stmoment})  and second (\ref{eq:2ndmoment})  moments of the Boltzmann equation, together with the matching condition (\ref{eq:matching}).

\subsection{Numerical solution in the 0+1d limit}

\begin{figure}[t]
\centerline{\includegraphics[angle=0,width=0.6\textwidth]{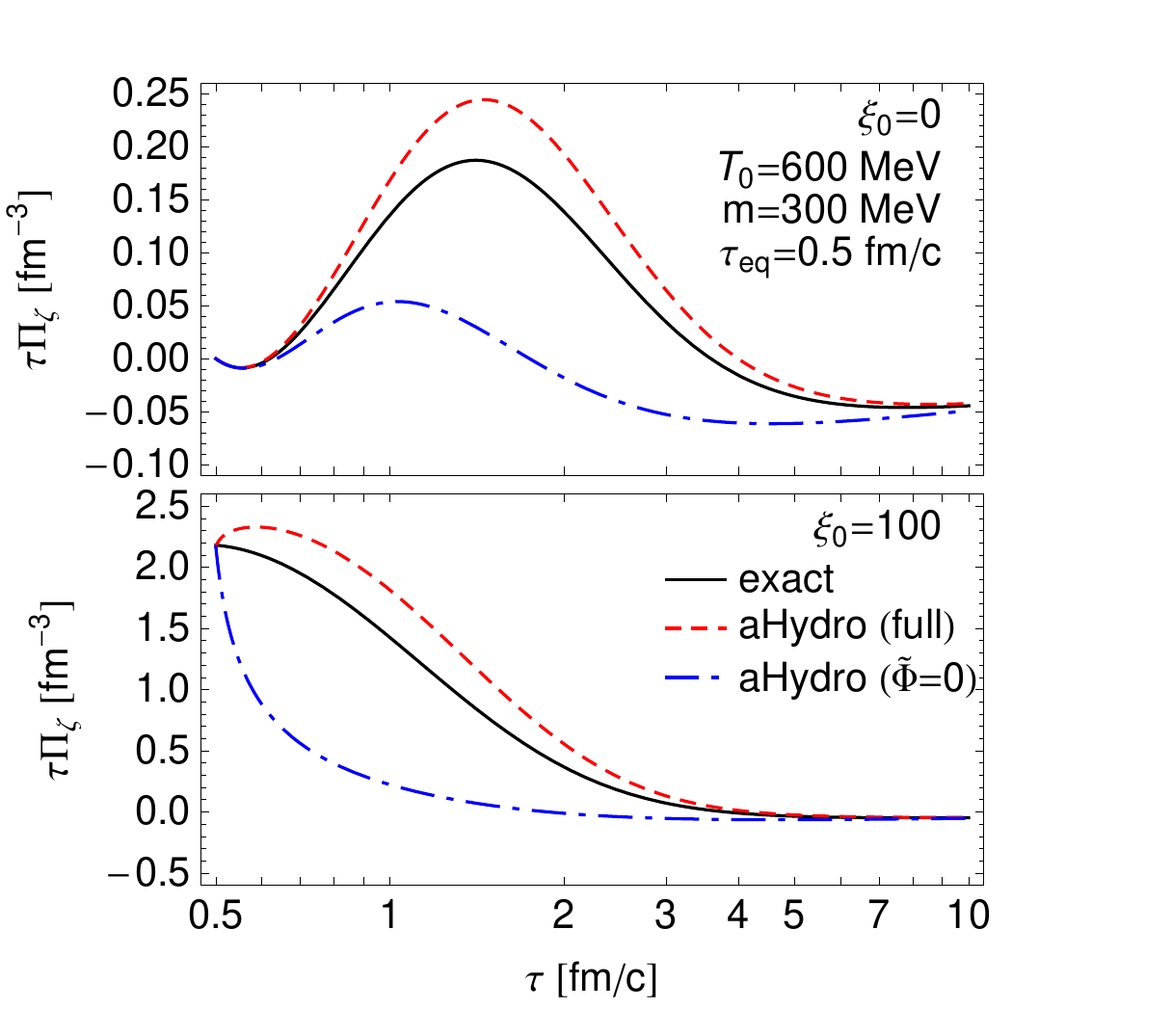}}
\caption{(Color online) Proper-time evolution of the bulk pressure correction using three different approaches: (i) exact solution of the Boltzmann equation \cite{Florkowski:2014sfa} (black solid line), (ii) the full aHydro equations including the bulk degree of freedom (red dashed line), and (iii) the aHydro equations with the ellipsoidal bulk degree of freedom set to zero (blue dot-dashed line).  For both panels we used $m=$ 300 MeV, $\tau_0$ = 0.5 fm/c, $\tau_{\rm eq}$ = 0.5 fm/c, and $T_0$ = 600 MeV.  In the top panel, we fixed the initial spheroidal anisotropy parameter $\xi_0=0$ and, in the bottom panel, we chose $\xi_0 = 100$.  Figure used with permission from Ref.~\cite{Nopoush:2014pfa}.}
\label{fig:bulk_300}
\end{figure}

In Fig.~\ref{fig:bulk_300} we compare the proper-time evolution of the bulk pressure 
\be
\Pi_\zeta(\tau) = \frac{1}{3}
\left[P_L(\tau) + 2 P_T(\tau)
- 3 P_{\rm eq}(\tau) \right] ,
\label{eq:PIkz}
\ee
obtained by solving the dynamical equations in the case of a transversally-homogeneous and boost-invariant (0+1d) system.  We compare the results obtained by solving the non-conformal aHydro equations specified above and labelled as ``aHydro (full)'' to (a) the exact solution of the RTA Boltzmann equation with massive particles \cite{Florkowski:2014sfa} and (b) the spheroidal prescription without an explicit bulk degree of freedom \cite{Nopoush:2014pfa}.  As we can see from this figure, the inclusion of the bulk variable $\Phi$ results in much better agreement with the exact solution of the Boltzmann equation.  The agreement can be further improved by going to second-order as we will demonstrate in a forthcoming section.  In addition, we will discuss how the exact solution shown in Fig.~\ref{fig:bulk_300} was obtained in a Sec.~\ref{sec:exactsolutions}, where we present three exact solutions to the RTA Boltzmann equation and the lessons learned from comparisons of various aHydro and vHydro schemes to them.

\section{3+1d quasiparticle anisotropic hydrodynamics}
\label{sec:ahydroqp}

In the last section, we introduced aHydro for a non-conformal QGP, but we considered the mass to be constant in the discussion. In this section, we will introduce {\em quasiparticle anisotropic hydrodynamics} (aHydroQP) to take into account the non-conformality of QCD in a self-consistent manner. As many perturbative methods suggest, the QGP can be described as a system of massive quasiparticles with temperature-dependent masses.  For example, in the hard thermal loop (HTL) resummation studies \cite{Braaten:1989mz,Braaten:1991gm,Andersen:1999fw,Andersen:2003zk,Andersen:2004fp,Andersen:2011sf,Haque:2014rua} one finds that both quarks and gluons acquire temperature-dependent quasiparticle masses and non-trivial dispersion relations. In this section, we will review the quasiparticle method presented originally  in Ref.~\cite{Alqahtani:2015qja} by summarizing the main points.

\subsection{The realistic equation of state}
\label{subsec:EoS}

At high temperatures, quarks and gluons can be considered as an ideal gas due to asymptotic freedom, but at lower temperatures, mainly around the critical temperature, one should consider non-perturbative methods, i.e. lattice QCD to obtain the energy density and pressure as a function of temperature. In this section, we will introduce the realistic equation of state (EoS) taken from the Wuppertal-Budapest collaboration \cite{Borsanyi:2010cj}. Then, we will show how one can obtain the thermal mass from the realistic EoS. Finally, we will specify the transport coefficients used here in aHydroQP in RTA.

To obtain the energy density and pressure we use an analytic parameterization for the trace anomaly obtained from Ref.~\cite{Borsanyi:2010cj}
\be
\frac{{\cal I}_{\rm eq}(T)}{T^4}=\bigg[\frac{h_0}{1+h_3 t^2}+\frac{f_0\big[\tanh(f_1t+f_2)+1\big]}{1+g_1t+g_2t^2}\bigg]\exp\!\Big(\!-\!\frac{h_1}{t}-\frac{h_2}{t^2}\Big) ,
\label{eq:trace}
\ee
where the trace anomaly is defined by ${\cal I}_{\rm eq} = \epsilon_{\rm eq} - 3 P_{\rm eq}$, and the parameters introduced above in Eq.~(\ref{eq:trace}) are  $t\equiv T/(0.2 \; \rm GeV)$, $h_0=0.1396$, $h_1=-0.1800$, $h_2=0.0350$, $f_0=2.76$, $f_1=6.79$, $f_2=-5.29$, $g_1=-0.47$, $g_2=1.04$, and $h_3=0.01$. 

Once one  has the trace anomaly,  the pressure can be obtained by integrating the trace anomaly
\be
\frac{P_{\rm eq}(T)}{T^4}=\int_0^T \frac{dT}{T}\frac{{\cal I}_{\rm eq}(T)}{T^4} \, ,
\label{eq:P_lattice}
\ee
then the energy density can be obtained using the definition of the trace anomaly
\be
\epsilon_{\rm eq}(T) = 3 P_{\rm eq}(T) + {\cal I}_{\rm eq}(T) \, ,
\label{eq:E_lattice}
\ee
%
\begin{figure}[t!]
\centering
\includegraphics[width=.47\linewidth]{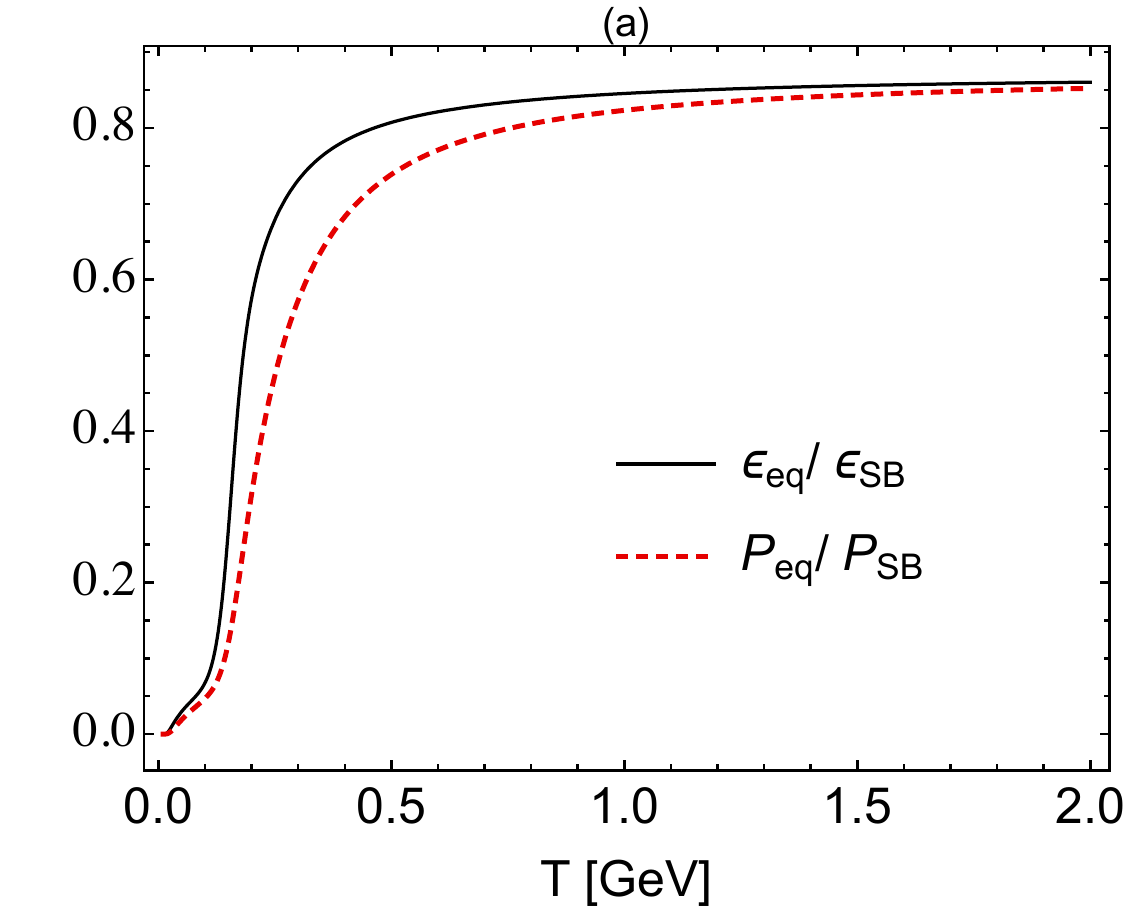}
$\;\;$
\includegraphics[width=0.46\linewidth]{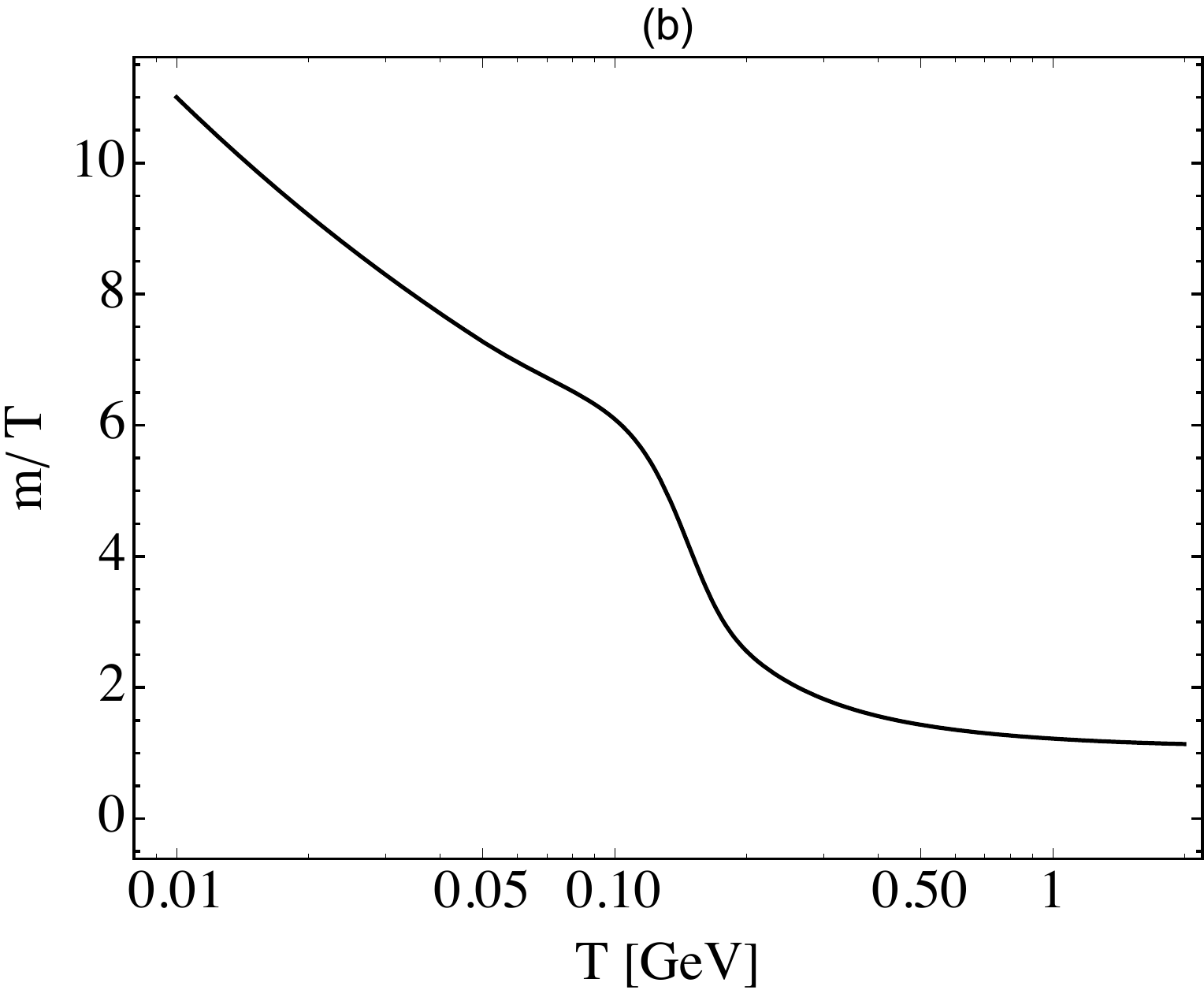}
\caption{(Color online) In panel (a)  we show the energy density scaled by the Stefan-Boltzmann limit $\epsilon/{\epsilon_{\rm SB}}$ (solid line) and the pressure scaled by the Stefan-Boltzmann limit $P/{P_{\rm SB}}$ (red dashed line) where both as a function of temperature.  In panel (b) we show the temperature dependence of the quasiparticle mass scaled by the temperature. Figure used with permission from Ref.~\cite{Alqahtani:2015qja}.}
\label{fig:eos}
\end{figure}
At high temperatures, both the energy density and pressure approach  the ideal limit (Stefan-Boltzmann limit) given by
\be 
P_{\rm SB}= \frac{\epsilon_{\rm SB}}{3}  = \frac{N_{\rm dof} T^4}{\pi^2}  \, ,
\ee
with
\be 
N_{\rm dof} = \frac{\pi^4}{45} \left( N_c^2-1+\frac{7}{4} N_c N_f \right) \, ,
\ee
where $N_f$ and $N_c$ are the number of flavors and colors respectively taken here to be $N_f=N_c=3$.

Next, to find the thermal mass, one can use the following  thermodynamic identity to solve for $m(T)$ knowing the energy density and pressure from lattice QCD
\be
\epsilon_{\rm eq}+P_{\rm eq}=Ts_{\rm eq} = 4 \pi \tilde{N} T^4 \, \hat{m}_{\rm eq}^3 K_3\left( \hat{m}_{\rm eq}\right) \, ,
\label{eq:meq}
\ee
 where $s_{\rm eq} $ here is calculated for an equilibrium massive  gas using a Boltzmann distribution function. 
 
 In Fig.~\ref{fig:eos}-a, we show the temperature dependence of the energy density and pressure obtained from lattice QCD scaled by $\epsilon_{\rm SB}$ and ${P_{\rm SB}}$, respectively. At high temperatures as expected both the energy density and pressure approach the ideal limit, but have large corrections to the ideal limit at lower temperatures.  In Fig.~\ref{fig:eos}-b, we show the temperature dependence of the thermal mass scaled by the temperature.  As can be seen from this figure, $m \sim T$ at high temperatures, indicating that the quasiparticle mass is consistent with expectations from hard-thermal-loop perturbation theory \cite{Braaten:1989mz,Braaten:1991gm,Andersen:1999fw,Andersen:2003zk,Andersen:2004fp,Andersen:2011sf,Haque:2014rua} in this regime.
 
\begin{figure}[t!]
\centering
\includegraphics[width=0.6\linewidth]{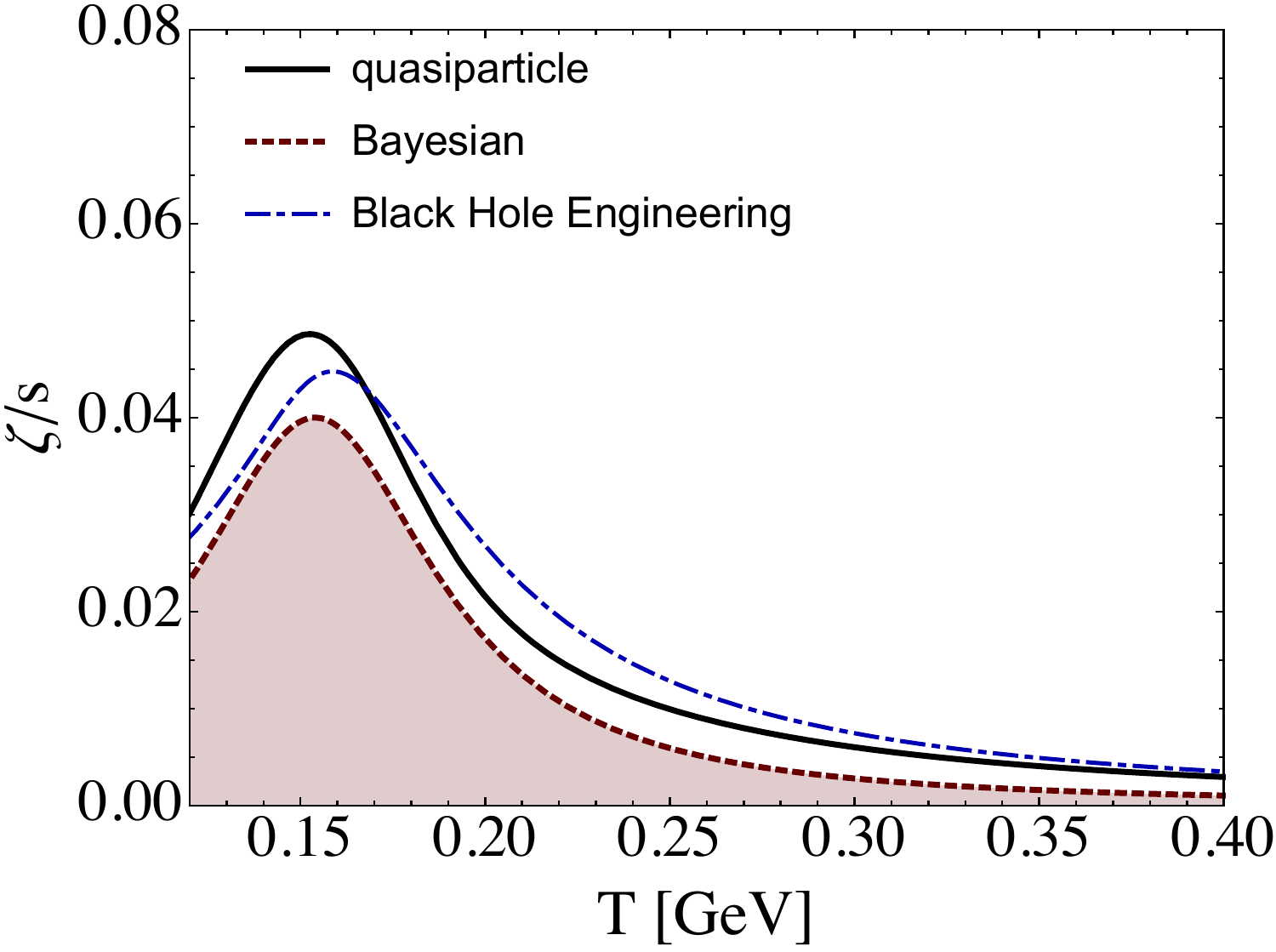}
\caption{(Color online) The bulk viscosity scaled by the entropy density $\zeta/s$ as a function of $T$ is shown for the quasiparticle model (black solid line), Bayesian analysis (red short-dashed line) \cite{Bass:2017zyn}, and Black hole engineering (blue long dot-dashed line) \cite{Rougemont:2017tlu}. }
\label{fig:zeta}
\end{figure}

We now turn to the transport coefficients used in the aHydroQP in RTA.  The quasiparticle expressions for the shear viscosity and the bulk viscosity taken from Refs.~\cite{Romatschke:2011qp} and \cite{Tinti:2016bav} are
\ba 
\frac{\eta}{\tau_{\rm eq}}&=& \frac{1}{T} I_{3,2}(\hat{m}_{\rm eq}) \,\, ,\label{eq:eta} \\
\frac{\zeta}{\tau_{\rm eq}} &=& 
\frac{5}{3 T} I_{3,2} (\hat{m}_{\rm eq})
- c_s^2 (\epsilon +P) 
+ c_s^2 m \frac{dm}{dT} I_{1,1}(\hat{m}_{\rm eq}) \, ,
\label{eq:zeta}
\ea
where $c_s^2$ is the speed of sound squared given by  ($c_s^2 = \partial P_{\rm eq}/\partial\epsilon_{\rm eq}$), and the special functions are given by
\ba 
I_{3,2}(x) &=& \frac{N_{\rm dof} T^5\, x^5}{30 \pi^2} \bigg[ \frac{1}{16} \Big(K_5(x)-7K_3(x)+22 K_1(x) \Big)-K_{i,1}(x) \bigg]  \, ,  \\ 
I_{1,1}(x) &=& \frac{N_{\rm dof} T^3 \, x^3}{6 \pi^2} \Bigg[ \frac{1}{4} \Big( K_3(x)-5K_1(x)\Big)+K_{i,1} (x)\Bigg] \, , \\ 
K_{i,1}(x)&=&\frac{\pi}{2}\Big[1-x K_0(x) s_{-1}(x)-xK_1(x) s_0(x)\Big] \, ,
\ea
where $K_n$ are the modified Bessel functions of the second kind, and $s_n$ are the modified Struve functions.
Using Eq.~(\ref{eq:eta}), the relaxation time can be written as 
\ba 
\tau_{\rm eq}(T)= \bar{\eta} \, \frac{\epsilon+P}{I_{3,2}(\hat{m}_{\rm eq})} \, .
\ea
As a result, by using the RTA, one has no control on all transport coefficients, i.e., once $\eta/s$ is fixed, the bulk viscosity ($\zeta/s$) is not a free parameter anymore. In Fig.~\ref{fig:zeta}, we show the temperature dependence of bulk viscosity scaled by the entropy density $\zeta/s$ predicted by the quasiparticle model compared to other two methods' predictions: Bayesian analysis \cite{Bass:2017zyn} and black hole engineering \cite{Rougemont:2017tlu}.  From this figure, one can see that all results presented are quite similar. Later on, we will comment on the bulk viscosity predicted by other vHydro studies which is quite large compared to these results.

\subsection{Thermodynamic consistency in aHydroQP}
\label{subsec:thermQP}
\begin{figure}[t!]
\centering
\includegraphics[width=0.5\linewidth]{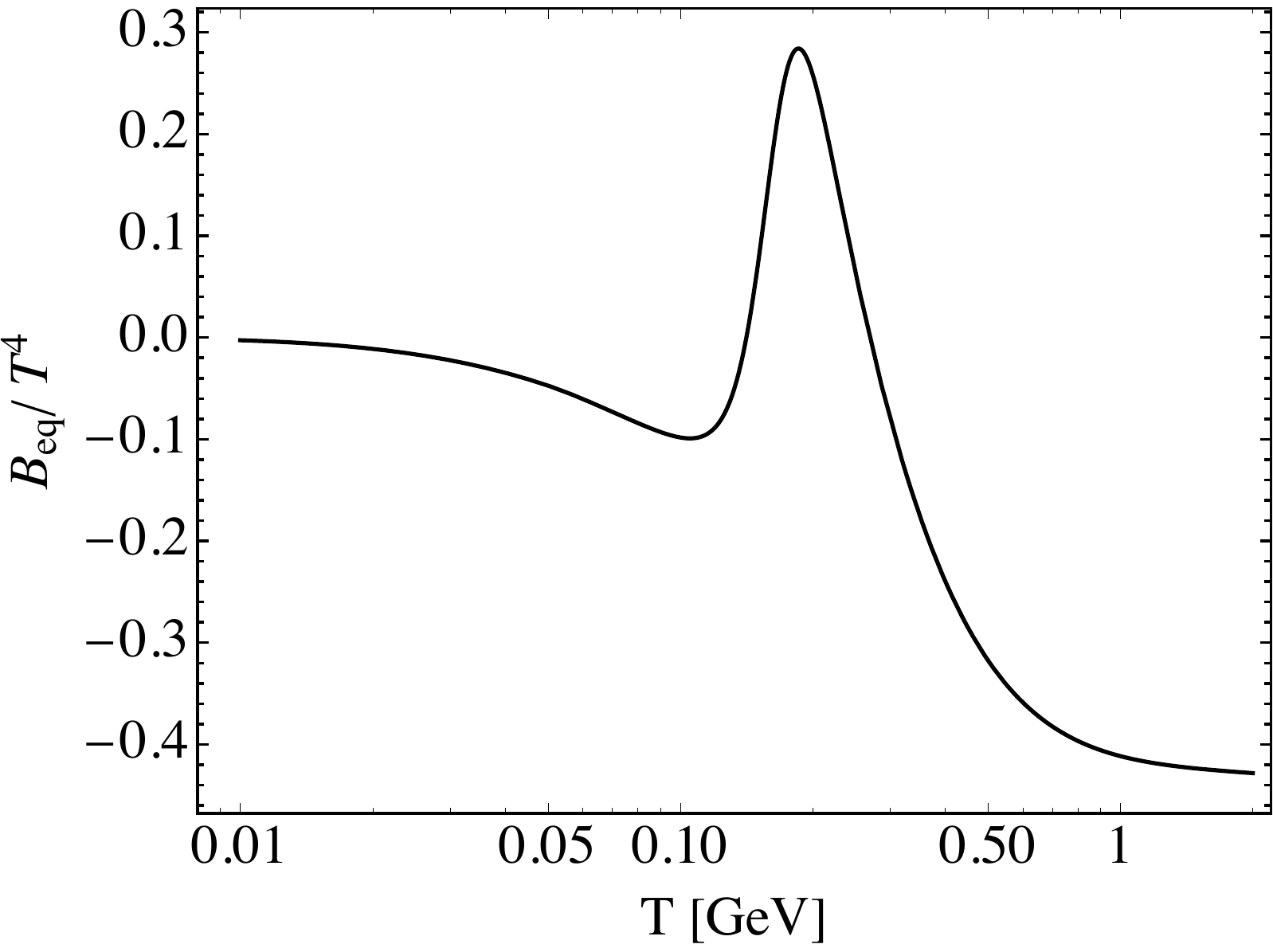}
\caption{This figure shows the temperature dependence of $B_{\rm eq}$ scaled by $T^4$. Figure used with permission from Ref.~\cite{Alqahtani:2015qja}.}
\label{fig:beq}
\end{figure}

For a massive gas of particles using Boltzmann distribution function, the equilibrium energy density, pressure, and entropy density are given by 
\ba
\epsilon_{\rm eq}(T,m) &=& 4 \pi \tilde{N} T^4 \, \hat{m}_{\rm eq}^2
 \Big[ 3 K_{2}\left( \hat{m}_{\rm eq} \right) + \hat{m}_{\rm eq} K_{1} \left( \hat{m}_{\rm eq} \right) \Big] \, , 
\label{eq:EeqConstantM} \\
 P_{\rm eq}(T,m) &=& 4 \pi \tilde{N} T^4 \, \hat{m}_{\rm eq}^2 K_2\left( \hat{m}_{\rm eq}\right)\, ,\\
 s_{\rm eq}(T,m) &=&4 \pi \tilde{N} T^3 \, \hat{m}_{\rm eq}^2 \Big[4K_2\left( \hat{m}_{\rm eq}\right)+\hat{m}_{\rm eq}K_1\left( \hat{m}_{\rm eq}\right)\Big] \, .
\label{eq:SeqConstantM}
\ea
However, if one simply inserts temperature-dependent masses into the bulk variables above, one finds that the resulting expressions are not thermodynamically consistent.  To see how this problem arises, the entropy density can be obtained in two different ways, one by $ s_{\rm eq} = (\epsilon_{\rm eq}+ P_{\rm eq})/T$ and second by $ s_{\rm eq} =\partial P_{\rm eq}/\partial T$.  Using the expressions above, one finds that these two methods do not give the same answer when we have a temperature-dependent mass $m(T)$. To fix this, one introduces a background field in the energy-momentum tensor definition 
\be
T^{\mu\nu}_{\rm eq} = T^{\mu\nu}_{\rm kinetic,eq} + g^{\mu\nu} B_{\rm eq}(T)  \, .
\label{eq:Tmunu}
\ee
As a result, bulk thermodynamic variables for an equilibrium Boltzmann gas of quasiparticles become
\ba
\epsilon_{\rm eq}(T,m) &=& 4 \pi \tilde{N} T^4 \, \hat{m}_{\rm eq}^2
 \Big[ 3 K_{2}\left( \hat{m}_{\rm eq} \right) + \hat{m}_{\rm eq} K_{1} \left( \hat{m}_{\rm eq} \right) \Big]+B_{\rm eq} \, , 
\label{eq:Eeq} \\
 P_{\rm eq}(T,m) &=& 4 \pi \tilde{N} T^4 \, \hat{m}_{\rm eq}^2 K_2\left( \hat{m}_{\rm eq}\right)-B_{\rm eq} \, , 
\label{eq:Peq} \\
s_{\rm eq}(T,m) &=&4 \pi \tilde{N} T^3 \, \hat{m}_{\rm eq}^2 \Big[4K_2\left( \hat{m}_{\rm eq}\right)+\hat{m}_{\rm eq}K_1\left( \hat{m}_{\rm eq}\right)\Big] .
\label{eq:Seq} 
\ea
 As we can see, the energy density is modified by $\epsilon_{\rm kinetic} +B_{\rm eq}$, and the pressure is modified as well by $P_{\rm kinetic} -B_{\rm eq}$. Now, to determine $B_{\rm eq}$, one should requires the thermodynamic consistency 
\be
T s_{\rm eq} = \epsilon_{\rm eq} + P_{\rm eq} = T \frac{\partial P_{\rm eq}}{\partial T} \, .
\label{eq:thermoid}
\ee
For this identity to hold, using Eqs.~(\ref{eq:Eeq}), (\ref{eq:Peq}), and (\ref{eq:thermoid}) one obtains 
\ba
\frac{dB_{\rm eq}}{dT} &=& - \frac{1}{2} \frac{dm^2}{dT} \intdP \, f_{\rm eq}(x,p) \, ,\nonumber \\  
&=& -4\pi \tilde{N}m^2 T K_1(\hat{m}_{\rm eq}) \frac{dm}{dT} \, .
\label{eq:BM-matching-eq-1}
\ea
One can obtain the full $B_{\rm eq}$ by integrating Eq.~(\ref{eq:BM-matching-eq-1}). In Fig.~\ref{fig:beq}, we show the temperature dependence of $B_{\rm eq}$ scaled by $T^4$ .

\subsection{3+1d dynamical equations in aHydroQP}
\label{subsec:dHydroQP}

In the case of having thermal quasiparticles, one should use the Boltzmann equation in its general form which accounts for spatial variation of the particle mass~\cite{Jeon:1995zm,Romatschke:2011qp,Alqahtani:2015qja}
\be
p^\mu \partial_\mu f(x,p)+\frac{1}{2}\partial_i m^2\partial^i_{(p)} f(x,p)=-\mathcal{C}[f]\,,
\label{eq:boltzmanneq}
\ee
where at leading-order in the aHydro expansion $f(x,p)$ is the anisotropic distribution function specified in  Eq.~(\ref{eq:fform}). Below we will only list the differences between what we presented before in Sec.~\ref{sec:lonoconformal} and the quasiparticle method. As we discussed before, to keep thermodynamic consistency one should modify the energy-momentum tensor by adding a background field
\be
T^{\mu\nu} \equiv \intdP \, p^\mu p^\nu f(x,p)+B({\boldsymbol\alpha},\lambda) g^{\mu\nu} \, \label{eq:T-int} 
\ee
where $B({\boldsymbol\alpha},\lambda)$ is the non-equilibrium background field which generalizes the modification of the equilibrium energy-momentum tensor.  This results to modification of the  previous definitions of the energy density and pressure components   defined in Eqs.~(\ref{eq:E_P_difenitions}), by including the effect of the background field as follows
\ba
\epsilon &=& {\cal H}_3({\boldsymbol\alpha},\hat{m}) \, \lambda^4+B \, ,\nonumber \\
P_i &=& {\cal H}_{3i}({\boldsymbol\alpha},\hat{m}) \, \lambda^4-B \, ,
\label{eq:bulkcom}
\ea
where $\hat{m} \equiv m/\lambda$. As explained before, for the equilibrium state, in order to conserve the energy-momentum tensor, there must be a relation between the background field and the thermal mass that should be satisfied.  In a non-equilibrium situation, this relation is given by
\be
\partial_\mu B = -\frac{1}{2} \partial_\mu m^2 \intdP  f(x,p)\,.
\label{eq:B}
\ee
The first moment equations in this case are the same as the ones presented before in Eqs.~(\ref{eq:1stmoment}) with $\epsilon$ and $P$ obtained from the quasiparticle-modified forms. However, the second moment is modified by an extra term corresponding to the thermal mass 
\be
\partial_\alpha  I^{\alpha\nu\lambda}- J^{(\nu} \partial^{\lambda)} m^2 =-\intdP \, p^\nu p^\lambda{\cal C}[f]\, \label{eq:I-conservation} ,
\ee
where $J^\nu$ is the number current.
The term involving the particle current has no effect on the $X_\mu X_\nu \partial_\alpha I^{\alpha \mu \nu}$, $Y_\mu Y_\nu \partial_\alpha I^{\alpha \mu \nu}$, and $Z_\mu Z_\nu \partial_\alpha I^{\alpha \mu \nu}$ projections, but it results in an additional contribution for some other projections such as $u_\mu X_\nu \partial_\alpha I^{\alpha \mu \nu}$ \cite{Alqahtani:2015qja}.\footnote{This is due to the fact that the particle current $J^\mu$ is purely timelike at leading order.}
  In summary, in 3+1d systems, we have eight independent variables  $\alpha_i$, $u_x$, $u_y$, $\vartheta$, $T$, $\lambda$, so we need eight equations which are: four from the first moment presented in Eqs.~(\ref{eq:1stmoment}), three from the second moment shown in Eqs.~(\ref{eq:2ndmoment}) and one from the matching condition in Eq.~(\ref{eq:matching}). Later on, we will show the phenomenological results of 3+1d aHydroQP.

\subsection{Standard anisotropic hydrodynamics}
\label{subsec:aHydros}

In this section, we will try to emphasize the importance of the way that the EoS is implemented in aHydro. In a prior approach to implement a realistic EoS, dubbed standard aHydro, a straightforward method to implement the EoS in the context of aHydro was proposed, however it took the non-conformality of the QGP into account only in an approximate way \cite{Nopoush:2015yga,Alqahtani:2015qja}. Since,  later in this review, we will show some phenomenological results using this method, we give a very brief introduction to it and show some comparisons with aHydroQP in some simple systems.

In standard aHydro, one assumes the system to be massless and to implement a realistic EoS, one relates the energy density and pressure components by hand similar to how it is done in standard vHydro. In standard aHydro, where $m \rightarrow 0$, the energy density and pressure components can be written as
\ba
\epsilon &=& {\cal H}_3({\boldsymbol\alpha}) \, \lambda^4 \, ,\nonumber \\ 
P_i &=& {\cal H}_{3i}({\boldsymbol\alpha}) \, \lambda^4 \, , 
\label{eq:EPaHydro}
\ea
where we have ${\cal H}_3({\boldsymbol\alpha}) \equiv {\cal H}_3({\boldsymbol\alpha},m=0) $ and  ${\cal H}_{3i}({\boldsymbol\alpha}) \equiv {\cal H}_{3i}({\boldsymbol\alpha},m=0)$.\footnote{Equation~(\ref{eq:bulkcom}) reduces to (\ref{eq:EPaHydro}) in the limit $B\rightarrow0$ and $m\rightarrow0$.}  We notice that the energy density and pressure components are factorized into two functions where one of them is the isotropic energy density and pressure components and the other one carries all the anisotropy information, i.e., ${\cal H}$-functions. Standard aHydro uses the above relations, relating the factor of $\lambda^4$ to the equilibrium energy density and pressure components as follows  
\ba
\epsilon_{\rm eq}(\lambda) = 24 \pi \tilde{N}\lambda^4 \, , \nonumber \\
P_{\rm eq}(\lambda) = 8\pi \tilde{N}\lambda^4\, ,
\label{eq:massless2}
\ea
where $\tilde{N}$ is defined below Eq.~(\ref{eq:hfuncs3}).  Substituting Eq.~(\ref{eq:massless2}), one can rewrite Eqs.~(\ref{eq:EPaHydro})  as 
\ba
\epsilon &=& \hat{{\cal H}}_3({\boldsymbol\alpha}) \, \epsilon_{\rm eq}(\lambda) \, ,\nonumber \\ 
P_i &=& \hat{{\cal H}}_{3i}({\boldsymbol\alpha}) \,P_{\rm eq}(\lambda) \, , 
\label{eq:EPaHydros}
\ea
with $\hat{{\cal H}}_3({\boldsymbol\alpha}) \equiv {\cal H}_3({\boldsymbol\alpha})/(24 \pi  \tilde{N})$ and  $\hat{{\cal H}}_{3i}({\boldsymbol\alpha}) \equiv {\cal H}_{3i}({\boldsymbol\alpha})/(8 \pi  \tilde{N})$.
Hence, in this approximate way one can impose the realistic EoS  by replacing $\epsilon_{\rm eq}(\lambda)$ and $P_{\rm eq}(\lambda)$ using  lattice QCD calculations.

\begin{figure}[t!]
\centering
\hspace{-6mm}
\hspace{-6mm}
\includegraphics[width=.47\linewidth]{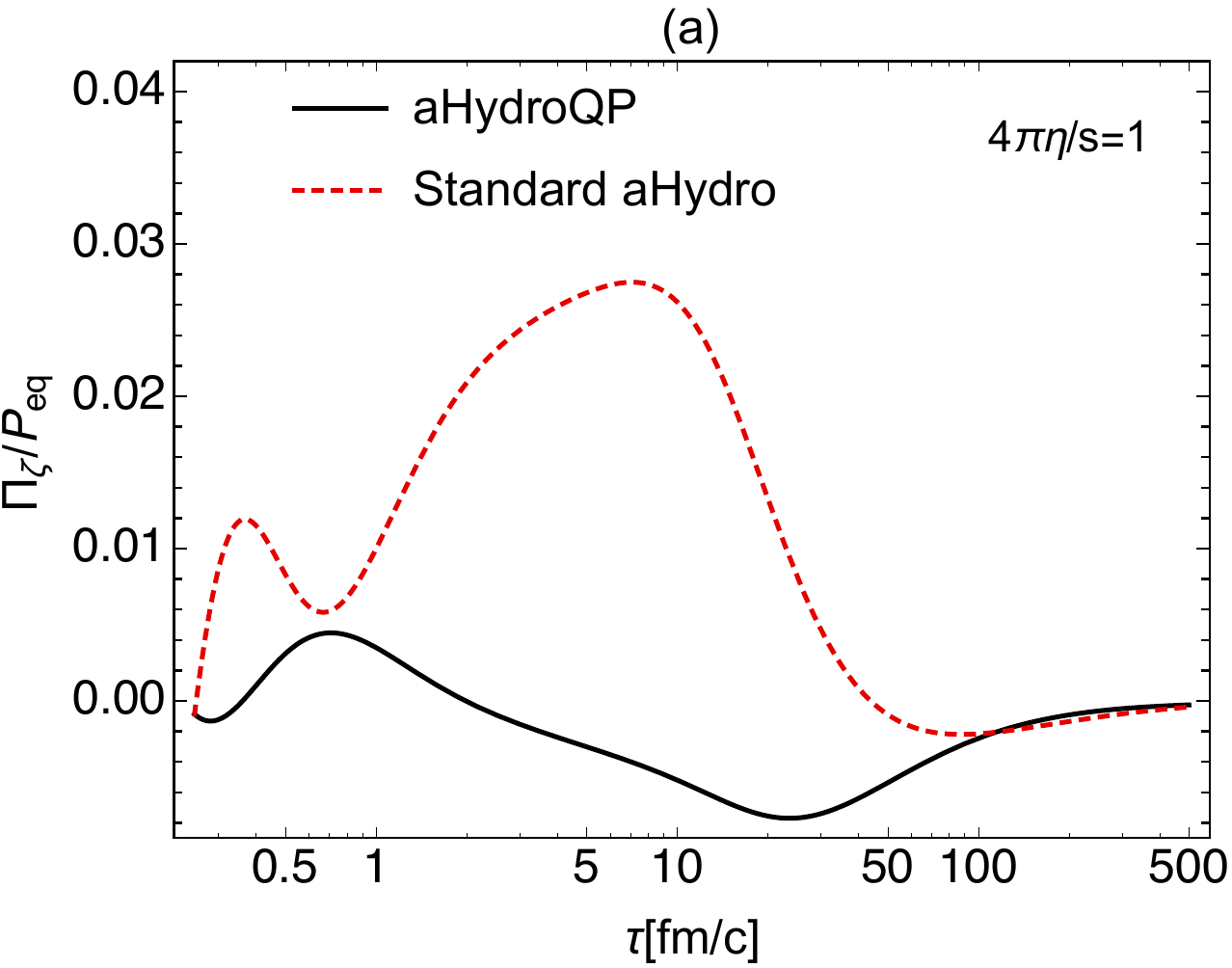}
$\;\;$
\includegraphics[width=0.48\linewidth]{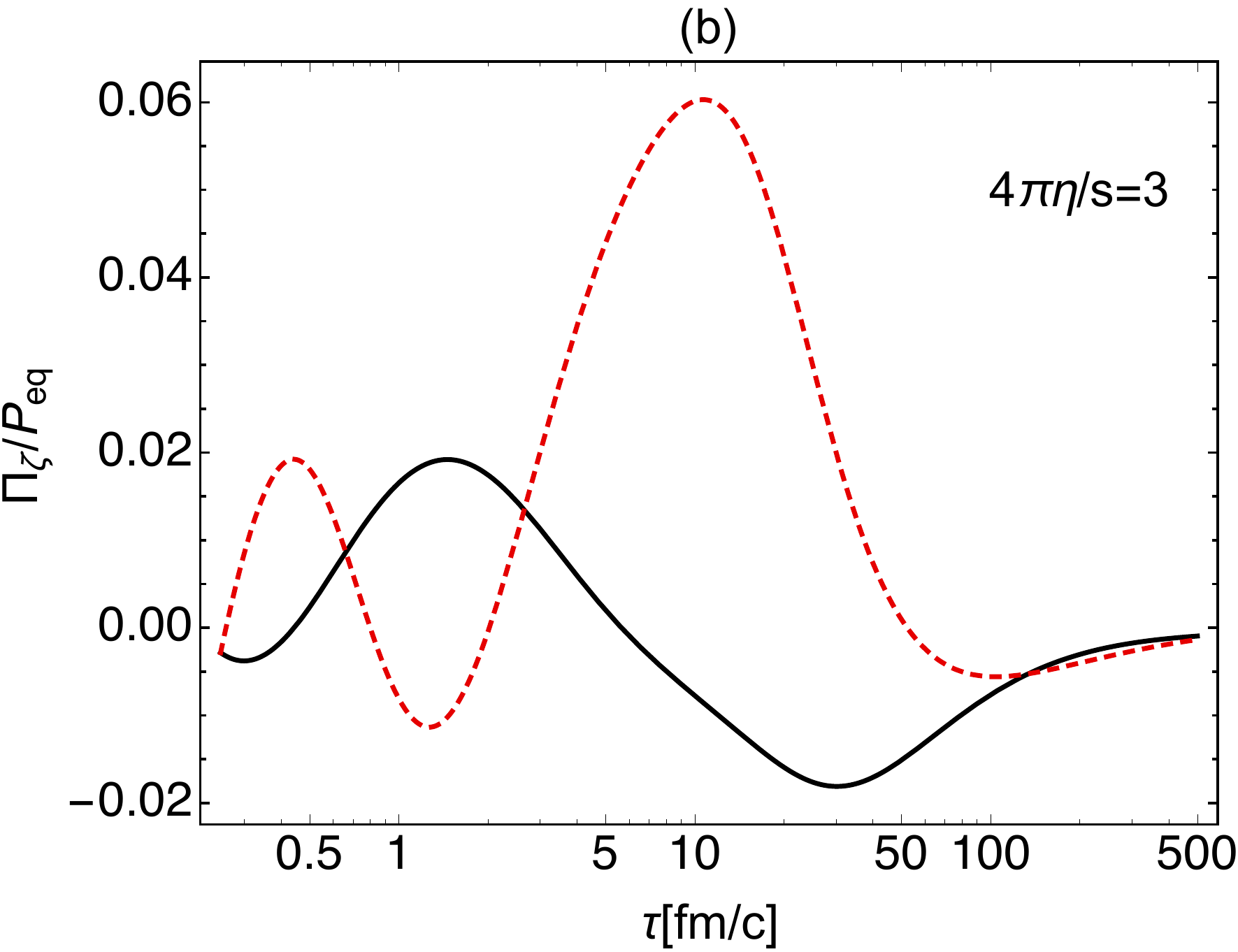}
\caption{(Color online) In panel (a) and (b) we show the bulk pressure scaled by the equilibrium pressure  predicted by aHydroQP (black solid line) and aHydro (red dashed line) for $4 \pi \eta/s=1$ and $3$, respectively. Figure used with permission from Ref.~\cite{Alqahtani:2015qja}. }
\label{fig:bulk0+1}
\end{figure}

Next, we turn to some comparisons between standard aHydro and aHydroQP in simple systems. To begin, we consider a boost-invariant  and transversally-homogeneous system (0+1d). In Fig.~\ref{fig:bulk0+1}, we show the bulk pressure scaled by the equilibrium pressure for $4 \pi \eta/s=1$ and $3$ in panel (a) and (b) respectively. As clearly can be seen from these comparisons, we see a substantial difference between these two methods depending on the way that the EoS is implemented. The large difference seen in the evolution of the bulk pressure correction stems from the consistent inclusion of conformal symmetry breaking effects in the quasiparticle model, as opposed to the standard implementation which only partially accounts for conformal symmetry breaking.  For example, in the standard implementation although one includes conformal symmetry breaking effects at the level of the equation of state, non-conformal corrections to transport coefficients are not properly treated.  As can be seen from Fig.~\ref{fig:bulk0+1}, the standard EoS implementation results in a larger bulk pressure correction which naively would translate into larger radial flow.  We also point out that, for both the quasiparticle and standard EoS implementations, one observes that it takes quite a long time for the system to approach the Navier-Stokes behavior, $\Pi_{\rm NS} =  - \zeta \theta$.  This is due to the shear-bulk coupling which emerges in second-order viscous hydrodynamics \cite{Denicol:2014mca} and anisotropic hydrodynamics \cite{Nopoush:2014pfa}.

\begin{figure}[t!]
\centering
\includegraphics[width=1.\linewidth]{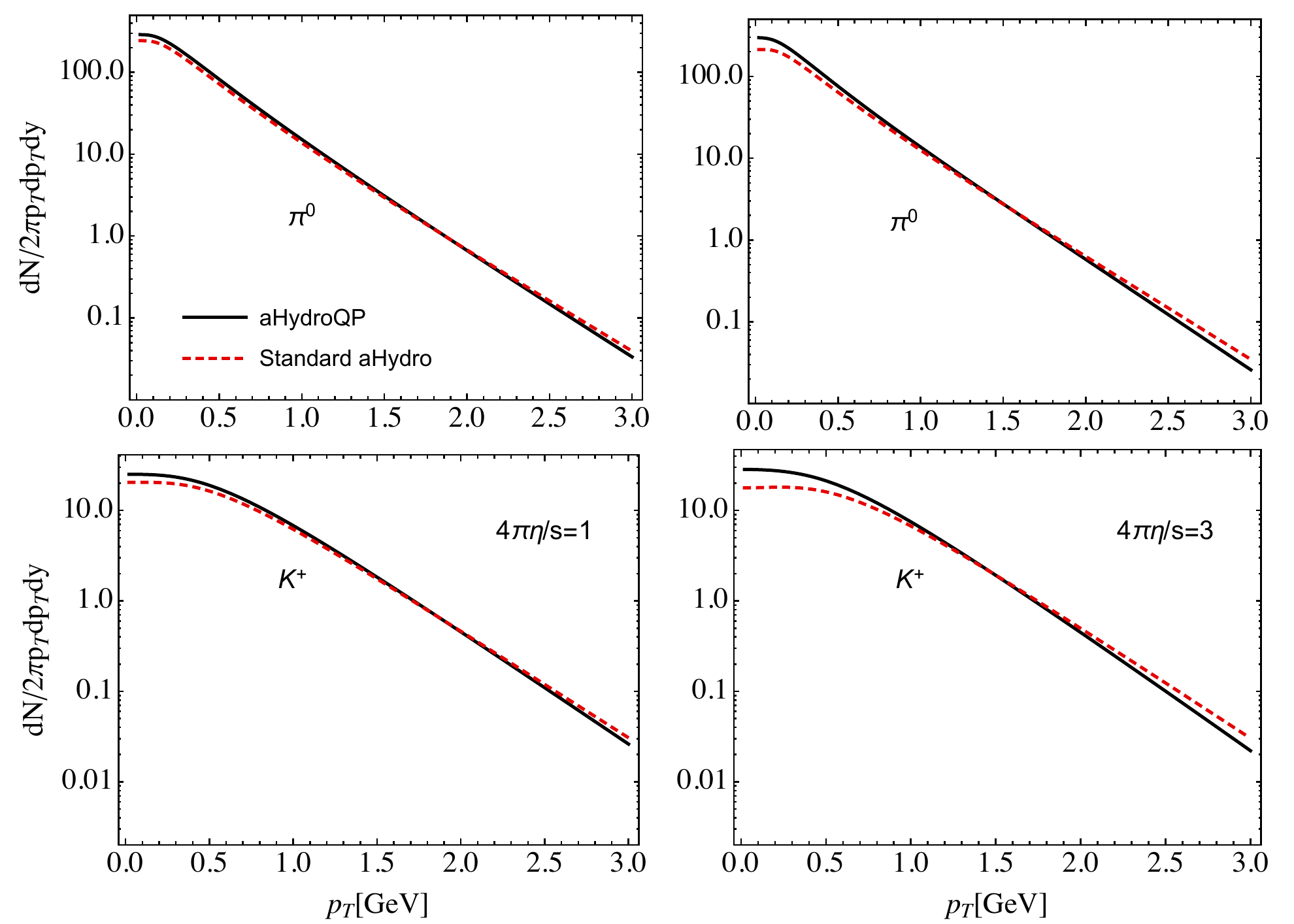}
\caption{(Color online) The pions and kaons spectra are shown as a function of transverse momentum $p_T$ in 1+1d systems for Pb-Pb collisions obtained from aHydroQP (black solid line) and standard aHydro (red dashed line) for $4\pi \eta/s=1$ and $3$ in left panel and right panels respectively. Figure used with permission from Ref.~\cite{Alqahtani:2016rth}.}
\label{fig:spectra1+1}
\end{figure}

To illustrate this we next consider a boost-invariant  and azimuthally-symmetric system (1+1d). We show the primordial  spectra of pions and kaons for $4 \pi \eta/s=1$ and $3$ in Fig.~\ref{fig:spectra1+1}.  For this figure, we implemented fixed energy-density freeze-out at an energy density which corresponds to an effective temperature of $T_{\rm FO} = 150$ MeV. Although both methods agree quite well at high $p_T$, they clearly disagree at low $p_T$ where standard aHydro obviously underestimates the spectra. As we will see later on, this will cause some disagreement with the experimental data in 3+1d systems.  Note, however, that the expectation that the standard method results in stronger radial flow is demonstrated clearly in Fig.~\ref{fig:spectra1+1}.

\newcommand{\rs}{{\rm RS}}
\newcommand{\dft}{\delta\tilde{f}}
\newcommand{\Pit}{\tilde{\Pi}}
\newcommand{\pit}{\tilde{\pi}}
\newcommand{\I}{\tilde{{\cal I}}}
\newcommand{\J}{\tilde{{\cal J}}}

\section{Second-order anisotropic hydrodynamics}
\label{sec:vaHydro}

As mentioned previously, the conceptual starting point for second-order aHydro is an expansion of the one-particle distribution function around an ellipsoidally-deformed form (\ref{eq:genf}).  In this section, we will review efforts to systematically include the corrections stemming from the non-ellipsoidal terms captured by $\delta \tilde{f}$.  There have been a few works along these lines, see e.g. Refs.~\cite{Bazow:2013ifa,Bazow:2015cha,Molnar:2016vvu,Molnar:2016gwq,Martinez:2017ibh} and here we would like to provide a brief presentation of the method and some key results, focussing on the non-conformal framework laid out in Refs.~\cite{Bazow:2015cha} since it includes the conformal case as a special case.  

\subsection{Setup and thermodynamic moments}

When one includes the corrections from $\delta \tilde{f}$ the framework has been dubbed ``viscous anisotropic hydrodynamics'' (vaHydro) in order to distinguish it from leading-order aHydro.  In this framework, one takes a simple form for the leading-order one-particle distribution function which accounts for non-spheroidal momentum-space anisotropies in a perturbative manner.  As a result, in vaHydro one can assume that the background anisotropy tensor is spheroidal and of Romatschke-Strickland form generalized to a non-conformal system \cite{Martinez:2012tu}.  For a non-conformal spheroidally-anisotropic system, the anisotropy tensor $\Xi^{\mu\nu}$ can be expanded as~\cite{Martinez:2012tu}
\be
\Xi^{\mu\nu}=u^\mu u^\nu - \Phi\Delta^{\mu\nu}+\xi z^{\mu}z^{\nu} \, .
\ee
In the LRF, the non-conformal extension of the Romatschke-Strickland form, $f_\rs$, is 
\be
f_\rs=f_{\rm iso}\left(\frac{1}{\Lambda}\sqrt{m^2{+}(1{+}\Phi)p^2_T{+}(1{+}\Phi{+}\xi)p^2_z}\right) 
\equiv f_{\rm iso}\left(\frac{E_\rs}{\Lambda}\right) , 
\label{eq:lo_rs2}
\ee
where $E_\rs^2{\,\equiv\,}(1{+}\Phi)m_T^2\cosh^2y + \xi m_T^2\sinh^2y - \Phi m^2$, with $m_T^2{\,=\,}m^2{+}p_T^2$.
Above we have assumed vanishing chemical potential and that $f_0$ has the functional form of a local thermal equilibrium distribution as discussed previously.

The starting point for the derivation of the vaHydro dynamical equations is the relativistic Boltzmann equation for a system of massive particles (\ref{eq:boltzmann1}).  The particle four-current and the energy-momentum tensor can be obtained from the first and second moments of $f$
\ba
J^\mu &=\langle p^\mu\rangle \, , \\
T^{\mu\nu} &=\langle p^\mu p^\nu\rangle \, ,
\ea
where \mbox{$\langle{\cal O}\rangle \equiv \int dP\,{\cal O}(p)f(x,p)$} with $dP$ being the Lorentz invariant momentum-space measure introduced in Eq.~(\ref{eq:invphase}). 

If the system is approximately spheroidal in momentum-space, one can expand $f$ as
\be
f(x,p)=f_\rs(x,p)+\dft \, ,
\ee
and $J^\mu$ and $T^{\mu\nu}$ can be tensor decomposed as
\ba
\label{eq:JmuTmunu}
J^\mu&=& n u^\mu+v^{\langle\mu\rangle}\;,\\
T^{\mu\nu}&=& \epsilon u^\mu u^\nu-(P_T+\Pit)\Delta^{\mu\nu}
+\left(P_L-P_T\right)z^\mu z^\nu+\pit^{\mu\nu}\;.
\ea
Above $v^{\langle\mu\rangle}$ is the particle current transverse to $u^\mu$.  The tilde variables collect the non-spheroidal corrections to $J^\mu$ and $T^{\mu\nu}$, with $\Pit$ being the residual bulk viscous pressure, and $\pit^{\mu\nu}$ being the residual shear stress tensor defined by
\begin{equation} 
\begin{array}{lll}
\label{dissipativeCurrents}
{n}\equiv\langle E\rangle_\rs \, , & v^\mu\equiv \langle p_i\rangle_{\tilde\delta}X^\mu_i \, , & {}\\
\epsilon\equiv \langle E^2 \rangle_\rs \, , & P_T\equiv \langle p^2_T \rangle_\rs \, , & P_L\equiv \langle p^2_z \rangle_\rs 
\, ,\\
\Pit\equiv -\frac{1}{3}\langle \Delta^{\alpha\beta}p_\alpha\ p_\beta\rangle_{\tilde\delta} \, ,  & \tilde\pi^{\mu\nu}\equiv\langle p^{\langle \mu}p^{\nu \rangle}\rangle_{\tilde\delta} \, .
\end{array}  
\end{equation}
Above, $\langle\cdots\rangle_\rs\equiv\int dP\,(\cdots)f_\rs$ and $\langle\cdots\rangle_{\tilde\delta}\equiv\int dP\,(\cdots)\dft$, and we have made use of the generalized Landau matching conditions $\langle E\rangle_{\tilde\delta}=\langle E^2\rangle_{\tilde\delta}=0$. We note that, in this context, the total bulk viscous pressure $\Pi$ can be obtained from
\begin{equation}
\Pi = \frac{2P_T+P_L}{3}-P_\mathrm{eq}+\Pit \, ,
\end{equation}
and the total shear stress tensor can be obtained from
\be
\pi^{\mu\nu} =  \left(P_L{-}P_T\right) \left(\frac{\Delta^{\mu\nu}}{3}+z^\mu z^\nu\right) + \pit^{\mu\nu} \, .
\ee

\subsection{14-moment expansion}

In the 14-moment approximation, the deviation $\dft$ from the locally anisotropic state appearing in Eq.~(\ref{eq:lo_rs2}) is expanded to second order in momenta as~\cite{Bazow:2013ifa,Bazow:2015cha}
\begin{equation} \label{q_expanded}
\begin{split}
\frac{\dft}{f_\rs\tilde{f}_\rs} &= - \beta E + w E^2 - \frac{w}{3} \Delta^{\mu \nu} p_{\mu} p_{\nu}
+ w_{\langle \mu \nu \rangle} p^{\langle\mu} p^{\nu\rangle}\, ,
\end{split}
\end{equation}
where $\tilde{f}_\rs \equiv 1-af_\rs$, $\beta \equiv 1/T$, and $w \equiv u_\mu u_\nu w^{\mu\nu}$ with $w^{\mu\nu}$ being a general rank-two tensor.  Since so far the equations of motion have only been presented for zero chemical potential, there is no heat current $\tilde{v}^\mu$ and the coefficients of any terms linear in $p^{\langle\mu\rangle}$ in Eq.~(\ref{q_expanded}) vanish. Plugging Eq.~(\ref{q_expanded}) into Eqs.~(\ref{dissipativeCurrents}), the 14-moment coefficients can be expressed in terms of macroscopic quantities (moments of the distribution) by inverting a matrix equation~\cite{Bazow:2013ifa,Bazow:2015cha}.

\subsection{Leading-order moment equations}

From the zeroth moment of the Boltzmann equation one obtains
\be
D_u n + n\theta_u={\cal C}_0 \, ,
\label{eq:vazero}
\ee
where, as previously, $D_u \equiv u^\mu \partial_\mu$ and $\theta_u = \partial_\mu u^\mu$.  From the projections of the first moment of the Boltzmann equation along and transverse to $u^\mu$ one obtains~\cite{Bazow:2013ifa,Bazow:2015cha}
\be
D_u\epsilon + (\epsilon{+}P_T{+}\Pit)\theta_u
+ (P_L{-}P_T)u_\nu D_z z^\nu-\pit^{\mu\nu}\sigma_{\mu\nu}=0 \, , \label{eq:vafirst1}
\ee
and
\ba
&& (\epsilon{+}P_T{+}\Pit)D_u u^\alpha 
 - \nabla^\alpha (P_T{+}\Pit)
+ \Delta^\alpha_{\ \nu}\partial_\mu\pit^{\mu\nu}
\nonumber\\
&&\qquad+ z^\alpha D_z (P_L{-}P_T)
+ z^\alpha (P_L{-}P_T)(\partial_{\mu}z^{\mu}) 
+  (P_L{-}P_T)D_z z^\alpha 
-(P_L{-}P_T)u^\alpha u_\nu D_z z^\nu=0 \, , \label{eq:vafirst2}
\hspace{1.2cm}
\ea
where $\sigma^{\mu\nu}\equiv\nabla^{\langle\mu}u^{\nu\rangle}$.  Finally, from the equation of motion for the second moment of the Boltzmann equation projected with $\Delta_{\mu\nu}^{\alpha\beta}$ in order to isolate the traceless and transverse components, one obtains~\cite{Bazow:2015cha}
\ba
&&\hspace{-5mm}X^{\langle\alpha}_i X^{\beta\rangle}_j \left[D_u{\I}^{ij}_{10}
+\psi^{ij}_\Pi D_u{\Pit}+\psi^{ij\mu\nu}_\pi D_u{\pit}_{\mu\nu}
+\Pit D_u{\psi}^{ij}_\Pi+\pit_{\mu\nu}D_u{\psi}^{ij\mu\nu}_\pi
\right] \nonumber \\
&& \hspace{5mm}
+X^{\langle\alpha}_i X^{\beta\rangle}_j \left[
D_u{\I}^{ij}_{10}
+\psi^{ij}_\Pi\Pit+\psi^{ij\mu\nu}_\pi\pit_{\mu\nu}
\right]\theta_u \nonumber \\
&&\hspace{1cm}
+2\left[
D_u{\I}^{ij}_{10}
+\psi^{ij}_\Pi\Pit+\psi^{ij\mu\nu}_\pi\pit_{\mu\nu}
\right]\Delta^{\alpha\beta}_{\nu\lambda}
\left(X^{\nu}_i D_u X^{\lambda}_j
+X^{\nu}_i D_j u^{\lambda}\right)
={\cal C}^{\langle\alpha\beta\rangle}_0\;. \label{eq:vasecond}
\ea
The definitions of the various special functions appearing above can be found in Ref.~\cite{Bazow:2015cha}.

\subsection{Equations of motion for the non-spheroidal corrections}

The moment equations above involve both $\pit^{\mu\nu}$ and $\Pit$.  In order to close the system, one must obtain equations of motion for their evolution similar to how the standard viscous shear tensor and bulk evolution equations are obtained.  The starting points for this procedure
are the kinetic definitions of $\Pit$ and $\pit^{\mu\nu}$~\cite{Denicol:2012cn,Denicol:2010xn,Bazow:2013ifa}:
\ba
D_u{\Pit}&=&-\frac{m^2}{3}\int dP\,D_u{\dft}\;,\label{eq:evolve_res_moments}\\
D_u{\pit}^{\langle\mu\nu\rangle}&=&\Delta^{\mu\nu}_{\alpha\beta}\int dP\, p^{\langle\alpha}p^{\beta\rangle}D_u{\dft}\;.\label{eq:evolve_res_moments2}
\ea
To proceed, one writes the Boltzmann equation as \mbox{$D_u\delta{\tilde{f}}=-D_u{f}_\rs - ( p{\cdot}\nabla (f_\rs{+}\dft)-C[f] )/E$}.  Plugging this into Eqs.~(\ref{eq:evolve_res_moments}) and (\ref{eq:evolve_res_moments2}) and computing the necessary moments of the one-particle distribution function gives~\cite{Bazow:2015cha}
\be
\label{eq:Pi1}
-\frac{3}{m^2}D_u\Pit-{\cal C}_{-1} = {\cal W}-{\cal X}\theta_u-{\cal Y}^{\mu\nu}\sigma_{\mu\nu}+\frac{3}{m^2}\Pit\theta_u
-\delta_{\Pi\Pi}^{\mu\nu}\Pit\nabla_\mu u_\nu-\pit_{\alpha\beta}\delta_{\Pi\pi}^{\mu\nu\alpha\beta}\nabla_\mu u_\nu \, ,
\ee
\ba
\label{eq:pimunu1}
D_u{\pit}^{\left\langle \mu \nu \right\rangle }
-{\cal C}_{-1}^{\left\langle\mu \nu \right\rangle }
&=& {\cal K}^{\mu\nu}+{\cal L}^{\mu\nu}+{\cal M}^{\mu\nu}
+{\cal H}^{\mu\nu\lambda}\left(D_u{z}_\lambda+u^\alpha\nabla_\lambda z_\alpha\right)
+(1+\Phi){\cal Q}^{\mu\nu\lambda\alpha}\nabla_\lambda u_\alpha \nonumber\\
&& \hspace{1cm}
-\frac{5}{3}\pit^{\mu\nu}\theta_u
-2\pit^{\langle\mu}_\lambda\sigma^{\nu\rangle\lambda} 
+2\pit^{\langle\mu}_\lambda\omega^{\nu\rangle\lambda}
+2\Pit\sigma^{\mu\nu} \nonumber\\
&& \hspace{2cm}
-\Pit\delta_{\pi\Pi}^{\mu\nu\alpha\beta}\nabla_\alpha u_\beta
-\delta_{\pi\pi}^{\mu\nu\alpha\beta\sigma\lambda}\pit_{\sigma\lambda}\nabla_\alpha u_\beta
\;.
\ea
Expressions for the various dissipative forces (e.g. ${\cal W}$ and ${\cal Y}^{\mu\nu}$) and transport coefficients (e.g. $\delta_{\Pi\Pi}^{\mu\nu}$ and $\delta_{\Pi\pi}^{\mu\nu\alpha\beta}$) appearing above can be found in Ref.~\cite{Bazow:2015cha}.  Eqs.~(\ref{eq:vazero}), (\ref{eq:vafirst1}), (\ref{eq:vafirst2}), (\ref{eq:vasecond}), (\ref{eq:Pi1}), and (\ref{eq:pimunu1}) are the final closed set of evolution equations for vaHydro in the 14-moment approximation.  In closing, we note that it is possible to go beyond the 14-moment approximation in vaHydro.  For more information about this possibility see Ref.~\cite{Molnar:2016vvu}.

\subsection{0+1d Bjorken expansion}

Assuming that the system is describable using kinetic theory, undergoing 0+1d Bjorken expansion, possessing classical statistics, and subject to an RTA collisional kernel, the equations above reduce to the following~\cite{Bazow:2015cha}
\ba
&& \hspace{-7mm}
(\partial_\xi\epsilon) \partial_\tau \xi + (\partial_\Lambda\epsilon) \partial_\tau \Lambda 
+(\partial_\Phi\epsilon) \partial_\tau \Phi 
=
- \frac{1}{\tau}\left(\epsilon + P_L+\Pit-\pit\right) ,
\nonumber \\
&&\frac{\partial_\tau\xi}{1{+}\Phi{+}\xi}-2\left(3{+}\hat{m}\frac{K_1(\hat{m})}{K_2(\hat{m})} \right)\frac{\partial_\tau\Lambda}{\Lambda}
+\left(\frac{2}{1{+}\Phi}{+}\frac{1}{1{+}\Phi{+}\xi}\right)\partial_\tau{\Phi} 
\nonumber \\
&&\hspace{6cm} =
\frac{2}{\tau}+\frac{2}{\tau_{\rm eq}}\left(1-\frac{T}{\Lambda}\frac{K_2(\hat{m}_{\rm eq})}{K_2(\hat{m})}(1{+}\Phi)\sqrt{1{+}\Phi{+}\xi}\right), \hspace{1cm}
\ea
which come from the first and second moments, respectively.  The evolution equations for the non-spheroidal corrections are
%
\begin{figure}[t!]
\centerline{\includegraphics[width=17cm]{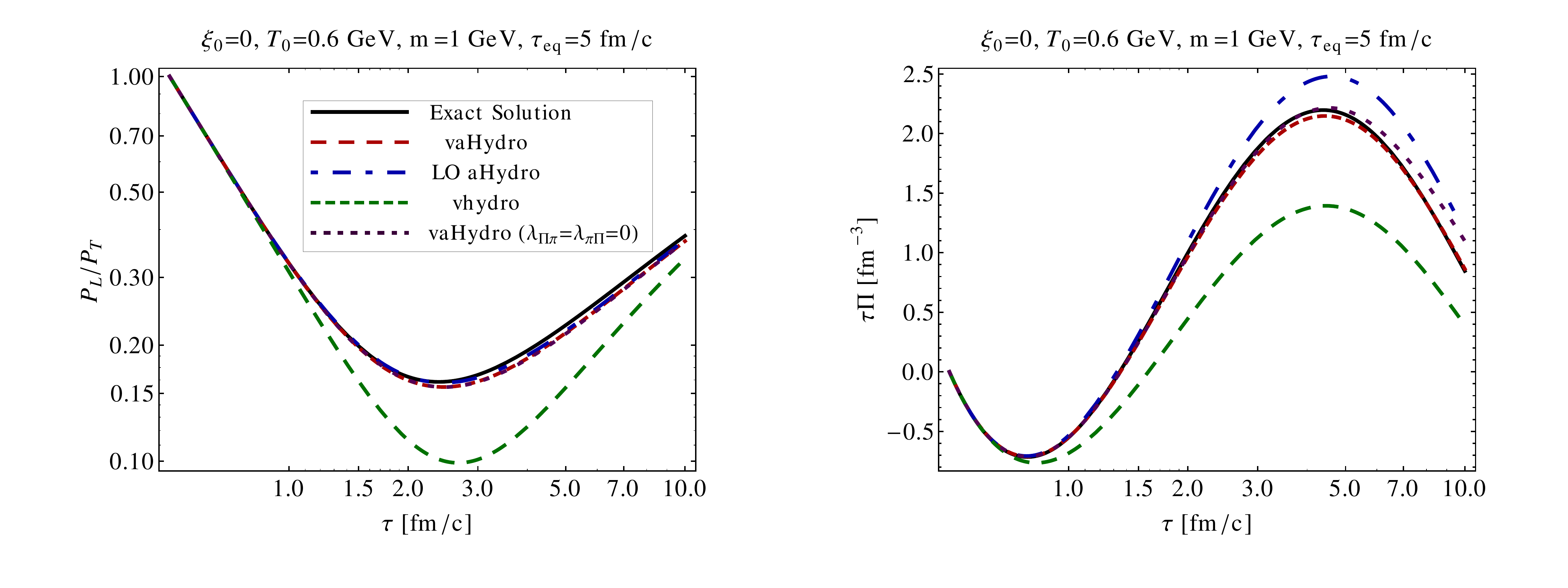}}
\vspace{-8mm}
\caption{(Color online) The left panel shows the pressure anisotropy $P_L/P_T$ as a function of proper time and the right panel shows the bulk viscous correction multiplied by the proper time, $\tau \Pi$, again as a function of proper time.  The exact solution to the Boltzmann equation is shown as a solid black line \cite{Florkowski:2014sfa}, vaHydro as a red long-dashed line, LO aHydro as a blue long dot-dashed line \cite{Nopoush:2014pfa}, second-order vHydro as a green short-dashed line \cite{Denicol:2010xn,Denicol:2012cn,Denicol:2012es}, and finally, vaHydro with the shear-bulk couplings turned off as a purple short-dashed line. Initial conditions, masses used, etc are indicated above each panel. Figure used with permission from Ref.~\cite{Bazow:2015cha}.} 
\label{fig:vahydroTEQ}
\end{figure}
%
\ba
\partial_\tau {\Pit}&=&-\Gamma\left(\frac{2P_\perp{+}P_L}{3}{-}P_\mathrm{eq}{+}\Pit\right)
\nonumber\\
&&
\hspace{-1cm}{+}\frac{m^2}{3\Lambda}\left(\J_{0,0,1}\frac{\partial_\tau\Lambda}{\Lambda} -\frac{1}{2}\J^{zz}_{0,0,-1}\partial_\tau{\xi}
-\frac{3}{2}\J_{2,1,-1}\partial_\tau{\Phi}
+\frac{1{+}\Phi{+}\xi}{\tau}\J^{zz}_{0,0,-1}\right) 
-\lambda_{\Pi\Pi}\frac{\Pit}{\tau}-\lambda_{\Pi\pi}\frac{\pit}{\tau}\,, \nonumber \\
\\
\partial_\tau{\pit}&=&-\Gamma\left(\pit-\frac{2}{3}(P_L-P_\perp)\right)
\nonumber\\
&&\hspace{-13mm}{+}\frac{1}{\Lambda}\left[\left(\J^{zz}_{0,0,1}-\J_{2,1,1}\right)\!\frac{\partial_\tau\Lambda}{\Lambda}+\left(\frac{1{+}\Phi{+}\xi}{\tau}-\frac{\partial_\tau\xi}{2}\right)\left(\J^{zzzz}_{0,0,-1}{-}\J^{zz}_{2,1,-1}\right)
-\frac{3}{2}\left(\J^{zz}_{2,1,-1}{-}\frac{5}{3}\J_{4,2,-1}\right)\!\partial_\tau\Phi
\right] 
\nonumber\\
&&+\lambda_{\pi\Pi}\frac{\Pit}{\tau}+\lambda_{\pi\pi}\frac{\pit}{\tau}\,.
\ea
Note that for 0+1d Bjorken expansion there is only one independent component of the shear stress tensor and above we see that both $\xi$ and $\tilde\pi$ appear as independent variables.  In this context, $\tilde\pi$ encodes the corrections to the pressure anisotropy which are not captured by the leading-order evolution of $\xi$.  It is possible to subsume all shear corrections into the leading-order anisotropy parameter using the so-called anisotropic matching principle which we will discuss in the next section. In Fig.~\ref{fig:vahydroTEQ}, we present a comparison presented originally in Ref.~\cite{Bazow:2015cha} between various hydrodynamical frameworks applied to the non-conformal Boltzmann equation subject to Bjorken flow.  As can be seen from this figure, the aHydro framework most accurately reproduces both the pressure anisotropy (related to the shear correction) and the bulk viscous correction when compared to the exact solution~\cite{Florkowski:2014sfa}.  Additionally, we see that vaHydro provides a clear quantitative improvement over the leading-order aHydro results.  For more information about the exact solution presented in Fig.~\ref{fig:vahydroTEQ}, see Sec.~\ref{sec:exactsolutions}.


\section{The anisotropic matching principle}
\label{sec:anisomatching}

As mentioned earlier, the most widely used method for obtaining the dynamical equations in the framework of aHydro is taking moments of the Boltzmann equation with a specific prescription for the collisional kernel.  However, this method suffers from an ambiguity in the choice of dynamical equations, because, it provides us with an uncontrolled number of equations which often does not match the number of dynamical variables. While this ambiguity has been successfully treated  in the framework of second-order vHydro \cite{Denicol:2014loa,Denicol:2014mca,Jaiswal:2014isa}, this treatment cannot be naively extended to aHydro, because in aHydro we have at least one more degree of freedom (e.g. longitudinal momentum-space anisotropy parameter).

An alternative method for obtaining the dynamical equations for the pressure corrections, e.g. $\pi^{\mu \nu}$ and $\Pi$, directly from the Boltzmann equation \cite{Tinti:2015xwa} is the so-called {\it anisotropic matching} principle, which is based on a generalization of the Landau matching condition. In conventional hydrodynamics approaches, the Landau matching condition allows one to fix the LRF of the fluid  and accordingly defines the effective temperature in such a way as to enforce energy conservation. In anisotropic matching, we basically use the new degrees of freedom (e.g. anisotropy parameters and bulk degree of freedom) to fix the dissipative pressure corrections.

\subsection{Extending Landau matching}

For elaboration of the idea, we remind the reader that in the general formulation of aHydro we have ten degrees of freedom, i.e. three independent components of  four velocity, momentum scale $\lambda$, bulk degree of freedom, and five independent anisotropy tensor components. This is exactly the same as the number of independent parameters of energy-momentum tensor 
\ba
T^{\mu\nu}=\epsilon\, u^\mu u^\nu-(P_{\rm eq}+\Pi)\Delta^{\mu\nu} +\pi^{\mu\nu}\,.
\ea  
Through implementation of the Landau matching condition in traditional hydrodynamics, where $f(x,p)=f_{\rm eq}(x,p)+\delta f(x,p)$, we fix only the fluid four-velocity, i.e.
\ba
u_\mu T^{\mu\nu}=\epsilon u^\nu \,,
\ea
or 
\ba
\Delta^\mu_\alpha u_\beta T^{\alpha \beta} =\int dP\, (p\cdot u)p^{\langle\mu\rangle} \delta f=0\,,
\ea
which also implies that
\ba
\epsilon= \int dP\, (p\cdot u)^2 f = \int dP\, (p\cdot u)^2 f_{\rm eq} \,.
\ea
One can generalize this method to aHydro, with $f(x,p)=f_{\rm a}(x,p)+\delta{\tilde f}$, by assuming that all components of $T^{\mu\nu}$ (including dissipative pressure corrections) can be reproduced using only  the leading-order anisotropic distribution function, i.e.
\ba
T^{\mu \nu}= \int dP\, p^\mu p^\nu f=\int dP\, p^\mu p^\nu f_{\rm a}\,,
\ea
with $f_a$ given by Eq.~(\ref{eq:fform}), which implies
\ba 
\int dP\, p^\mu p^\nu \delta{\tilde f}=0\,.
\ea

This anisotropic matching prescription \cite{Tinti:2015xwa} can be used to determine equations of motion for the dissipative corrections. Calculating the first moment of the Boltzmann equation one obtains the equations for evolution of the energy density and the fluid four-velocity, which are similar to those obtained using standard vHydro
\ba
D_u \epsilon &=& -\left( \epsilon +P_{\rm eq} + \Pi \right)\theta + \sigma_{\mu\nu}\pi^{\mu\nu}, 
 \label{eq:1st-mom-u} \\ 
 \left( \epsilon +P_{\rm eq} +\Pi \right)D_u u^\alpha &=& -\nabla^\alpha\left(P_{\rm eq} + \Pi \right) -\Delta^\alpha_\mu\partial_\nu \pi^{\mu\nu}\,.
 \label{eq:1st-mom}
\ea
In order to derive the equations for evolution of pressure corrections, one starts from  
\ba
D_u T^{\mu\nu}=\int  dP\, p ^\mu p^\nu D_u f \,,
\ea
to obtain the exact convective derivatives of the pressure corrections as
\ba
 D_u \pi^{\langle\mu\nu\rangle} &=& \int dP \, p^{\langle\mu}p^{\nu\rangle} \, D_u f \, , \\
 D_u \Pi &=& -\frac{1}{3}\int dP \left(p\cdot\Delta\cdot p \right) D_u f - D_u P_{\rm eq} \, .
\ea
Next, by using the definition of four-index projection operator and the Boltzmann equation itself, one has
\ba
  D_u \pi^{\langle\mu\nu\rangle} -{\cal C}^{\langle\mu\nu\rangle}_{-1} \!&=&\! -\Delta^{\mu\nu}_{\rho\sigma} \nabla_\alpha  \int dP \,\frac{ p^\rho p^\sigma p^\alpha  \,  f}{(p\cdot u)} \, -  \left(\sigma_{\rho \sigma} +\frac{1}{3} \, \theta \, \Delta_{\rho\sigma} \right) \int dP \,\frac{ p^{\langle\mu} p^{\nu\rangle} p^\rho p^\sigma  \,  f}{(p\cdot u)^2} \, , \nonumber \\
D_u \Pi + \frac{1}{3} \Delta_{\mu\nu}{\cal C}^{\mu\nu}_{-1} \!&=&\!-D_u P_{\rm eq} + \frac{1}{3}\Delta_{\mu\nu} \nabla_\rho  \int dP \,\frac{ p^\mu p^\nu p^\rho  \,  f}{(p\cdot u)} \, \nonumber\\&&\qquad\qquad\qquad\qquad\quad+ \frac{1}{3} \left(\sigma_{\rho \sigma} +\frac{1}{3} \, \theta \, \Delta_{\rho\sigma} \right)  \int dP \,\frac{ (p\cdot \Delta \cdot p) p^\rho p^\sigma  \,  f}{(p\cdot u)^2} \, , \;\;
\label{eq:matching-pi1}
\ea
with ${\cal C}^{\mu_1 \dots \mu_n}_r$ defined in Eq.~(\ref{eq:calcdef}).  
So far no approximation has been made and, correspondingly, the set of equations is not closed. Writing $f=f_a+\delta \tilde f$, the anisotropic matching principle tells us that $\delta \tilde f$ does not contribute to any of the terms appearing in Eqs.~(\ref{eq:1st-mom-u}) and (\ref{eq:1st-mom}) \cite{Tinti:2015xwa}. It is therefore natural to close the system of equations (\ref{eq:1st-mom-u})-(\ref{eq:matching-pi1}) by ignoring $\delta\tilde f$ everywhere, i.e. by substituting $f=f_a$ also in the collision terms and in all the integrals on the right-hand-side of Eqs. (\ref{eq:matching-pi1}).

To further simplify the equations one can assume RTA for the collisional kernel (\ref{eq:rtadef}).  Equations (\ref{eq:matching-pi1}) then become
\ba
  D_u \pi^{\langle\mu\nu\rangle} + \frac{1}{\tau_{\rm eq}} \pi^{\mu\nu} =  -  \left(\sigma_{\rho \sigma} +\frac{1}{3} \, \theta \, \Delta_{\rho\sigma}  \right) \int dP \,\frac{ p^{\langle\mu} p^{\nu\rangle} p^\rho p^\sigma  \,  f_a}{(p\cdot u)^2} - 2 \, \pi_\alpha^{<\mu}\sigma^{\nu>\alpha} \nonumber \\ 
  + 2\, P \, \sigma^{\mu\nu} -\frac{5}{3}\, \theta \, \pi^{\mu\nu} + 2 \, \pi_\alpha^{<\mu}\omega^{\nu >\alpha},  \label{eq:matching-1d-pi1} \\ 
D_u P + \frac{1}{\tau_{\rm eq}}\left( P-P_{\rm eq} \right) =  \frac{1}{3} \left( \sigma_{\rho \sigma} +\frac{1}{3} \, \theta \, \Delta_{\rho\sigma} \right)  \int dP \,\frac{ (p\cdot \Delta \cdot p) p^\rho p^\sigma  \,  f_a}{(p\cdot u)^2} 
 +\frac{2}{3} \, \pi_{\mu\nu}\sigma^{\mu\nu} - \frac{5}{3} \, P \,  \theta\,,
\label{eq:matching-1d-pi2}
\ea
with $P=P_{\rm eq}+\Pi$. 

\subsection{0+1d limit}

Although above we presented the anisotropic matching principle as it was originally introduced by Tinti \cite{Tinti:2015xwa}, an alternative derivation which yielded the same result was presented in Ref.~\cite{Molnar:2016gwq}.  Here we will demonstrate explicitly that the two methods are equivalent for a 0+1d conformal system, i.e. $\Pi=0$, $\pi^{\mu\nu}={\rm diag}\{0,\pi/2,\pi/2,-\pi\}$, $\sigma^{\mu\nu}={\rm diag}\{0,1,1,-2\}/(3\tau)$, $\omega^{\mu\nu}=0$, and $\theta=1/\tau$. Using these simplifications and adding the $zz$-projection of Eq.~(\ref{eq:matching-1d-pi1}) to Eq.~(\ref{eq:matching-1d-pi2}), one obtains
\ba
\partial_\tau P_L+\frac{1}{\tau}(3P_L-I_{240})=-\frac{1}{\tau_{\rm eq}}(P_L-P_{\rm eq} )\,,
\ea
with $P_L=P-\pi$ and
\ba
I_{240}=\int dP\,E^{-2} {p}_z^4  f(x,p)\,.
\ea 
The above equation is precisely that obtained in~Ref.~\cite{Molnar:2016gwq}.  We will present comparisons of the anisotropic matching principle with the method of taking momentum-moments of the Boltzmann equation in the next section.


\newcommand{\piti}{{\hat{\tilde{\pi}}}}
\newcommand{\besp}{\begin{split}}
\newcommand{\Ih}{\hat{I}}
\newcommand{\hint}{\hat{\mathcal{H}}}
\newcommand{\pp}{{\hat{p}}}
\newcommand{\It}{\tilde{I}}
\newcommand{\po}{\hat{p}_\Omega^2}
\newcommand{\R}{\hat{\mathcal{R}}}

\section{Testing against exact solutions to the Boltzmann equation}
\label{sec:exactsolutions}

One way to assess which hydrodynamical formalism, e.g. Israel-Stewart vHydro, DNMR vHydro, or aHydro, is the best is to make phenomenological predictions for typical experimental observables and then fit any free parameters, e.g. initial central temperature, freeze-out temperature(s), etc., and compare the fit quality among the various approaches.  However, since in practice the number of free parameters, e.g. transport coefficients, might be quite large (even infinite), this procedure would be replete with uncertainty, making it difficult to draw firm conclusions from such theory/data comparisons.  Another way to proceed is to find some cases in which exact solutions to a given microscopic model are possible and then to apply each hydrodynamical formalism to this microscopic model and determine which one best reproduces the exact solution.  This line of inquiry was followed by several authors in recent years in the context of the Boltzmann equation in RTA subject to some flow profiles that are similar to what is generated in URHICs, e.g. Bjorken and Gubser flows \cite{Florkowski:2013lza,Florkowski:2013lya,Florkowski:2014sfa,Denicol:2014tha,Denicol:2014xca}.

The starting point for all of the solutions presented in this section is the relativistic RTA Boltzmann equation specified by Eqs.~(\ref{eq:boltzmann1}) and (\ref{eq:rtadef}).  For the purposes of this review, we will assume a single component fluid and take the background equilibrium distribution function $f_{\rm eq}$ to be a classical Boltzmann distribution
\begin{equation}
f_{\rm eq} = \frac{2}{(2\pi)^3} \exp\left(- \frac{p \cdot u}{T} \right) .
\label{Boltzmann}
\end{equation}
However, it is possible to obtain the exact solution to the RTA Boltzmann equation for quantum distributions \cite{Florkowski:2014sda} and multi-component fluids \cite{Florkowski:2017jnz}.

\subsection{0+1d Bjorken Flow}

In 0+1d, the general solution of the RTA Boltzmann equation (\ref{eq:boltzmann1}) and (\ref{eq:rtadef}) can be expressed as~\cite{Florkowski:2013lza,Florkowski:2013lya,
Baym:1984np,Baym:1985tna,Heiselberg:1995sh,Wong:1996va}
\begin{equation}
f(\tau,w,p_T) = D(\tau,\tau_0) f_0(w,p_T)  + \int_{\tau_0}^\tau \frac{d\tau^\prime}{\tau_{\rm eq}(\tau^\prime)} \, D(\tau,\tau^\prime) \, 
f_{\rm eq}(\tau^\prime,w,p_T) \, ,  \label{solG}
\end{equation}
where
\begin{equation}
w \equiv  tp_L - z E \, ,
\label{w}
\end{equation}
and we have introduced the damping function
\begin{equation}
D(\tau_2,\tau_1) \equiv \exp\left[-\int_{\tau_1}^{\tau_2}
\frac{d\tau^{\prime\prime}}{\tau_{\rm eq}(\tau^{\prime\prime})} \right].
\end{equation}

\subsubsection{Exact solution to the 0+1d conformal RTA Boltzmann equation }

\begin{figure}[t]
\centerline{
\includegraphics[width=.475\linewidth]{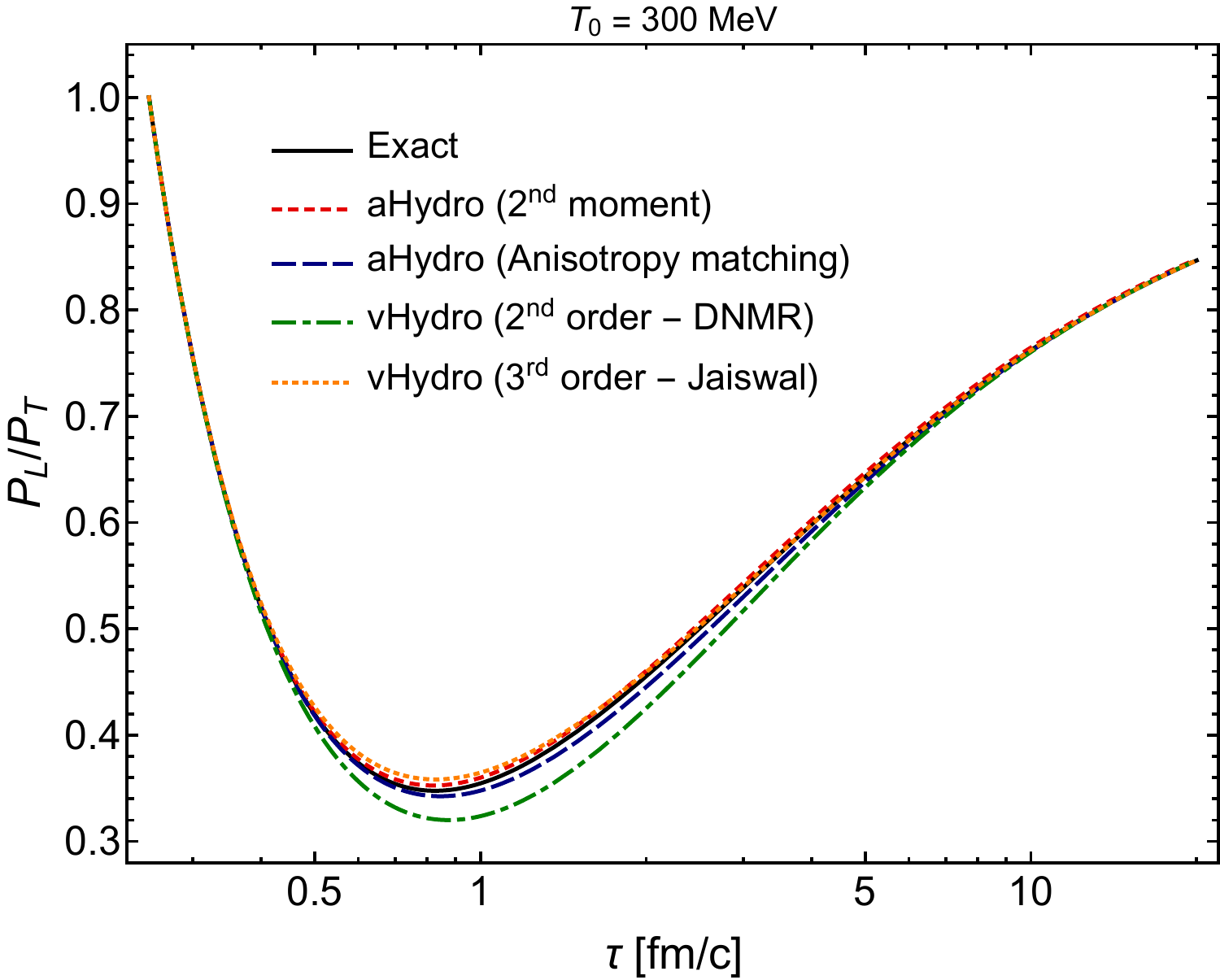}
\hspace{3mm}
\includegraphics[width=.475\linewidth]{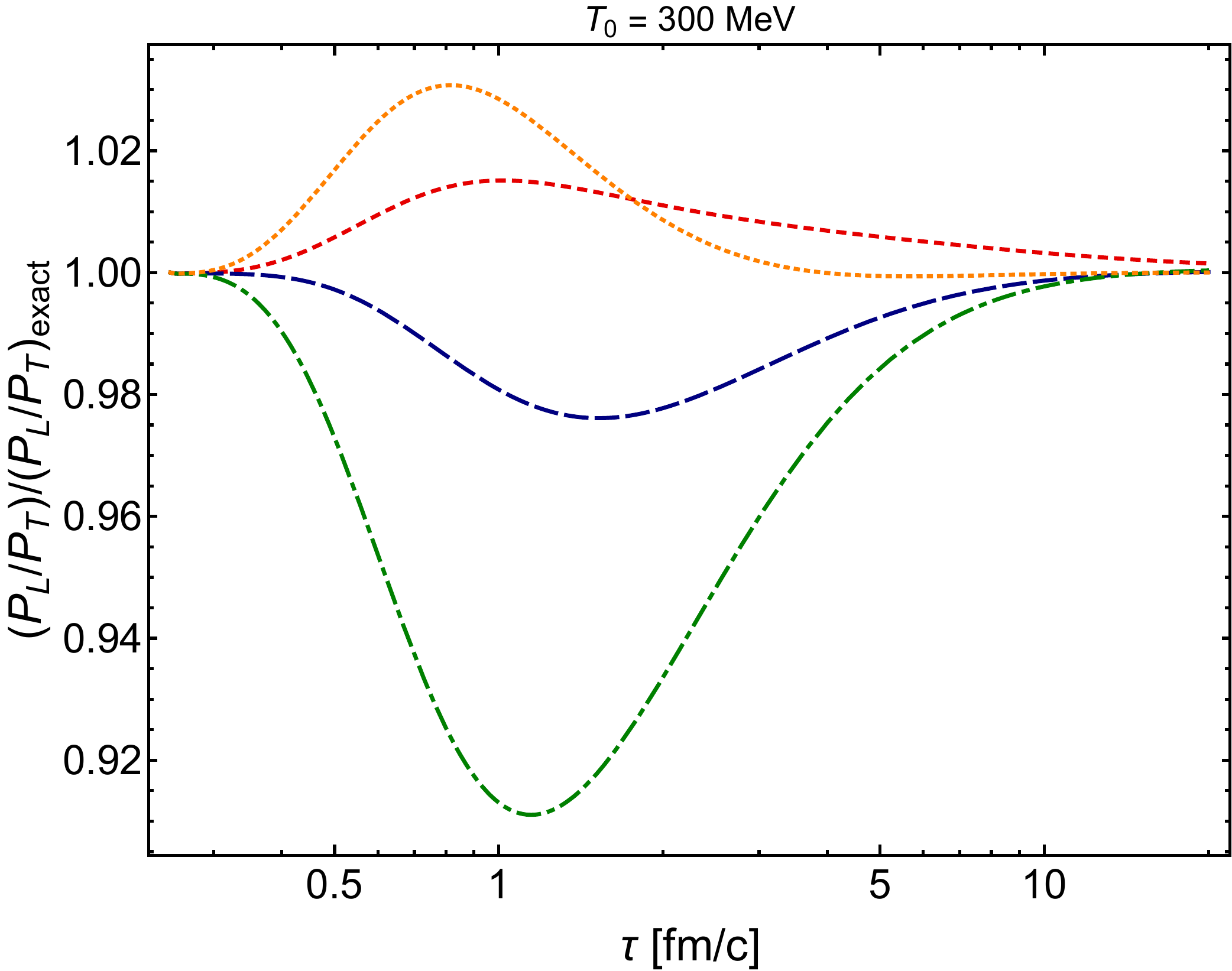}
}
\caption{(Color online) Comparisons of results obtained using various aHydro and vHydro dynamical equations and the exact solution of the 0+1d RTA Boltzmann equation for for a conformal system undergoing Bjorken flow.  A detailed description of the figure can be found in the text.}  
\label{fig:exactcomp1}
\end{figure}

Multiplying the left- and right-hand-sides of Eq.~(\ref{solG}) by $E^2$ and integrating both sides using the Lorentz-invariant integration measure $\int dP$ and assuming that the initial distribution function is of Romatschke-Strickland form (\ref{eq:rsform}), one obtains an integral equation for the energy density~\cite{Florkowski:2013lza,Florkowski:2013lya}
\be
\bar{\epsilon}(\tau) = D(\tau,\tau_0) \,
\frac{{\cal R}\big(\xi_{\rm FS}(\tau)\big)}{{\cal R}\left(\xi_0\right)}
+ \int_{\tau_0}^{\tau} \! \frac{d\tau^\prime}{\tau_{\rm eq}(\tau^\prime)} \, D(\tau,\tau^\prime) \, 
\bar{\epsilon}(\tau^\prime) \, {\cal R}\!\left( \! \left(\frac{\tau}{\tau^\prime}\right)^2 - 1 \right) ,
\label{eq:inteqconf}
\ee
where ${\bar{\epsilon} = {\epsilon}/{\epsilon}_0}$ is the energy density scaled by its initial value, ${\cal R}$ is defined in Eq.~(\ref{eq:Rfuncs}),  $\xi_0$ is the initial momentum-space anisotropy, ${\xi_{\rm FS}(\tau) = (1+\xi_0)(\tau/\tau_0)^2-1}$, and
\be
{D(\tau_2,\tau_1) = \exp\!\left[-\int_{\tau_1}^{\tau_2} d\tau^{\prime\prime} \, \tau^{-1}_{\rm eq}(\tau^{\prime\prime})\right]} ,
\ee
is the damping function.  This integral equation can be solved using the method of iteration:  one makes an initial guess for the energy density as a function of proper time in a given interval and then evaluates the right-hand-side of Eq.~(\ref{eq:inteqconf}) using discrete quadratures.  Once this step is done, one uses the result obtained from this procedure for the next iteration and keeps iterating until a desired convergence level is achieved in the interval considered.  In practice, this method converges reasonably quickly, allowing one to numerically obtain the exact solution.  
A public code which efficiently solves Eq.~(\ref{eq:inteqconf}) and computes the most relevant momentum-moments can be found here~\cite{MikeCodeDB}.  

Once the energy density as a function of proper time is obtained using this procedure one automatically obtains the effective temperature via the scaled fourth root of the energy density.  From this, one can compute the full distribution function using Eq.~({\ref{solG}}).  If one is interested in, for example, the transverse and longitudinal pressures, one can proceed taking moments of this equation, all relevant moments of the distribution function, e.g. once $T(\tau)$ is obtained, the longitudinal pressure can be computed using
\be
P_L(\tau) = D(\tau,\tau_0) \,
\frac{{\cal R}_L\big(\xi_{\rm FS}(\tau)\big)}{{\cal R}_L\left(\xi_0\right)}
+ \int_{\tau_0}^{\tau} \! \frac{d\tau^\prime}{\tau_{\rm eq}(\tau^\prime)} \, D(\tau,\tau^\prime) \, 
\bar{\epsilon}(\tau^\prime) \, {\cal R}_L\!\left( \! \left(\frac{\tau}{\tau^\prime}\right)^2 - 1 \right) .
\label{eq:plexact}
\ee 

In Fig.~\ref{fig:exactcomp1} we show comparisons of results obtained using various aHydro and vHydro dynamical equations and the exact solution of the 0+1d RTA Boltzmann equation for a conformal system undergoing Bjorken flow.  The left panel shows the pressure ratio $P_L/P_T$ as a function of proper time obtained from the exact solution (solid black line), the aHydro solution obtained from the trace-subtracted second moment (red dashed line), the aHydro solution obtained using the anisotropic matching principle i.e.~Eq.~(\ref{eq:matching-pi1}) with $f=f_a$ (blue long-dashed line), the second-order vHydro equations of DNMR (green dot-dashed line) \cite{Denicol:2012cn}, and the third-order vHydro equations of Jaiswal \cite{Jaiswal:2013vta} (red dotted line).  The right panel shows the ratio of the result obtained in each scheme to the exact result.  As can be seen from this figure, the two aHydro prescriptions perform the best compared to both 2nd-order and 3rd-order vHydro even though they are formally only leading-order in the aHydro expansion.  At late times, the 3rd-order result gives slightly higher accuracy in this specific case, however, we note that a systematic study varying both the initial conditions and value of $\eta/s$ shows that aHydro provides the most accurate dynamics compared to the exact solution, particularly for initial conditions which deviate strongly from isotropy and/or large values of $\eta/s$.  Finally, we note that the accuracy of aHydro can be even further increased by going to second-order using vaHydro \cite{Bazow:2013ifa,Bazow:2015cha}.  We will present some comparisons of the vaHydro result with exact solutions in the subsequent discussion.

\subsubsection{Exact solution to the 0+1d non-conformal RTA Boltzmann equation}

\begin{figure}[t]
\centerline{\includegraphics[width=.65\linewidth]{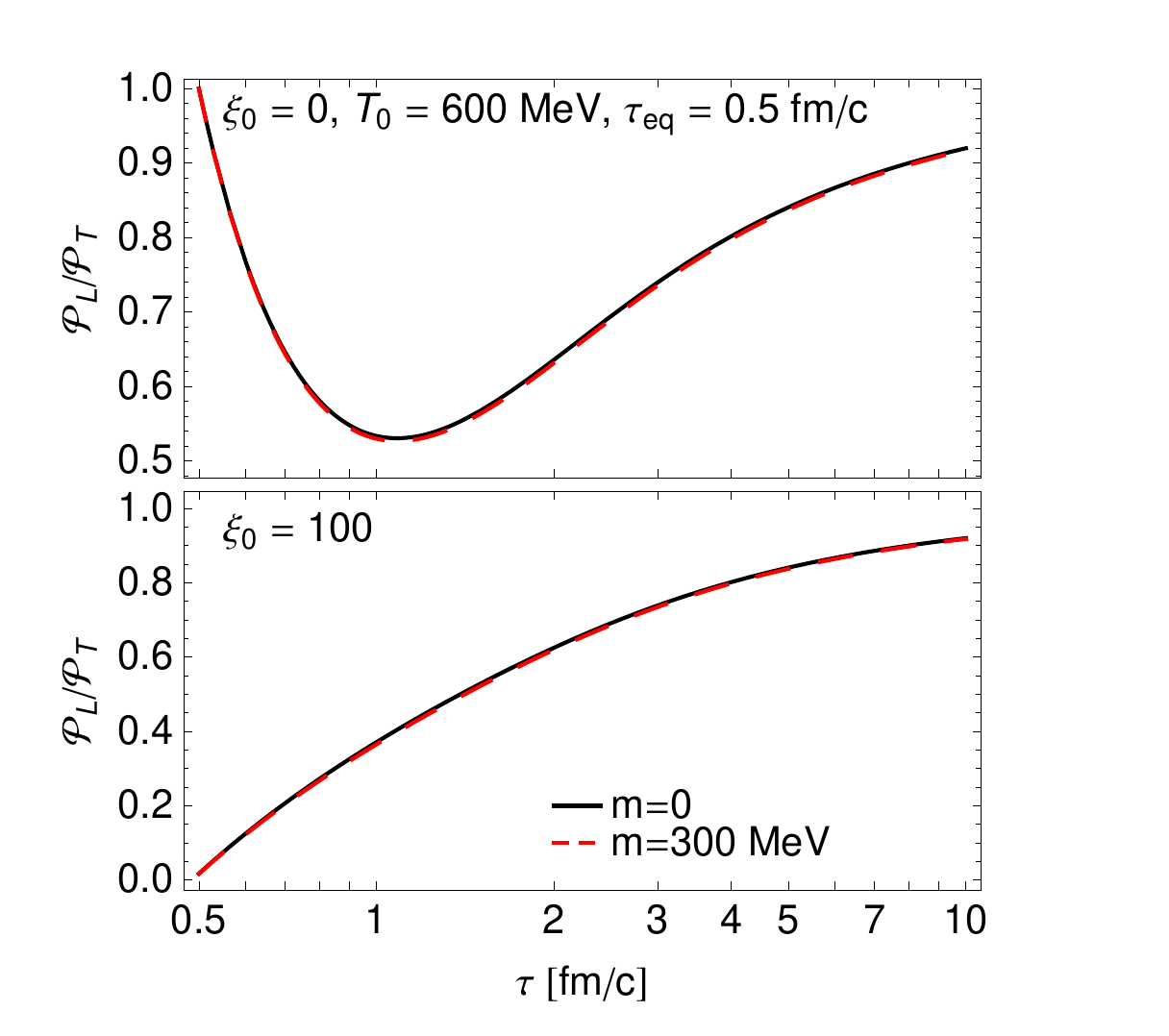}}
\caption{(Color online) Comparison of the evolution of the pressure anisotropy for two different initial momentum anisotropies obtained from numerical solution to Eq.~(\ref{eq:LM2}).  The top and bottom panels show the result for isotropic and oblate initial momentum-space anistropies, respectively.  In both panels, the solid black line indicates the massless solution and the red dashed line indicates the solution obtained assuming $m = 300$ MeV.    Figure used with permission from Ref.~\cite{Florkowski:2014sfa}. 
}  
\label{fig:PL2PT}
\end{figure}

\begin{figure}[t!]
\centerline{\includegraphics[width=0.95\linewidth]{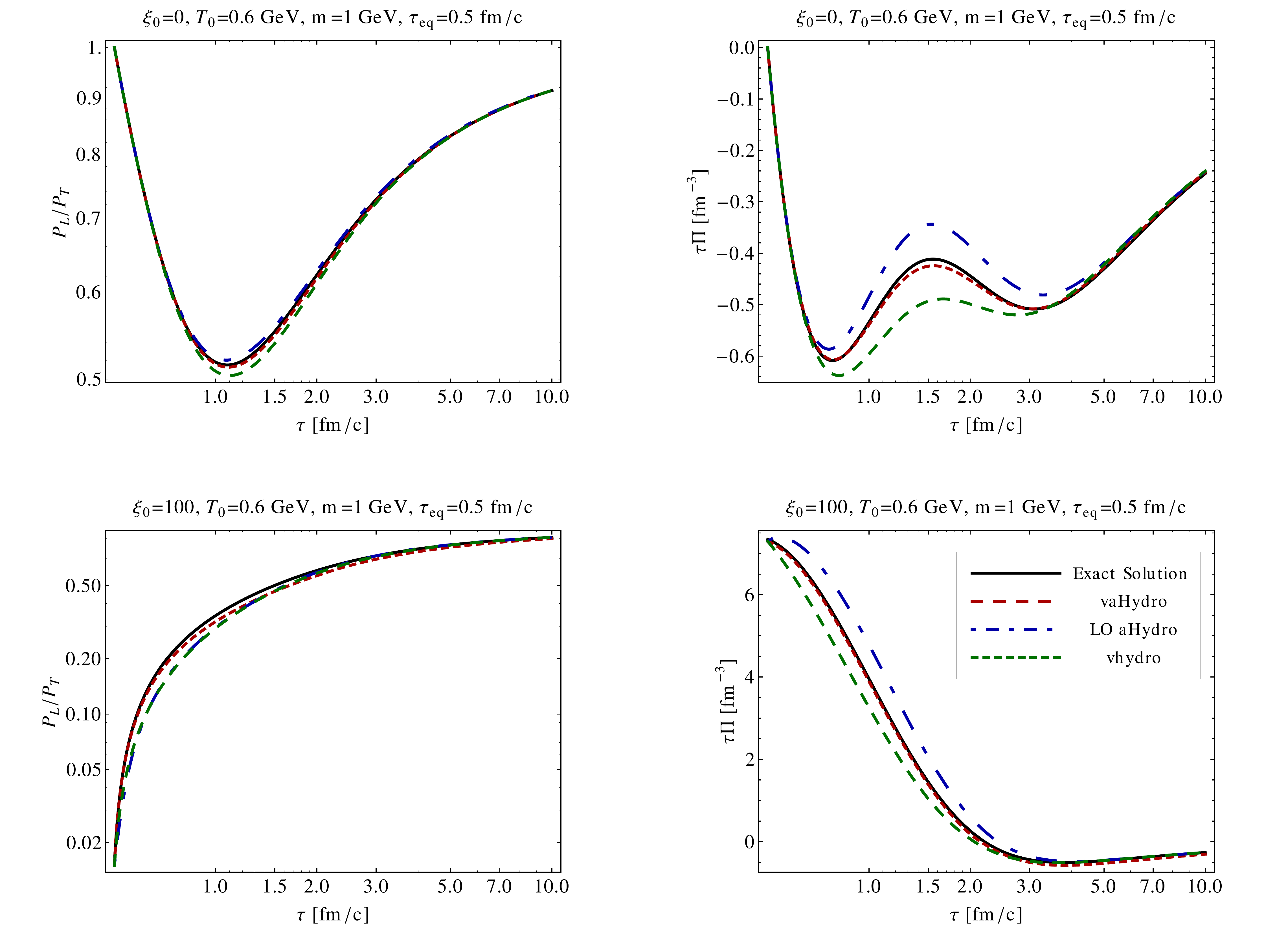}}
\vspace{-5mm}
\caption{(Color online) Comparison of the exact solution obtained using Eq.~(\ref{eq:LM2}) (solid black line), vHydro (green long-dashed line) \cite{Jaiswal:2014isa}, leading-order aHydro (blue dot-dashed line) \cite{Nopoush:2014pfa}, and vaHydro (green dashed line) \cite{Bazow:2015cha}. The left panels show the pressure anisotropy as a function of proper time assuming an initially isotropic state (top) or oblate anisotropic state (bottom).  The right panels show the bulk correction to the pressure times the proper time as  function of proper time.  Once again the top and bottom panels show two different initial momentum-space anisotropies.  Figure used with permission from Ref.~\cite{Bazow:2015cha}.}  
\label{fig:exactcomp2}
\end{figure}

A similar procedure can be used to solve the 0+1d RTA Boltzmann equation for massive particles allowing one to study the impact of the breaking of conformal symmetry and to assess which dissipative hydrodynamics framework provides the most accurate description of the bulk viscous correction to the pressure.  The resulting integral equation is \cite{Florkowski:2014sfa}
\begin{eqnarray}
&&  \hspace{-5mm}
2 m^2 T(\tau) \left[ 3 T(\tau) K_2\left(\frac{m}{T(\tau)}\right)
+ m K_1\left(\frac{m}{T(\tau)}\right) \right]
\nonumber \\
&& 
 = D(\tau,\tau_0) \Lambda^4_0 \tilde{\cal H}_2\left[ \frac{\tau_0}{\tau \sqrt{1+\xi_0}},\frac{m}{\Lambda_0}\right]  + \int\limits_{\tau_0}^\tau 
\frac{d\tau^\prime}{\tau_{\rm eq}(\tau')} D(\tau,\tau^\prime)
T^4(\tau^\prime) 
\tilde{\cal H}_2\left[ \frac{\tau^\prime}{\tau},\frac{m}{T(\tau^\prime)}\right] ,
\label{eq:LM2}
\end{eqnarray}
where
\begin{equation}
{\cal H}_2(y,\zeta) 
= y \left( \sqrt{y^2+\zeta^2} + \frac{1+\zeta^2}{\sqrt{y^2-1}}
\tanh^{-1} \sqrt{\frac{y^2-1}{y^2+\zeta^2}} \, \right) .
\label{eq:H2an}
\end{equation}
In this case the integral equation is written in terms of the effective temperature $T(\tau)$. It can, once again, be solved using the iterative method \cite{Banerjee:1989by}. In the massless limit ($m \to 0$), Eq.~(\ref{eq:LM2}) reduces to Eq.~(\ref{eq:inteqconf}).

In Fig.~\ref{fig:PL2PT} we show a comparison of the evolution of the pressure anisotropy for two different initial momentum anisotropies obtained from numerical solution to Eq.~(\ref{eq:LM2}).  The top and bottom panels show the results for isotropic and oblate initial momentum-space anistropies, respectively.  In both panels, the solid black line indicates the massless solution and the red dashed line indicates the solution obtained assuming $m = 300$ MeV.  As can be seen from this figure, the mass of the particle does not have any appreciable effect on the level of momentum-space anisotropy emerging during the evolution.  Next, in Fig.~\ref{fig:exactcomp2} we show comparison of the exact solution obtained using Eq.~(\ref{eq:LM2}) (solid black line), vHydro (green long-dashed line) \cite{Jaiswal:2014isa}, leading-order aHydro (blue dot-dashed line) \cite{Nopoush:2014pfa}, and vaHydro (green dashed line) \cite{Bazow:2015cha}. In this figure, we show the evolution of the pressure anisotropy in the left column and the bulk viscous correction to the pressure in the right column.  As can be seen from this figure, all schemes show a good qualitative agreement with the exact solution, however, vaHydro proves to be the most accurate prescription as might be expected since it is second-order in the aHydro expansion.\footnote{Further comparisons of different leading-order aHydro schemes with the exact solution in the non-conformal case can be found in Ref.~\cite{Tinti:2015xra}.}  Note, importantly, that the non-conformal Israel-Stewart equations fail to describe the exact solution even qualitatively because this, now rather dated, formalism neglects the coupling between the shear and bulk corrections to the pressures in the dynamical equations \cite{Denicol:2014mca}.

\subsection{1+1d Gubser flow}

In order to have a test case that is more closely connected to heavy-ion collisions, it is desirable to have exact solutions to the Boltzmann equation in a case in which there is also transverse expansion.  Such a profile is provided by the so-called Gubser flow~\cite{Gubser:2010ui,Gubser:2010ze}.\footnote{In this section, we use the mostly plus convention for the Minkowski metric, i.e. $g^{\mu\nu} = {\rm diag}(-1,1,1,1)$, in order to connect more easily to existing literature on Gubser flow.}  Gubser flow is established based on symmetries:  one assumes that the system is boost invariant, cylindrically symmetric with respect to the beam line at all times, and reflection symmetric about the xy-plane.  With these assumptions, one can construct a flow with SO(3)$_q\,{\otimes}\,$SO(1,1)$\,{\otimes}\,$Z$_2$ symmetry~\cite{Gubser:2010ui,Gubser:2010ze}.  The requirement of SO(3)$_q$ symmetry couples the temporal and radial dependence, resulting in a non-trivial radial flow pattern which makes Gubser flow unique.  As a result of the imposed symmetries, all dynamical variables only depend on $\tau = \sqrt{t^2-z^2}$ and $r = \sqrt{x^2+y^2}$ through the dimensionless combination $G(\tau,r)=(1-q^2{\tau}^2+q^2r^2)/(2 q{\tau})$, where $q$ is an arbitrary energy scale that sets the physical size of the system.  The resulting flow profile is $\tilde{u}^\mu = (\cosh\theta_\perp,\sinh\theta_\perp,0,0)$, with $\tanh\theta_\perp \equiv 2q^2\tau r/(1+q^2\tau^2+q^2r^2)$.  In this expression, the tilde indicates polar Milne coordinates with position four-vector $\tilde{x}^\mu=(\tau,\,r,\,\phi,\,\varsigma)$ and $\phi = \tan^{-1}(y/x)$ and $\varsigma$ is the spatial rapidity.  

To map this to a static flow, one performs a Weyl-rescaling and then a change of variables to de Sitter coordinates $\sinh{\rho} =  - (1-q^2{\tau^2}+q^2r^2)/(2q{\tau})$ and $\tan{\theta} = 2qr/(1+q^2{\tau}^2-q^2r^2)$. The variable $\rho$ is the ''de Sitter time'' and $\theta$ is an angular variable and, due to the symmetries of this flow, physical quantities only depend on $\rho$.  At fixed $r$, the limit $\tau \rightarrow 0^+$ corresponds to the limit $\rho \rightarrow -\infty$ and the limit $\tau\rightarrow\infty$ corresponds to the limit $\rho \rightarrow \infty$.  This means that the de Sitter map covers the entire forward light cone.  In the text that follows, Weyl-rescaled de Sitter-space quantities are indicated with a hat.

\begin{figure}
\centerline{\includegraphics[width=0.475\textwidth]{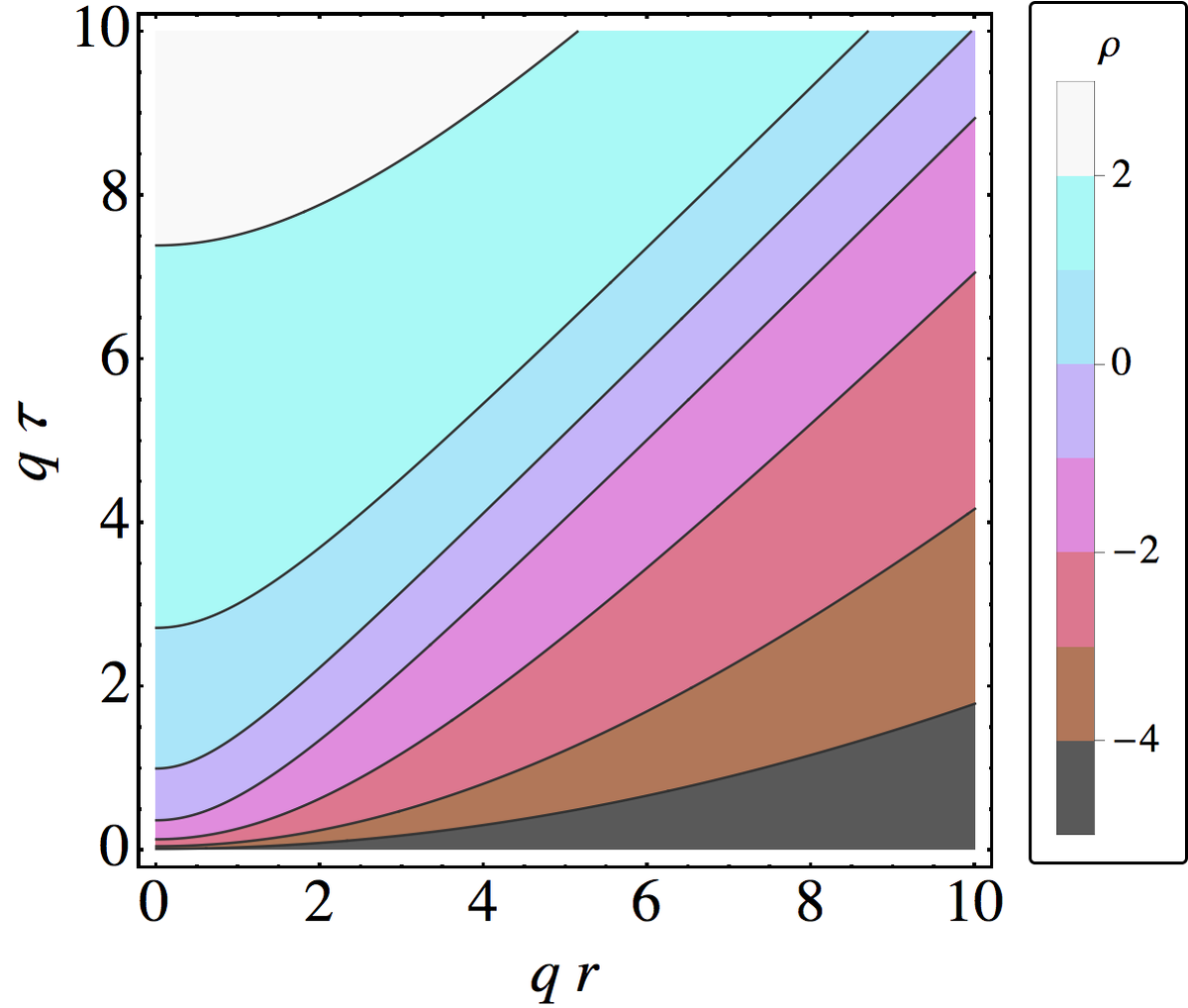}}
\caption{(Color online) Lines of constant $\rho$ in the $(\tau,r)$ 
plane. The origin in de Sitter time, $\rho{\,=\,}0$, corresponds to the line 
going through $(q\tau,qr){\,=\,}(1,0)$ and the upper right corner of the 
graph.  Figure used with permission from Ref.~\cite{Denicol:2014tha}.
}
\label{F1}
\end{figure}

Because, using this setup, the system effectively maps to a one-dimensional problem in Weyl-rescaled de Sitter-space, one can construct the exact solution in a similar manner as was done with Bjorken flow in the previous subsection.  One starts with the Boltzmann equation in Weyl-rescaled de Sitter-space, constructs the exact solution for the one-particle distribution function, and then takes Weyl-rescaled de Sitter-space momentum moments of the exact solution to obtain integral equations for the moments of interest.  The resulting one-dimensional integral equation for the Weyl-rescaled de Sitter-space energy density is \cite{Denicol:2014xca,Denicol:2014tha}
\be
{\hat{\varepsilon}}(\rho ) =\frac{3}{\pi ^{2}}\left[ D(\rho ,\rho _{0})\mathcal{H}\!\left( \frac{%
\cosh \rho _{0}}{\cosh \rho }\right) \hat{T}_{0}^{4}+\frac{1}{c}\int_{\rho
_{0}}^{\rho }d\rho ^{\prime }\,D(\rho ,\rho ^{\prime })\,\mathcal{H}\!\left( 
\frac{\cosh \rho ^{\prime }}{\cosh \rho }\right) \,\hat{T}^{5}(\rho ^{\prime
})\right] ,
\label{eq:exactene}
\ee
where $c = 5 \eta/s$,
\be
D(\rho_2,\rho_1)=\exp\!\left(-\int_{\rho_1}^{\rho_2} d\rho''\,
\frac{\hat T(\rho'')}{c} \right) ,
\label{defineD}
\ee
and
\be
\mathcal{H}(x)=\frac{1}{2}\left( x^{2}+x^{4}\frac{\tanh ^{-1}
\left( \sqrt{1{-}x^{2}}\right) }{\sqrt{1{-}x^{2}}}\right) .  
\label{eq:calh}
\ee
As before, one can solve this integral equation using the method of iteration. A code for this purpose can be downloaded here \cite{MikeCodeDB}.

\subsubsection{Leading-order aHydro for Gubser flow}

We now introduce our ansatz for the leading-order aHydro one-particle distribution function.  Since the system is cylindrically symmetric with respect to the beam line, the de Sitter space anisotropy tensor can be assumed to be diagonal.\footnote{Any off-diagonal contributions quickly relax to zero even if they are initially non-zero \cite{Marrochio:2013wla}.}   An ellipsoidally anisotropic distribution function can be constructed by introducing a tensor of the form $\hat{\Xi}^{\mu\nu}=\hat{u}^\mu \hat{u}^\nu+\hat{\xi}^{\mu\nu}$, where $\hat{u}^\mu$ is the de Sitter-space four-velocity and $\hat{\xi}^{\mu\nu}$ is a symmetric traceless anisotropy tensor \cite{Nopoush:2014qba,Strickland:2015utc}.  Expanding $\hat{\xi}^{\mu\nu}$ in the de Sitter-space basis gives $\hat{\xi}^{\mu\nu} = \hat{\xi}_\theta \hat{\Theta}^\mu \hat{\Theta}^\nu+\hat{\xi}_\phi \hat{\Phi}^\mu \hat{\Phi}^\nu+\hat{\xi}_\varsigma \hat{\varsigma}^\mu \hat{\varsigma}^\nu$, where $\hat{\Theta}^\mu$, $\hat{\Phi}^\mu$, and $\hat{\varsigma}^\nu$ are Weyl-rescaled de Sitter space basis vectors which obey $\hat{u}^\mu \hat{u}_\mu =-1$, $\hat{\Theta}^\mu\hat{\Theta}_\mu=1$, $\hat{\Phi}^\mu\hat{\Phi}_\mu=1$, $\hat{\varsigma}^\mu\hat{\varsigma}_\mu=1$.  The anisotropy tensor is traceless, i.e. $\hat{\xi}^{\mu}_{\ \mu} = 0$, and orthogonal to the flow, i.e. $\hat{u}_\mu \hat{\xi}^{\mu\nu} = 0$.  Using the tensor $\hat{\Xi}^{\mu\nu}$, one can construct an anisotropic distribution function for a conformal system~\cite{Nopoush:2014qba,Strickland:2015utc}
\be
f(\hat{x},\hat{p})=f_{\rm eq}\left(\frac{1}{\hat\lambda}\sqrt{\hat{p}_\mu\hat{\Xi}^{\mu\nu} \hat{p}_\nu}\right)\, ,
\label{eq:pdf}
\ee
where we have assumed vanishing chemical potential and $\hat\lambda$ is a non-equilibrium scale (transverse temperature) which can be identified with the de Sitter-space temperature, $\hat{T}$, only when $\hat\xi^{\mu\nu}=0$.

To determine the $\rho$-dependence of the scale $\hat\lambda$ and anisotropies $\hat\xi_i$, we take moments of the Boltzmann equation in RTA $\hat{p} \cdot D f = \hat{p}\cdot \hat{u} \, (f-f_{\rm eq})/\hat\tau_{\rm eq}$, where $D_\mu$ is the covariant derivative, and $\hat\tau_{\rm eq}$ is the Weyl-rescaled relaxation time. For a conformal system in RTA, one has $\hat\tau_{\rm eq} = 5\hat{\bar\eta}/\hat{T}$, where $\hat{\bar\eta}=\hat\eta/\hat{s}=\eta/s$ with $\hat\eta$ and $\hat{s}$ being the Weyl-rescaled shear viscosity and entropy density, respectively.  Taking the first and second moments of the Boltzmann equation in de Sitter coordinates, one obtains two coupled ordinary differential equations~\cite{Nopoush:2014qba}
\ba
4\frac{d\log\hat\lambda}{d\rho}+\frac{3 \hat\alpha_\varsigma^2\left(\frac{H_{2
   L}(\bar{y})}{H_2(\bar{y})}+1\right)-4}{3\hat\alpha_\varsigma^2-1} \, \frac{d\log\hat\alpha_\varsigma}{d\rho}+ \tanh\rho\left(\frac{H_{2T}(\bar{y})}{H_2(\bar{y})}+2\right) &=& 0\, ,
\label{eq:1st-mom-final2} \\
\frac{6\hat\alpha_{\varsigma }}{1-3 \hat\alpha _\varsigma ^2} \frac{d \hat\alpha_\varsigma}{d\rho} -\frac{3 \left(3 \hat\alpha_\varsigma^4-4\hat\alpha_\varsigma^2+1\right)}{4\hat\tau_{\rm eq} \hat\alpha _{\varsigma }^5} \left(\frac{\hat{T}}{\hat\lambda}\right)^5+2\tanh\rho=0 \, ,
\label{eq:2nd-mom-final2}
\ea
where $\hat\alpha_\varsigma \equiv (1+\hat{\xi}_\varsigma)^{-1/2}$, $\bar{y}^2 \equiv (3\hat\alpha_\varsigma^2-1)/2$, and $\hat{T}=\hat\alpha_\varsigma \hat\lambda \left(H_2(\bar{y})/2\right)^{1/4}\!/\bar{y}$. The definitions of the $H$-functions appearing above can be found in Ref.~\cite{Nopoush:2014qba}.

\begin{figure}[t]
\centerline{$\;\;$\includegraphics[width=1.05\linewidth]{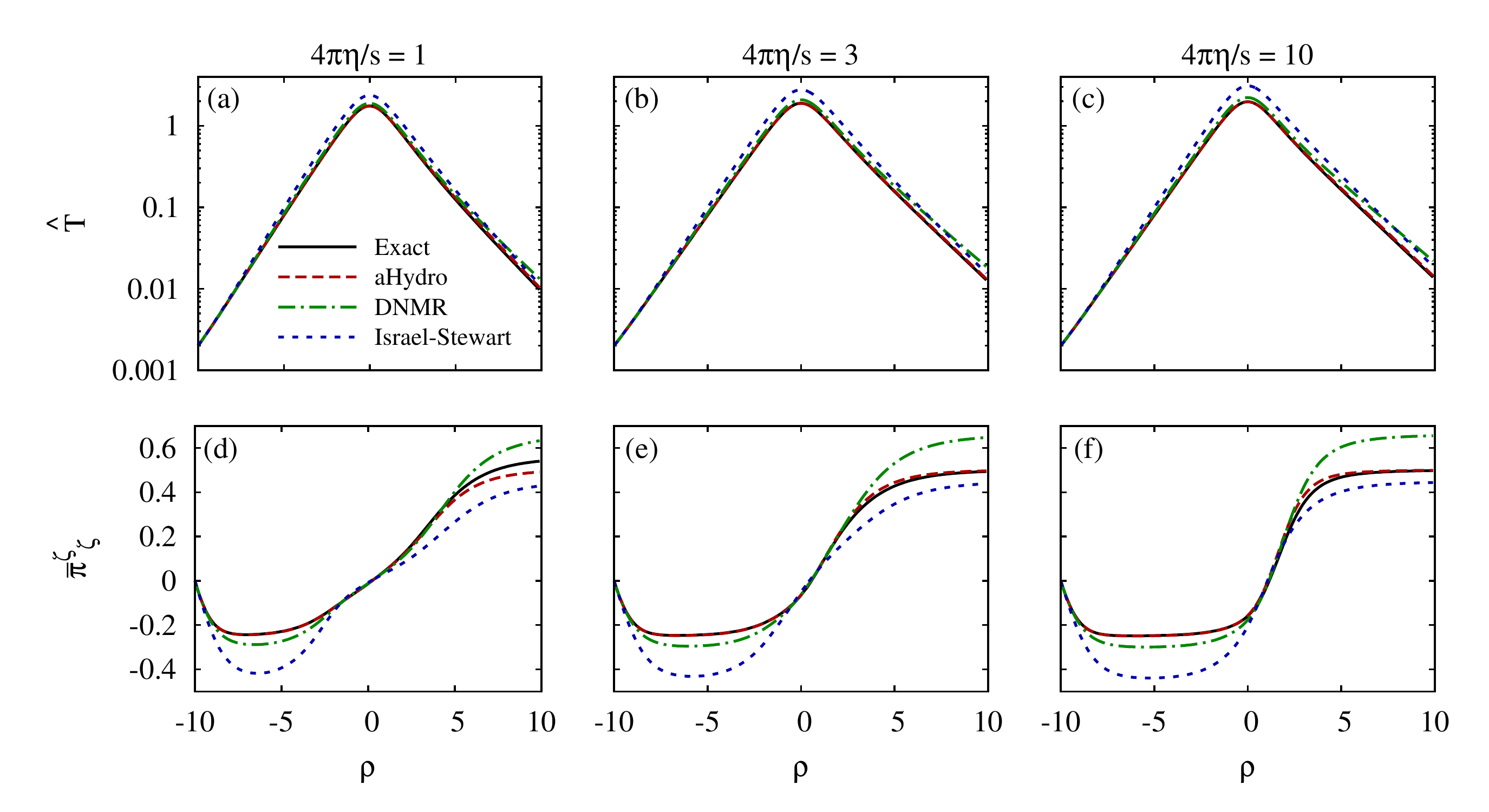}\hspace{5mm}}
\vspace{-4mm}
\caption{(Color online) In the top row, we compare the de Sitter-space effective temperature $\hat{T}$ obtained from the exact kinetic solution obtained in Refs.~\cite{Denicol:2014xca,Denicol:2014tha} (black solid line), the leading-order aHydro equations obtained in Ref.~\cite{Nopoush:2014qba} (red dashed line), DNMR second-order vHydro obtained in Ref.~\cite{Marrochio:2013wla} (green dot-dashed line), and Israel-Stewart second-order vHydro obtained in Ref.~\cite{Marrochio:2013wla} (blue dotted line).  Figure used with permission from Ref.~\cite{Denicol:2014tha}.
}
\label{fig:fig-rho-10-T0p002-az1}
\end{figure}

In Fig.~\ref{fig:fig-rho-10-T0p002-az1} we present a comparison of the exact solution of the RTA Boltzmann equation subject to Gubser flow with various dissipative hydrodynamics approaches.  The columns in Fig.~\ref{fig:fig-rho-10-T0p002-az1} (from left to right) correspond to three different choices of the shear viscosity to entropy density ratio with $4 \pi \eta/s \in \{1,3,10\}$, respectively.  In the top row, we compare results for the scaled effective temperature obtained using varies approaches and, in the bottom row, we compare results for the scaled shear $\bar{\pi}_{\varsigma}^{\varsigma} \equiv \hat{\pi}_\varsigma^\varsigma/(\hat{T}\hat{s})$ obtained with the same approaches.  The and values of $4 \pi \eta/s$ in the bottom row are the same as in the top row.  In all cases, at $\rho=\rho_0=-10$, we fixed the initial effective temperature to be $\hat{T}_0 = 0.002$ and the initial anisotropy to be  $\hat\alpha_{\varsigma,0}=1$, which corresponds to an isotropic thermal equilibrium initial condition in de Sitter space.  

As can be seen from this figure, the leading-order aHydro equations obtained in Ref.~\cite{Nopoush:2014qba} provide the best approximation to the exact result.  From the top row, we see that it is very difficult to distinguish the aHydro result for the effective temperature from the exact result.  From the bottom row, we see that the scaled shear $\bar{\pi}_{\varsigma}^{\varsigma} \equiv \hat{\pi}_\varsigma^\varsigma/(\hat{T}\hat{s})$ has visible differences between the aHydro solutions and the exact solution in the region above $\rho \gtrsim  0$, however, in all cases, at large $\rho$, one sees that aHydro has the correct asymptotic behavior unlike the other approaches considered.  The latter observation can be proven analytically~\cite{Nopoush:2014qba}.  In Ref.~\cite{Nopoush:2014qba} anisotropic initial conditions were also considered with the conclusion being the same.  Based on these findings the authors of Ref.~\cite{Nopoush:2014qba} concluded that leading-order aHydro described the spatio-temporal evolution of the system better than all dissipative hydrodynamics approaches known at the time.

\subsubsection{Second-order aHydro for Gubser Flow}

It is possible to further improve the agreement between aHydro and the exact solution by going to second-order (vaHydro) \cite{Martinez:2017ibh}.  In Ref.~\cite{Martinez:2017ibh} two different improved aHydro approaches were considered:  (1) Using the anisotropic matching principle at leading-order \cite{Tinti:2015xwa} and (2) computing the second-order (viscous) corrections to the Nopoush-Ryblewski-Strickland (NRS) solution presented in the previous subsection.  We would now like to summarize their main findings for both cases.

{\em Anisotropic Matching Principle:}~~This method uses the first moment of the Boltzmann equation to enforce energy conservation
\be
\partial_\rho\hat{\epsilon} + \tanh\rho\left(\frac{8}{3}\hat{\epsilon}{-}\hat{\pi}\right) = 0 \, .
\ee
The anisotropic matching principle \cite{Tinti:2015xwa} is then used to obtain the equation of motion for $\hat{\pi}$ \cite{Martinez:2017ibh}
\be
\partial_\rho\hat{\bar\pi} + \frac{\hat{\bar\pi}}{\hat{\tau}_r} =
   \frac{4}{3}\tanh\rho\left(\frac{5}{16} + \hat{\bar\pi} - \hat{\bar\pi}^2 - \frac{9}{16}\mathcal{F}(\hat{\bar\pi})\right) ,
\ee
where
\be
\label{eq:pibarxi}
   \hat{\bar\pi}(\xi) = \frac{3\hat{\pi}}{4\hat\epsilon}
   = \frac{1}{4}\left(\frac{3\,{\cal R}_{220}(\xi)}{{\cal R}_{200}(\xi)}-1\right) ,
\ee
and
\be
\label{eq:F}
   \mathcal{F}(\hat{\bar\pi}) \equiv \frac{{\cal R}_{240}\bigl(\xi(\hat{\bar\pi})\bigr)}{{\cal R}_{200}\bigl(\xi(\hat{\bar\pi})\bigr)} \, .
\ee
In the equations above, $\xi(\hat{\bar\pi})$ is understood to be the inverse of the function of $\hat{\bar\pi}(\xi)$ specified in Eq.~(\ref{eq:pibarxi}).

The special functions required are
\ba
{\cal R}_{200}(\xi) &=& {\cal R}(\xi) \, , \nonumber \\
{\cal R}_{220}(\xi) &=& -\frac{1}{\xi}\left[\frac{1}{1+\xi} - {\cal R}(\xi) \right] \, , \nonumber \\
{\cal R}_{240}(\xi) &=& \frac{1}{\xi^2} \left[\frac{3+\xi}{1+\xi} - 3 {\cal R}(\xi) \right] \, ,
\ea
with ${\cal R}(\xi)$ defined in Eq.~(\ref{eq:Rfuncs}).

{\em NRS scheme at NLO:}~~The leading-order NRS equations were obtained from the $zz$ projection minus one third of the sum of the $xx$ + $yy$ + $zz$ projections of the second moment of the Boltzmann equation.  At second-order one includes a viscous correction to the energy-momentum tensor such that $\hat{P}_L = \hat{P}_L^{\rm RS} + \piti$.  The differential equations necessary can once again be expressed in terms of moments of the distribution function, however, now the equations involve the residual viscous correction $\piti$.  The first-moment equation (energy conservation equation) becomes
\be
\partial_\rho\hat{\epsilon} + \tanh\rho\left(\frac{8}{3}\hat{\epsilon}{-}\hat{\pi}_{\rm RS}\right) = \piti\tanh\rho \, ,
\ee
where we have used  $\hat{P}_L^{\rm RS} = \hat{P}_0 + \hat{\pi}_{\rm RS} = \hat{\epsilon}/3 + \hat{\pi}_{\rm RS} $.

From the second-moment of the Boltzmann equation one can obtain an equation for $\xi$ using the NRS prescription
\be
  \partial_\rho\xi  + \frac{\xi(1+\xi)^{3/2}\,\R^{5/4}_{200}(\xi)}{\hat{\tau}_r}
  = - 2\tanh\rho\, (1{+}\xi) \, ,
\ee
and an equation of motion for the residual shear pressure~\cite{Martinez:2017ibh}
\ba
\label{eq:ressheareq2}
\partial_\rho\piti &=&
     - \frac{\hat{\pi}_\mathrm{RS}{+}\piti}{\hat{\tau}_{r}}
         - \tanh\rho\left[\frac{4}{3}\piti + \hat{\alpha}\,\Ih_{240} - \hat{\beta}\,\Ih_{340} 
                              + \frac{4}{3}\hat{\omega}\,\Ih_{440} 
                              + \frac{1}{2}\hat{\omega}_{\langle\eta\eta\rangle}\left(3\,\Ih_{460}{-}\Ih_{440}\right)\,\right]
\nonumber \\
&& \hspace{1cm}
    - \frac{\partial_\rho{\hat{\Lambda}}}{\hat{\Lambda}^2}\,\left(\hint_{221}{-}\frac{1}{3}\hint_{201}\right)	
      - \frac{\tanh\rho}{\hat{\Lambda}}\left(\frac{4}{3}\hint_{42-1}{-}\hint_{44-1}-\frac{1}{3}\hint_{40-1}\right)
\nonumber \\
&& \hspace{2cm}
      + \frac{\partial_\rho{\xi} }{2\hat{\Lambda}}\,\left(\hint_{44-1}{-}\frac{1}{3}\hint_{42-1}\right) .
\ee
The definitions of the various special functions and symbols above can be found in Ref.~\cite{Martinez:2017ibh}.

\begin{figure}[t]
\centerline{\includegraphics[width=0.95\linewidth]{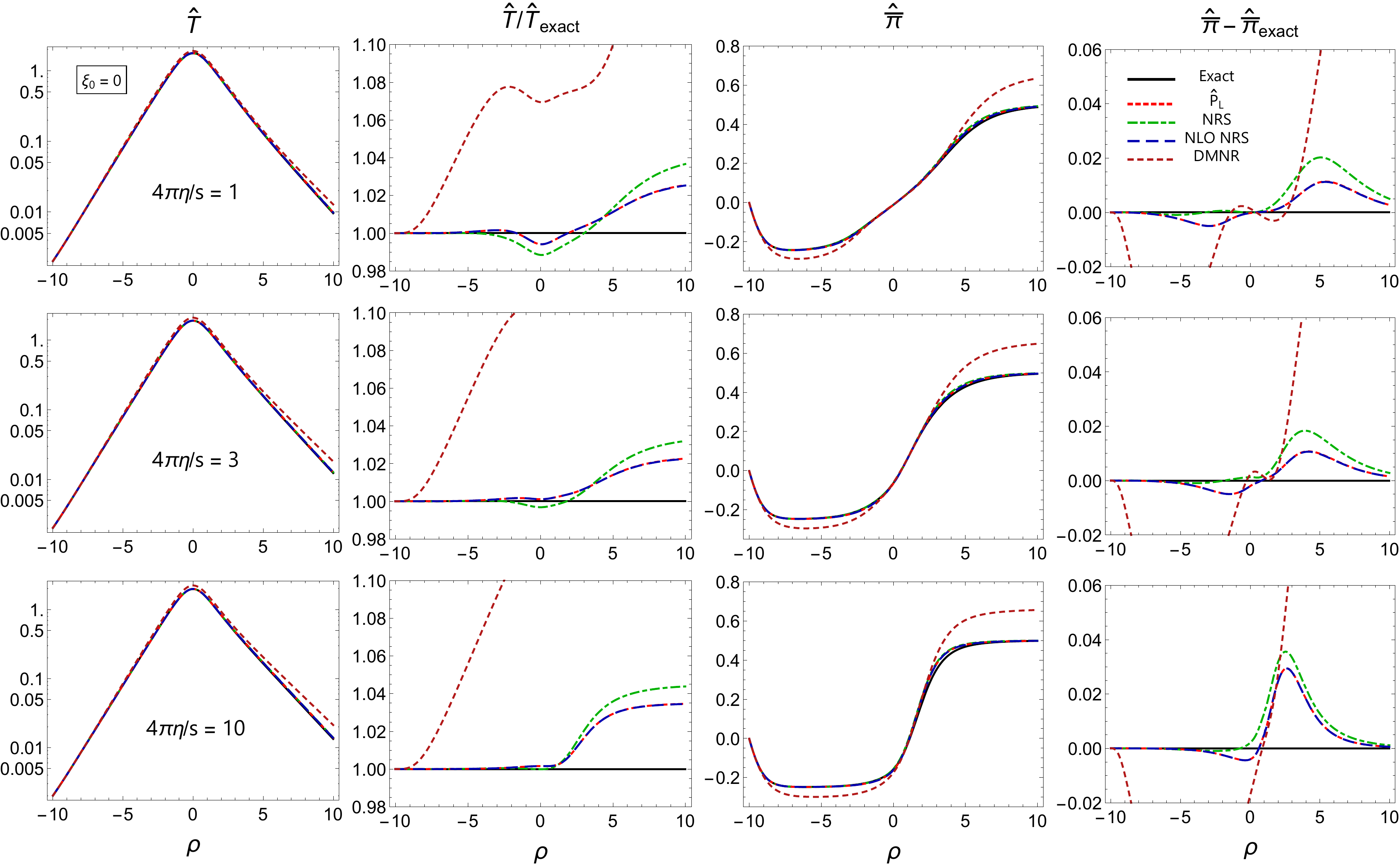}\hspace{5mm}}
\caption{(Color online) Scaled temperature and the normalized shear stress vs de Sitter time obtained using the exact solution of the RTA Boltzmann equation (black solid lines) and four different hydrodynamic approximations: second-order vHydro (DNMR theory, short-dashed magenta lines), aHydro using the anisotropic matching principle (dotted red lines), leading-order aHydro using the Nopoush-Ryblewski-Strickland scheme (NRS, dash-dotted green lines), and second-order aHydro in the NRS scheme now including the residual viscous corrections (NLO NRS, long-dashed blue lines).  Figure used with permission from Ref.~\cite{Martinez:2017ibh}.}
\label{fig:vaHydro-gubser}
\end{figure}

\subsubsection{Comparison of results}

In Fig.~\ref{fig:vaHydro-gubser} we present a figure from Ref.~\cite{Martinez:2017ibh} which compares various schemes:  second-order vHydro (DNMR), aHydro using the anisotropic matching principle, leading-order aHydro using the NRS prescription, and second-order NRS aHydro.  In the figure, the left column shows the Weyl-rescaled temperature $\hat{T}$, the second column shows the scaled temperature obtained using the various hydrodynamical schemes scaled by the exact solution, the third column shows the scaled shear correction $\hat{\bar\pi}$, and the fourth column shows the scaled shear correction obtained using the various hydrodynamical schemes scaled by the exact solution.  From this figure, we see that all aHydro implementations better reproduce the exact solution than standard second-order vHydro.  The maximum error in $\hat{T}$ is approximately 4\% in the range shown and the maximum deviation in $\hat{\bar\pi}$ is approximately 3\%.  Comparing the different aHydro results, we see that going to second-order marginally improves the agreement with the exact solution~\cite{Heinz:2015gka}.  We note also that the NLO NRS results and anisotropic matching condition results are almost indistinguishable, suggesting a connection between these two schemes which is further elaborated in Ref.~\cite{Martinez:2017ibh}.


\section{The anisotropic non-equilibrium attractor}
\label{sec:attractor}

As mentioned in the introduction, the QGP generated in URHICs is highly momentum-space anisotropic.  However, despite these large momentum-space anisotropies the system is seemingly well-described by relativistic dissipative hydrodynamics.  The timescale for the onset of hydrodynamical behavior in the QGP has been dubbed ``hydrodynamization'' time scale.  Many disparate theories find that this time scale is much shorter than the isotropization time \cite{Chesler:2008hg,Beuf:2009cx,Chesler:2009cy,Heller:2011ju,Heller:2012je,Heller:2012km,vanderSchee:2012qj,Casalderrey-Solana:2013aba,vanderSchee:2013pia,Heller:2013oxa,Keegan:2015avk,Chesler:2015bba,Kurkela:2015qoa,Chesler:2016ceu,Attems:2016ugt,Attems:2016tby,Attems:2017zam,Florkowski:2017olj}.  In recent years, we have learned that the process of hydrodynamization is driven by a {\em dynamical attractor} which varies a bit depending on the model one considers \cite{Heller:2015dha,Keegan:2015avk,Florkowski:2017olj,Romatschke:2017vte,Strickland:2017kux,Romatschke:2017acs,Behtash:2017wqg,Denicol:2017lxn}.  In a recent paper \cite{Strickland:2017kux} it was shown how to determine the dynamical attractor associated with aHydro and two different second-order vHydro frameworks: DNMR and Mueller-Israel-Stewart (MIS).  Here we will briefly review the method for obtaining the 0+1d attractor equation for conformal systems using aHydro and demonstrate that it provides the best approximation to the true attractor determined by exact solution of the Boltzmann equation.

\subsection{Attractor variables}

It is useful to introduce the dimensionless ``time'' \cite{Heller:2015dha}
\be
w \equiv \tau T(\tau) \, ,
\ee
where $T(\tau) \equiv \gamma \epsilon^{1/4}(\tau)$ is the effective temperature with $\gamma$ collecting all numerical factors associated number of degrees of freedom, etc.  From this, one readily obtains
\be
\varphi(w) \equiv \tau \frac{\dot w}{w} = 1 + \frac{\tau}{4} \partial_\tau\!\log \epsilon \, ,
\label{wdot1}
\ee
where, on the left, we have introduced the {\em amplitude} $\varphi$.\footnote{In the original literature the variable $\varphi$ was simply called ``$f$'', however, in order to avoid overlap with the one-particle distribution function, we have modified the notation slightly.}  From this we see that a solution for the proper-time evolution of the energy density uniquely specifies the $w$-dependence of the amplitude $\varphi$.  The amplitude $\varphi$ is bounded by energy positivity to the region $0 \leq \varphi \leq 1$~\cite{Janik:2005zt}.

\subsection{Conformal 0+1d aHydro attractor}

We begin by changing variables in the two conformal 0+1d aHydro dynamical equations to $\varphi$ and $w$.  The starting point are the first and second moments of the 0+1d conformal Boltzmann equation given in Eqs.~(\ref{eq:energydens}) and (\ref{eq:2ndmomf3}), respectively.  The first moment can be written as
\be
\tau\partial_\tau\!\log \epsilon = -\frac{4}{3}+\frac{\pi}{\epsilon} = 4(\varphi - 1) \, ,
\ee
where here $\pi$ is the single independent component of the 0+1d shear tensor with $P_T = P - \pi/2$ and $P_L = P + \pi$.  Using this, $\overline\pi \equiv \pi/\epsilon$ can be expressed solely in terms $\varphi$ 
\be
\overline\pi = 4\left(\varphi -\frac{2}{3}\right) . 
\ee
Note that the inverse Reynolds number is related to $\overline\pi$ (and hence $\varphi$) by $R_\pi^{-1} \equiv \sqrt{\pi^{\mu\nu}\pi_{\mu\nu}}/P_0 = 3 \sqrt{3/2} \, |\overline\pi|$.

From the first moment of the Boltzmann equation, by taking an additional derivative with respect to $\tau$ and changing variables to $\varphi$ and $w$, one can obtain the following first-order differential equation for $\varphi(w)$~\cite{Heller:2015dha}
\be
w \varphi \varphi' = -\frac{8}{3} + \frac{20}{3}\varphi - 4\varphi^2 + \frac{\tau}{4}\frac{\dot\pi}{\epsilon} \, ,
\label{eq:finalfirstmom}
\ee
where $\dot\pi = \partial_\tau \pi$ 
and from the aHydro second moment equation (\ref{eq:2ndmomf3}), one obtains~\cite{Strickland:2017kux}
\be
\frac{\tau}{4} \frac{\dot\pi}{\epsilon} =  \frac{8}{3} - \frac{20}{3} \varphi + 4 \varphi^2 + \left\{ \frac{1}{2} [1+\xi(\overline\pi)]  - \frac{w}{4 c_\pi} {\cal H}(\xi) \right\} \overline\pi' \, ,
\ee
where $c_\pi$ is the constant that relates the relaxation time and the inverse temperature  $\tau_{\rm eq}(T) = c_\pi/T$ which, for a RTA collisional kernel is $c_\pi = 5$, $\overline{\pi}'(\xi) = d\overline\pi(\xi)/d\xi$, and ${\cal H}(\xi) \equiv \xi (1+\xi)^{3/2}{\cal R}^{5/4}(\xi)$ with ${\cal R}(\xi)$ defined in Eq.~(\ref{eq:Rfuncs}).\footnote{In the final expression, $\xi$ should be understood to be a function of $\overline\pi$, and hence $\varphi$ since $\overline\pi = 4\varphi - 8/3$.  For further information, we point the reader to Sec.~\ref{sec:smallaniso}.}

Combining these two results gives us the aHydro ``attractor equation''
\be
\overline{w} { \varphi} \frac{\partial \varphi}{\partial \overline{w}}  = \left\{ \frac{1}{2} [1+\xi(\varphi)] - \frac{\overline{w}}{4} {\cal H} (\xi) \right\} \overline\pi' \, ,
\label{eq:ahydroattractoreq2}
\ee
where $\overline{w} = w/c_\pi$.  Note, importantly, that the right-hand-side of Eq.~(\ref{eq:ahydroattractoreq2}) resums an infinite number of terms in the inverse Reynolds number.  This is due to the fact that (a) the inverse function $\xi(\varphi)$ is nonlinear and (b)  the ${\cal H}$ function itself contains all orders in $\xi$.  As it stands the attractor equation is nothing but a rewriting of the equations of motion, however, by fixing the boundary condition at $\overline{w} \rightarrow 0$ one can obtain a unique solution which can be seen to be an attractor for the general dynamical equations. To obtain the aHydro attractor numerically, this differential equation must be solved with the boundary condition $\lim_{\overline{w} \rightarrow 0} \varphi(\overline{w}) =  3/4$ \cite{Strickland:2017kux}.  This boundary condition can be determined analytically by making a ``slow-roll'' approximation to the differential equation in the small-$\overline{w}$ limit.  The precise value of $\varphi(0)$ depends on the hydrodynamic framework being considered.  The value of $3/4$ obtained in aHydro is special because it guarantees that the resulting attractor solution possesses positive longitudinal pressure as $\overline{w} \rightarrow 0$.

\subsubsection*{Comparison of the aHydro attractor and the exact solution}
\vspace{2mm}

\begin{figure}[t!]
\centerline{
\includegraphics[width=0.47\linewidth]{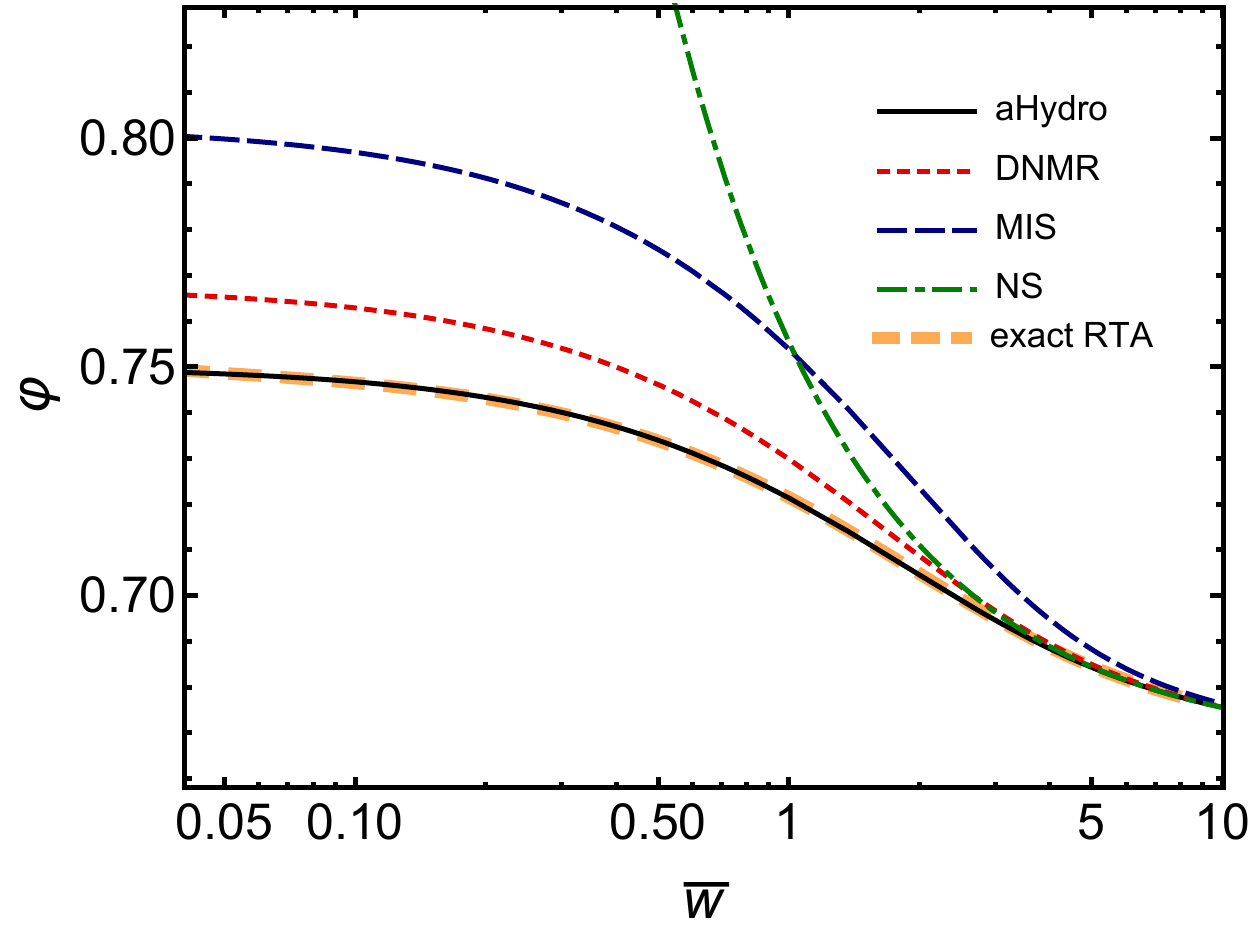}
$\;\;\;$
\includegraphics[width=.455\linewidth]{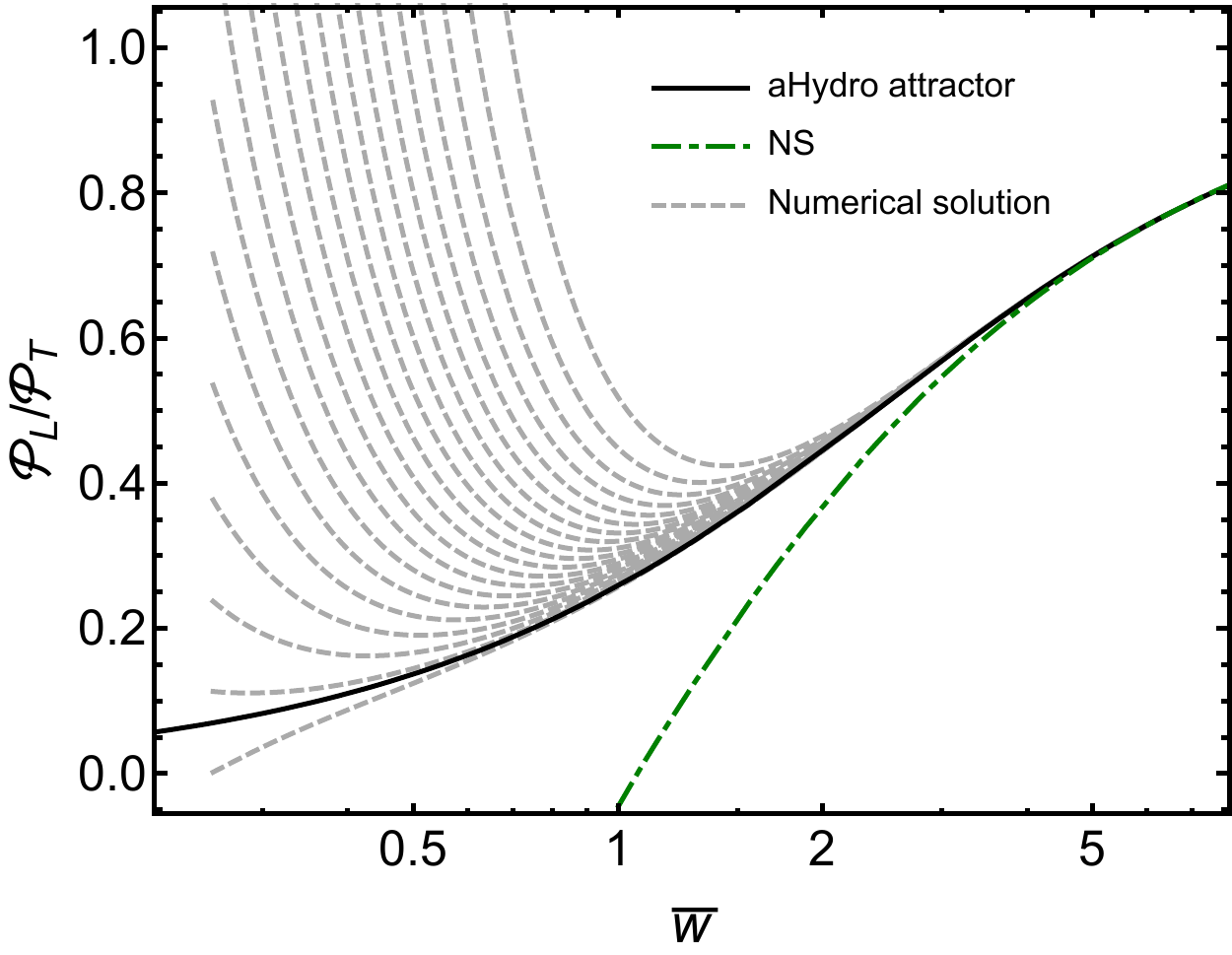}
$\;\;\;$
}
\caption{(Color online) In the left panel, we compare the solutions to the attractor equations for various models.  The aHydro (black solid line), DNMR (red short-dashed), MIS (blue long-dashed), and NS attractors (green dot-dashed) compared to the exact solution.  In the right panel, we plot the pressure anisotropy emerging from the aHydro attractor (solid black line) along with several direct numerical solutions to the aHydro dynamical equations (\ref{eq:firstmoment}) and (\ref{eq:2ndmomf}) (grey dashed lines) corresponding to a variety of initial conditions for $\pi$ assuming $4\pi\eta/s = 2$.
Figures used with permission from Ref.~\cite{Strickland:2017kux}.}
\label{fig:attractor}
\end{figure}

In Fig.~\ref{fig:attractor} we present two panels.  In the left panel, we compare the solutions to the attractor equations for various models for a system undergoing 0+1d boost-invariant Bjorken expansion.  The aHydro (black solid line), DNMR (red short-dashed), MIS (blue long-dashed), and Navier-Stokes (NS) attractors (green dot-dashed) compared to the attractor obtained via exact solution of the Boltzmann equation.\footnote{For more information about the method for obtaining the exact attractor solution, we refer the reader to the supplemental material supplied in Ref.~\cite{Romatschke:2017vte}.}

As can be seen from this figure, aHydro agrees best with the exact solution \cite{Florkowski:2013lza,Florkowski:2013lya} for the 0+1d conformal attractor.  In addition, one finds that the aHydro attractor possesses positive pressures for all values of $w$, whereas, the different second-order frameworks have attractors which result in negative pressures.  This is reflected in the left panel by the fact that $\varphi$ is bounded between 1/2 and 3/4 for aHydro whereas vHydro can violate these bounds.  In particular, if $\varphi > 3/4$ a negative longitudinal pressure results.

In the right panel of Fig.~\ref{fig:attractor}, we plot the pressure anisotropy emerging from the aHydro attractor (solid black line) along with several direct numerical solutions (grey dashed lines) corresponding to a variety of initial conditions for $\pi$ assuming $4\pi\eta/s = 2$.  This figure demonstrates the essential behavior of the attractor in that the black line divides the initial condition space in two, with all solutions with mildly oblate or prolate initial anisotropy approaching the attractor from above and those with extremely oblate initial anisotropy approaching from below.  The numerical solutions approach the attractor solution by a proper-time $\tau \sim \tau_{\rm attractor}$.  In LHC heavy-ion collisions, one expects initial temperatures \mbox{$T_0 \lesssim$ 500 MeV} at $\tau_0 = 0.25$ fm/c and $\eta/s \sim 0.2$, which translates into $\tau_{\rm attractor} \gtrsim $ 1.3 fm/c.\footnote{This number was determined by computing $T(\tau)$ using the aHydro dynamical equations and solving for the proper-time at which $\overline{w} = 2$.}  Prior to $\tau \sim \tau_{\rm attractor}$, each local region of the system is subject to the evolution of non-hydrodynamic modes \cite{Noronha:2011fi,Heller:2014wfa,Bazow:2015zca,Florkowski:2017olj} whose precise evolution depends on the microscopic theory being considered.  In addition, since the damping time of non-hydrodynamic modes scales inversely with the local effective temperature (determined from the energy density), one expects their effects to be more important in the dilute edges of the plasma.

Faced with such a situation it becomes critically important to identify the appropriate microscopic theory to describe the dynamics of the system.  In the center of the fireball, where the energy densities are the largest at early times, one would expect perturbative QCD approaches to be the most appropriate description, however, as one approaches the dilute edges a formulation in terms of hadronic kinetic theory would seem to be the most appropriate.  Since both regions could, in principle, be described in terms of the Boltzmann or Boltzmann-Vlasov equation and the same theories match smoothly onto the late-time hydrodynamical attractor,  this naturally leads one to focus on hydrodynamic theories that can be obtained from kinetic theory.  This further motivates continuing to develop aHydro into an accurate phenomenological tool.


\section{Anisotropic Freeze-out}
\label{sec:freeze-out}

In this section, we will review the freeze-out procedure in aHydro in  a pedagogical way following Ref.~\cite{Nopoush:2015yga}.
At low temperatures, the  QGP undergoes a transition from quarks and gluons to hadrons. Due to this transition one should deal with particles' momenta and energies which can be measured experimentally rather than the bulk variables such as the energy density and pressure components in the QGP phase. The standard way to do this is through ``Cooper-Frye freeze-out'' which is based on the equivalence of the energy-momentum tensor before and after the transition
\be 
T^{\mu \nu}_{\rm hydro}(x) = T^{\mu \nu}_{\rm kinetic}(x) =  \sum_i \int dP \, p^\mu p^\nu f_i(x,p) \, ,
\ee
where $i$ sums over all hadrons and $f_i$ has the microscopic anisotropy tensor $\Xi^{\mu\nu}$ and scale $\lambda$ as the local fluid element in which the matching occurs.  This matching is done on a freeze-out hypersurface where the system EoS is well-described by a non-interacting gas of hadrons.
As a result, one can only apply this procedure in dilute regions where interactions are expected to be small.
The quantity that we are looking for eventually is the distribution of particles over $p_T$, rapidity $y$, etc. We shall start by finding the number of particles in a small volume which can be written in a compact way as 
\be
dN=dV_\mu \int \frac{d^3p}{(2\pi)^3 E_p} p^\mu \, f(x,p) \, ,
\label{eq:dNdV}
\ee
where $dV_\mu$ is the vector volume which is the local three-volume times a four-dimensional unit vector orthogonal to the local volume.
By integrating over the entire volume one obtains the total number of particles produced from the fireball
\be 
N= \int dV_\mu (x) \int \frac{d^3 p}{(2\pi)^3 E_p} p^\mu f(x,p) \, ,
\ee
or, expressing $dV_\mu = d^3\Sigma_\mu$ where $d^3\Sigma_\mu$ is the four-dimensional vector element of volume, one has explicitly
\be 
N= \int \frac{d^3 p}{(2\pi)^3 E_p}  \int d^3\Sigma_\mu \, p^\mu f(x,p) \, ,
\label{eq:freezeJ}
\ee
where $\Sigma_\mu$ is a 3d hypersurface defining the 4d volume occupied by the fluid at the freeze-out temperature and $d^3\Sigma_\mu$ is volume vector. We will show later on some visualizations to make this understandable, but for now let us continue the formalism.  We should note here that Eq.~(\ref{eq:freezeJ}) is called the Cooper-Frye formula which is in principle counting the number of all particles emitted from the fluid through the hypersurface which can be parametrized in many ways, but it should be around all the fluid. 

\subsection{The Krakow parameterization}

One of the parameterizations, in Milne coordinates, that we used in phenomenological comparisons is \cite{Bozek:2009ty,Bozek:2009dw,Chojnacki:2011hb,Ryblewski:2015hea}
\ba
 \tau &=& \tau_0 + d(\zeta,\phi,\theta) \sin\theta \sin\zeta \nonumber  \, ,
 \nonumber   \\
 r &=&  d(\zeta,\phi,\theta) \sin\theta \cos\zeta  \, ,\nonumber \\
 \phi &=&\phi\, , \nonumber  \\
 \varsigma &=&  \frac{d(\zeta,\phi,\theta) \cos\theta}{\Lambda}  \, ,
  \label{eq:milnepar}
\ea
where $\Lambda$ is an arbitrary length scale which is introduced for dimensional purposes.
%
\begin{figure}[t!]
\centering
\includegraphics[width=0.5\linewidth]{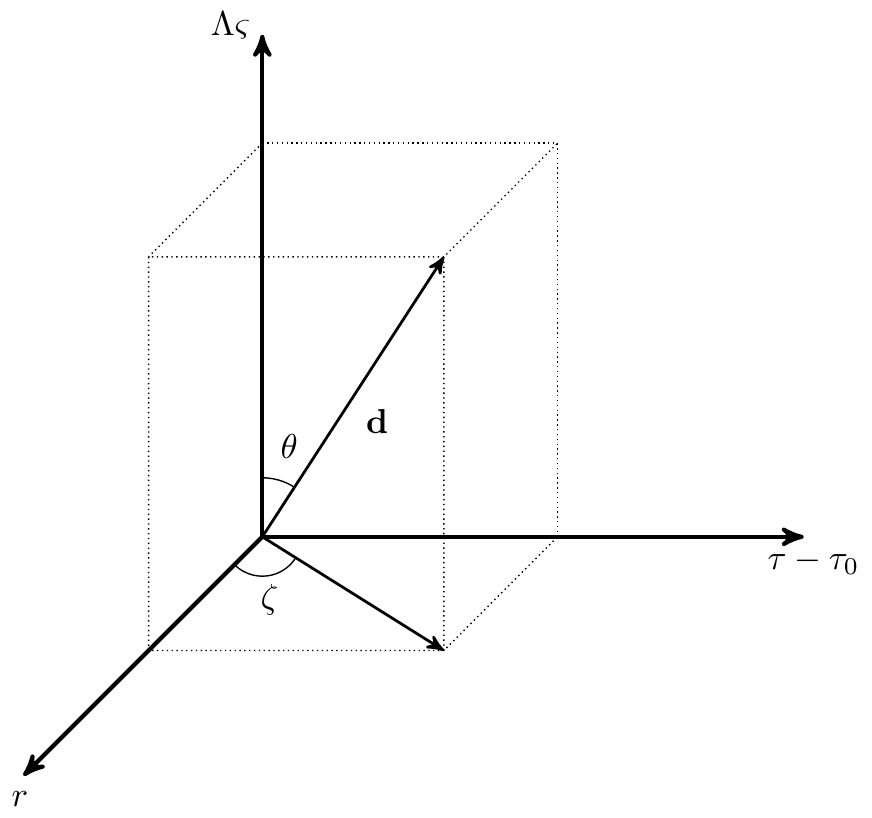}
\caption{A schematic diagram showing the system of coordinates $(\tau, r, \phi, \varsigma)$ used to parametrize the freeze-out hypersurface which is defined in Eq.~(\ref{eq:milnepar}). }
\label{fig:coord-freeze-out}
\end{figure}
In Fig.~\ref{fig:coord-freeze-out}, we show a schematic diagram to help visualizing this parametrization to understand the hypersurface structure. As can be seen from this sketch, $d(\zeta,\phi,\theta)$ is the distance between any point on the hypersurface and the origin where $\theta$ and $\zeta$ are the polar and azimuthal angles in the $(r,\tau-\tau_0,\Lambda\varsigma)$ three-dimensional subvolume.

\subsection{Implementing anisotropic Cooper-Frye freezeout}

To find the normal vectors to the hypersurface at any point one can find the ``Jacobian'' with respect to $\zeta$, $\phi$, and $\theta$
\be
d^3\Sigma_\mu = \epsilon_{\mu\alpha\beta\gamma} \frac{\partial x^\alpha}{\partial \zeta}\frac{\partial x^\beta}{\partial \phi}\frac{\partial x^\gamma}{\partial \theta}d\zeta d\phi d\theta\, ,
\label{eq:dSigma}
\ee
where $\epsilon_{\mu\alpha\beta\gamma}$ is the 4d Levi-Civita symbol defined as 
\ba 
\epsilon_{\mu\alpha\beta\gamma}=
\left\{ \begin{array}{lcc}
+1 & \mbox{For}
& \, {\rm even \, permutation} \, ,
 \\ -1 & \mbox{For}
& {\rm  odd \, permutation} \, , \\
0 & & \mbox {otherwise} \, .
\end{array}\right. \,\,\,\,\,\,\,\,\,\,\,\,\,
\ea 
As an example, let us take the zeroth-component of $d^3\Sigma_\mu$
\be
d^3\Sigma_0 = \epsilon_{0\alpha\beta\gamma} \frac{\partial x^\alpha}{\partial \zeta}\frac{\partial x^\beta}{\partial \phi}\frac{\partial x^\gamma}{\partial \theta}d\zeta d\phi d\theta\, .
\label{eq:dSigma0}
\ee
Now, one can sum over repeated indices or use the determinant method choosing $\epsilon_{0 1 2 3}=+1$ where both will give exactly the same answer. For illustration, let's do that for one case, the zeroth component
\begin{align}
\epsilon_{0\alpha\beta\gamma} \frac{\partial x^\alpha}{\partial \zeta}\frac{\partial x^\beta}{\partial \phi}\frac{\partial x^\gamma}{\partial \theta} =\left. 
\right. 
\begin{vmatrix}
 \frac{\partial x}{\partial \zeta} & \frac{\partial x}{\partial \phi} & \frac{\partial x}{\partial \theta}  \\ 
 \frac{\partial y}{\partial \zeta} & \frac{\partial y}{\partial \phi} & \frac{\partial y}{\partial \theta}  \\ 
 \frac{\partial z}{\partial \zeta} & \frac{\partial z}{\partial \phi} & \frac{\partial z}{\partial \theta} 
\end{vmatrix} \, .
\end{align}
By evaluating the determinant, one can obtain the necessary Jacobian. Since in the final state one has many different particles species $i$ with different properties, spin $(s_i)$, isospin $(g_i)$, and mass $(m_i)$ resulting in a different distribution function for each species ($f_i$), we will sum over all possible hadrons. Hence, the total number of particles is given by the sum over $i$ for all known hadrons and hadron resonances 
\be
N=\sum\limits_i \frac{(2s_i+1)(2g_i+1)}{(2\pi)^3}\int \! d^3p \frac{1}{\sqrt{m^2_i+p^2}}\int \! f_i(x,p) \, p^\mu d^3\Sigma_\mu \, ,
\label{eq:N}
\ee
where $(2s_i+1)(2g_i+1)$ is the degeneracy factor which depends on the spin $s_i$ and isospin $g_i$ of the hadron $i$.   In practice, we will assume that each of the hadronic distribution functions $f_i(x,p)$ is in generalized Romatschke-Strickland form (\ref{eq:genf}) and, in leading-order aHydro, one ignores the residual dissipative corrections $\delta \tilde{f}$.  This is a reasonable thing to do since the aHydro form incorporates the largest deviations from isotropic equilibrium, however, if one has a complete next-to-leading order aHydro scheme, then one should also include the corrections associated with $\delta \tilde{f}$ similarly to how $\delta f$ is included in standard viscous hydrodynamics.

Equation~(\ref{eq:N}) is all that is needed to compute the particle multiplicities, but we still have to calculate some quantities inside this formula, mainly $p^\mu \Xi_{\mu \nu} p^\nu $ which is the argument of the distribution function and $p^\mu d^3\Sigma_\mu$. Starting with $p^\mu \Xi_{\mu \nu} p^\nu $, we need to come up with a parametrization of $p^\mu$ which is convenient in our case   
\be
p^\mu\equiv(m_T \cosh y,p_T \cos\varphi,p_T \sin\varphi,m_T \sinh y)\, ,
\label{eq:ptl-mom}
\ee
%
\begin{figure}[t!]
\centering
\includegraphics[width=1.\linewidth]{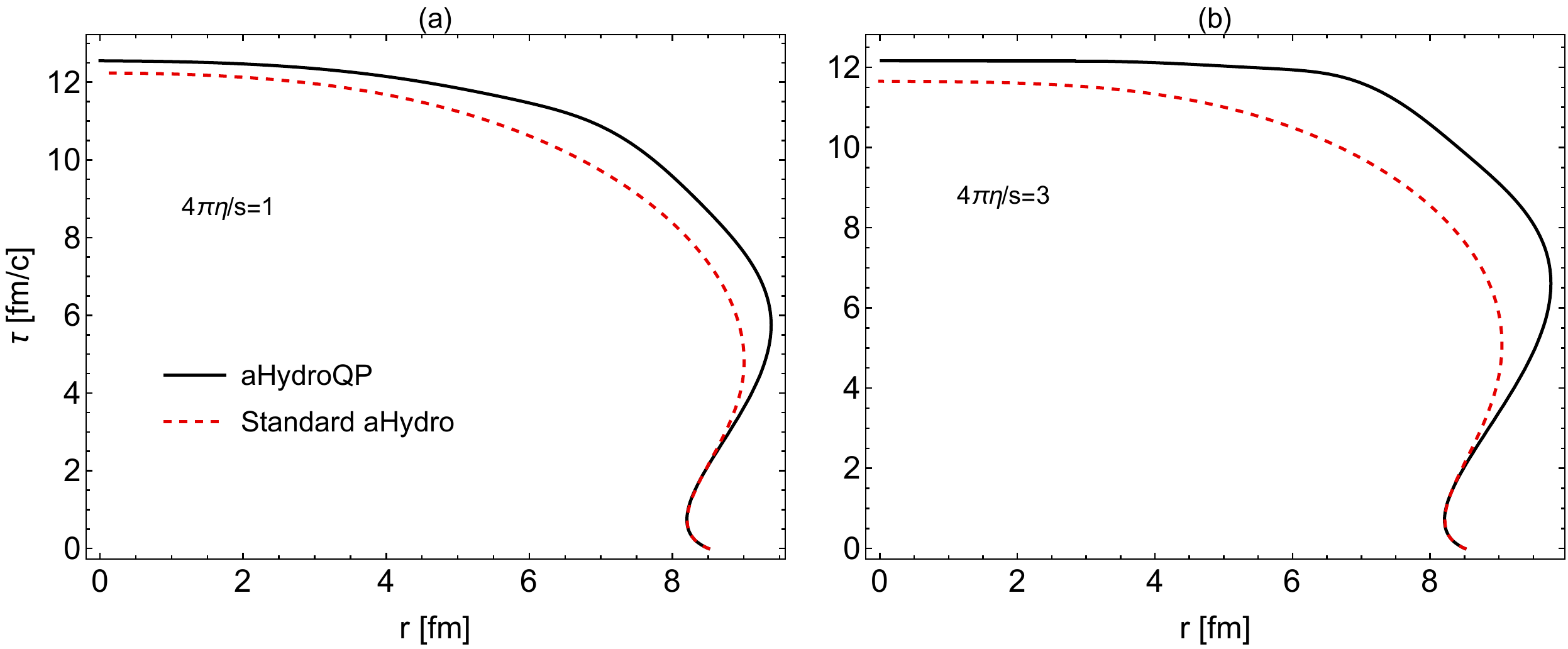}
\caption{(Color online) Comparisons of the freeze-out hypersurfaces in 1+1d systems between aHydroQP (black solid line) and standard  aHydro (red dashed line) for $4 \pi \eta/s=1$ in panel (a) and $4 \pi \eta/s=3$ in panel(b). Figure used with permission from Ref.~\cite{Alqahtani:2016rth}}
\label{fig:hypers1+1}
\end{figure}
%
\begin{figure}[t]
\centerline{
\includegraphics[width=.38\linewidth]{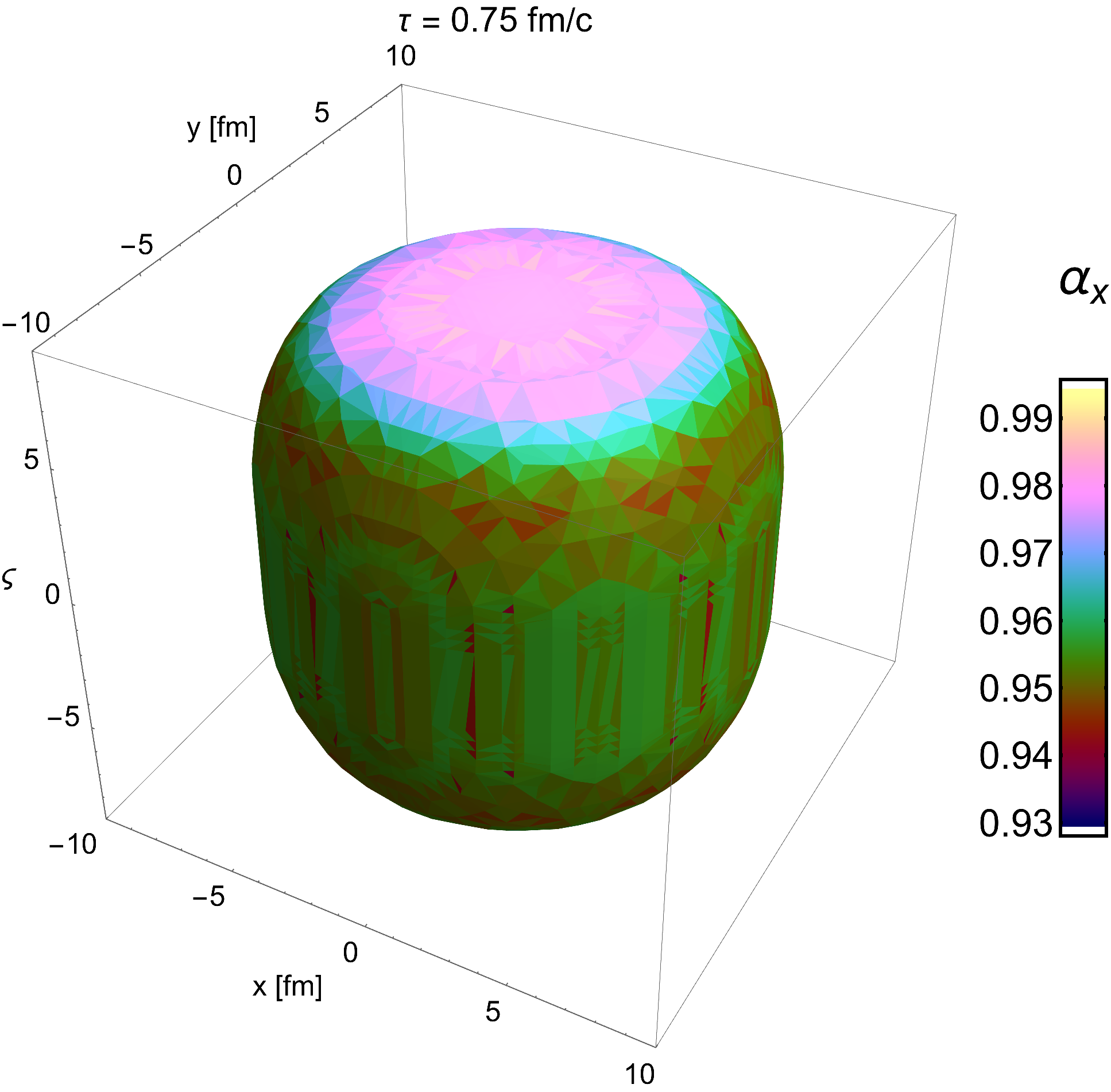}
\hspace{1cm}
\includegraphics[width=.38\linewidth]{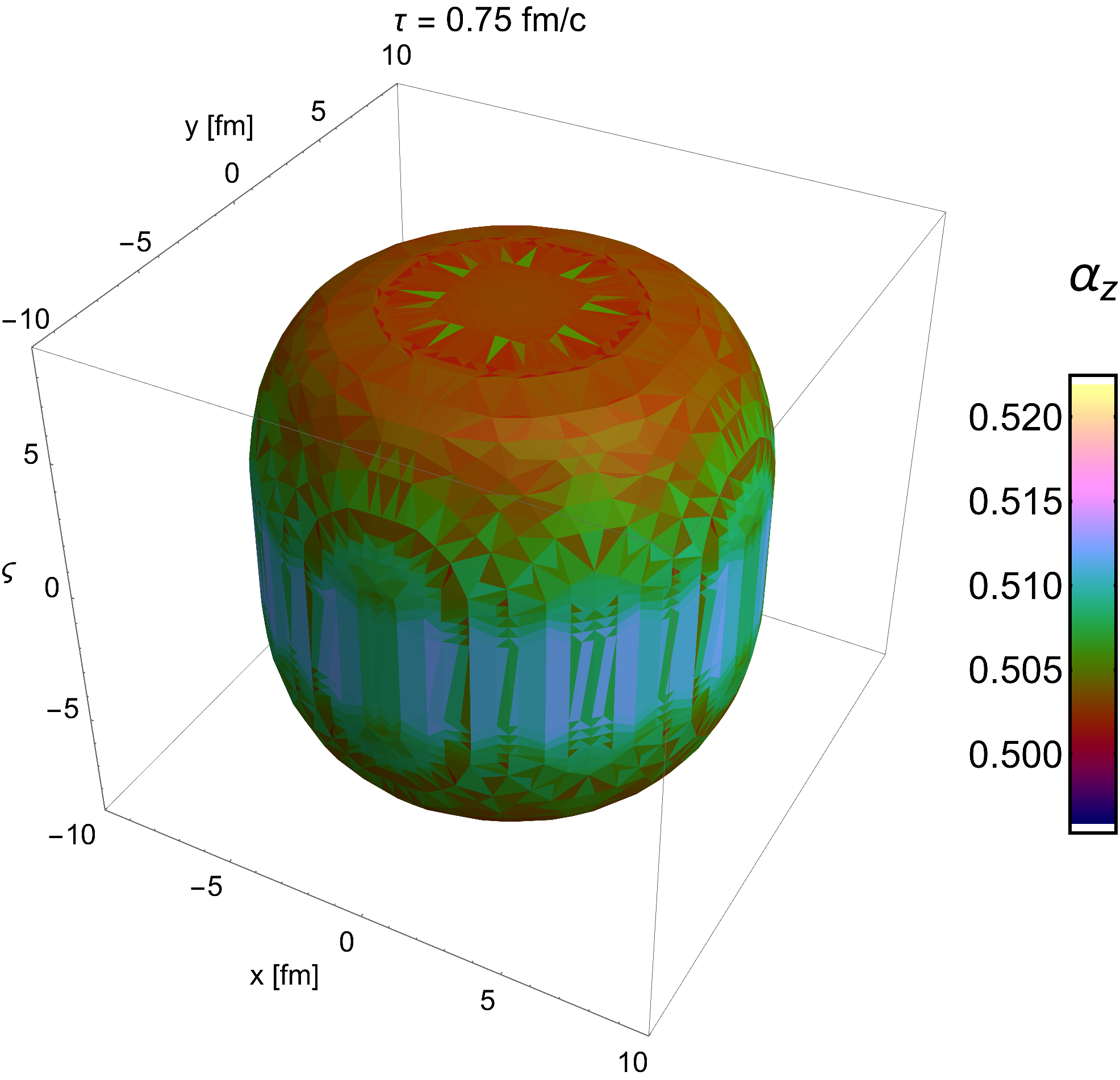}
}
\vspace{7mm}
\centerline{
\includegraphics[width=.38\linewidth]{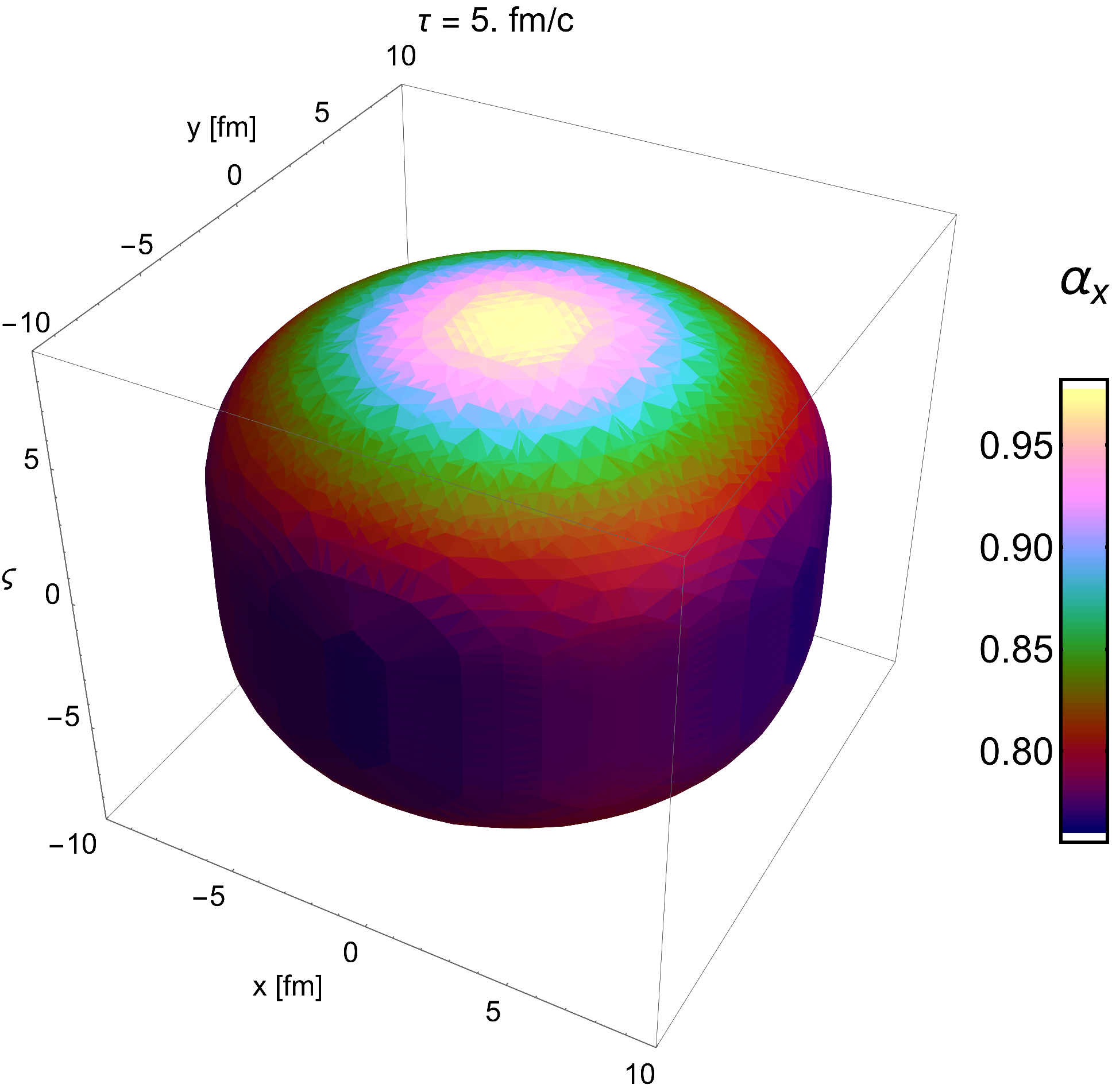}
\hspace{1cm}
\includegraphics[width=.38\linewidth]{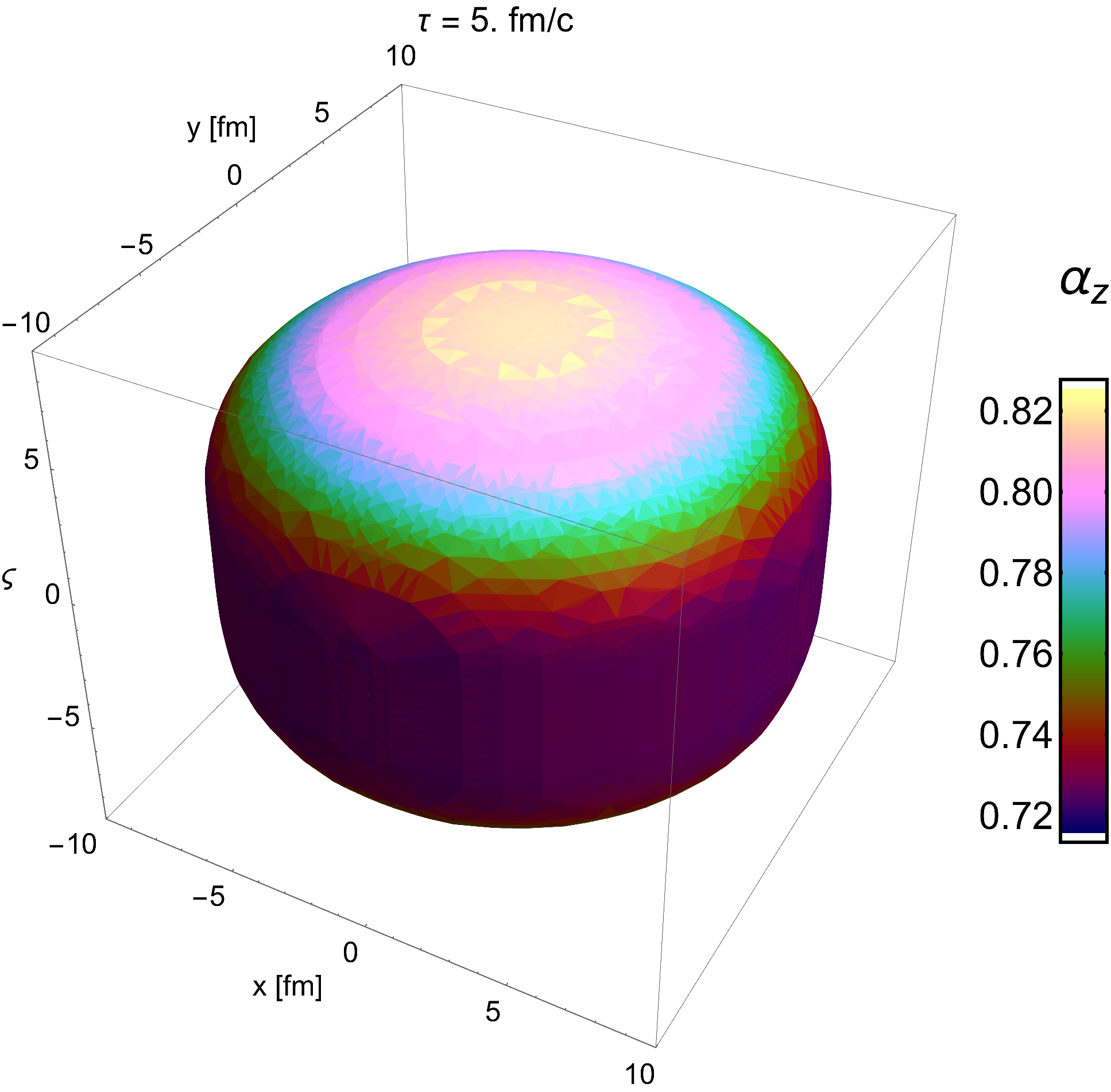}
}
\caption{Freeze-out hypersurface (FOHS) determined by $T_{\rm FO} = 130$ MeV, at two different proper times for a central Pb-Pb collision using optical Glauber initial conditions in the transverse plane and a central-rapidity plateau with Gaussian tails as described in Ref.~\cite{Alqahtani:2017tnq}.  The top row shows  $\tau = 0.75$ fm/c and the bottom row shows $\tau = 5$ fm/c.  The color-coding in the left and right columns shows $\alpha_x$ and $\alpha_z$ on the FOHS, respectively.}
\label{fig:FOHS3d}
\end{figure}
%
where $m_T$ is the transverse mass $m_T \equiv \sqrt{p_T^2+m^2}$, $y = \tanh^{-1}(p^z/p^0)$ is the particle's rapidity, and $\varphi$ is the particle's azimuthal angle. Now we have
\be
p_\mu \Xi^{\mu \nu} p_\nu = p_\mu(u^\mu u^\nu + \xi^{\mu\nu} - \Delta^{\mu\nu} \Phi) p_\nu \, .
\label{eq:ptl-mom2}
\ee
We need the contractions of $p_\mu$ and the basis vectors which are
\ba
p \cdot u &=& m_T \cosh (\theta _T) \cosh (y-\varsigma)-p_T \sinh (\theta _T)
   \cos (\phi -\varphi )\, ,\nonumber
    \\
p \cdot X &=& m_T \sinh (\theta _T) \cosh (y-\varsigma)-p_T \cosh (\theta _T)
   \cos (\phi -\varphi )
\, ,\nonumber \\
p \cdot Y &=& p_T \sin (\phi -\varphi )
\, ,\nonumber \\
p \cdot Z &=& -m_T \sinh (y-\varsigma)
\, .
\label{eq:pdotuxyz}
\ea
 Using all that above, one can show that $(p^\mu \Xi_{\mu\nu} p^\nu)$ is given by \cite{Alqahtani:2016rth}
\ba
p^\mu \Xi_{\mu\nu} p^\nu&=& 
(1+\Phi)\Big[m_T \cosh \theta _T\!\cosh
   (y-\varsigma)\!-\!p_T \sinh \theta _T\!\cos (\phi -\varphi )\Big]^2
   \nonumber \\ 
   &&+ \xi _x\,\Big[m_T \sinh \theta _T \cosh (y-\varsigma)-p_T \cosh \theta
   _T \cos (\phi -\varphi )\Big]^2
   \nonumber \\
   &&+\xi_y\,\, p_T^2  \sin^2(\phi -\varphi )+ \xi_z\, m_T^2  \sinh^2(y-\varsigma)
    - \Phi m^2 \, .
\label{eq:pxip}
\ea
Finally, we need to expand Eq.~(\ref{eq:dSigma}) for $x^\mu=(t,x,y,z)$ and contracting with $p^\mu$ to obtain \cite{Nopoush:2015yga}
\ba 
p^\mu d^3\Sigma_\mu  &=& \frac{\tau}{\Lambda}\sin\theta d^2 \Bigg[ p_T  \sin (\phi -\varphi ) \frac{\partial d}{\partial\phi} \notag \\ 
&+&\frac{\Lambda}{\tau}  m_T\cos\zeta \sin\theta \sinh(y-\varsigma) \bigg(d \cos\theta+\frac{\partial d}{\partial\theta}\sin\theta\bigg)\, \notag \\
&+&\cos\zeta \sin\theta\bigg(p_T \cos\zeta \cos (\phi -\varphi )+m_T \sin\zeta \cosh(y-\varsigma) \bigg)\!\bigg(d\sin\theta-\frac{\partial d}{\partial \theta}\cos\theta\bigg) \, \notag \\
&+&\cos\zeta\frac{\partial d}{\partial\zeta} \Big(p_T \sin \zeta \cos (\phi-\varphi) -m_T \cos\zeta \cosh(y-\varsigma)\Big) \Bigg]d\zeta d\phi d\theta\, .
\label{eq:pdSigma}
\ea 
The above formalism is for a general 3+1d system.  For a 1+1d system, we refer the reader to Ref.~\cite{Nopoush:2015yga}.
 
In Fig.~\ref{fig:hypers1+1}, we show the freeze-out hypersurface in 1+1d systems for different shear viscosity to entropy density ratios: $4 \pi \eta/s=1$ and $3$ in panels (a) and (b) respectively.  For this purpose, we initialized the system with a central temperature of 0.6 GeV at 0.25 fm/c using a Glauber wounded-nucleon profile and extracted the freeze-out hypersurface using a constant energy density corresponding to an effective temperature of $T_{\rm FO}$ = 150 MeV. First, we want to show the hypersurface shape in this simple case. We note here that the system undergoes the freeze-out at outer regions at very early times since these regions are much colder than central regions. However, at central regions the freeze-out happens at late times $\tau \sim 12$ fm/c. Second, we want to show the effect of the way that the EoS is implemented. As can be seen from this figure, standard aHydro predicts a smaller freeze-out volume which will result in few particles being generated.  We will return to this discussion when we present the spectra.

For illustration purposes, in Fig.~\ref{fig:FOHS3d} we show the 3d freeze-out hypersurface for central Pb-Pb collisions determined by $T_{\rm FO} = 130$ MeV at different proper times. In the top row we take $\tau = 0.75$ fm/c while in the bottom row we take  $\tau = 5$ fm/c. We also show the color-coding of $\alpha_x$ and $\alpha_z$ in the left and right columns, respectively.  As can be seen from this figure, at early times (top panels) the anisotropy parameter $\alpha_z$ is quite different than $\alpha_x$ indicating that there are large momentum-space anisotropies on the hypersurface.  However, at later times (bottom panels) we see $\alpha_z$ approaching $\alpha_x$, which shows that the system is evolving towards an isotropic state although it is still not perfectly isotropic on the freeze-out surface.  We also note that the transverse expansion of the material is clearly seen by comparing the top and bottom panels.


\section{3+1d anisotropic hydrodynamics phenomenology}
\label{sec:pheno}

The main motivation behind introducing aHydro is to be able to extract the QGP properties in a more reliable way.  However, at this point in time aHydro is not as well-developed as viscous hydrodynamics.  In this section, we will present some recent phenomenological results of aHydroQP using smooth Glauber-like initial conditions in order to perform an initial baseline study.  Although the aHydro results shown here do not include fluctuations, they do treat the deviations from isotropy more consistently and do not require arbitrary regulators to remove instabilities associated with large shear corrections. In all results presented here the comparisons were between aHydroQP predictions and ALICE data for Pb-Pb collisions at 2.76 TeV. Here, we will show only some comparisons and refer the reader for more details  to Refs.~\cite{Alqahtani:2017jwl,Alqahtani:2017tnq}. However, we would like to mention a few important things regarding some assumptions that we used in this model. First, we used smooth Glauber initial conditions which means that we neglect the effects of fluctuations of nucleons inside the nuclei which is an important ingredient for some observables like the elliptic  flow for very central collisions. We also assumed that $\eta/s={\rm const}$, but as we know from hadron resonance gas and the perturbative QCD methods, $\eta/s$ depends on the temperature.  The temperature dependence of $\eta/s$ is hard to determine from first principles. Nevertheless, one can propose models for the temperature dependence of $\eta/s$ and see which one the data favors more.  Another assumption is that we used isotropic initial conditions $\alpha_i(\tau=0)=1$.  We are planning to include these necessary ingredients to our model in the near future. After the freeze-out, the degrees of freedom change from quarks and gluons to hadrons and then we need another code to perform the hadronic freeze-out. For this purpose, we used a customized version of THERMINATOR 2 which allows for momentum-space anisotropies at freeze-out \cite{Chojnacki:2011hb,MikeCodeDB}.   By fitting to the pion, kaon, and proton spectra in the 0-5\% and 30-40\% centrality classes we found an initial central temperature of $T_0= 600$ MeV at $\tau_0=0.25$ fm/c, $\eta/s = 0.159$, and \mbox{$T_{\rm FO} = 130$ MeV}.  

\subsection{Standard anisotropic hydrodynamics}
\label{sec:spheno}
However, before presenting the aHydroQP results, we would like to show some theory/data comparisons resulting from initial attempts to implement the EoS by only taking into account the breaking of conformality in the EoS.  This comparison will set the stage for comparisons of the quasiparticle method which takes into account non-conformality in a self-consistent manner.
%
\begin{figure}[t!]
\centerline{
\includegraphics[width=0.5\linewidth]{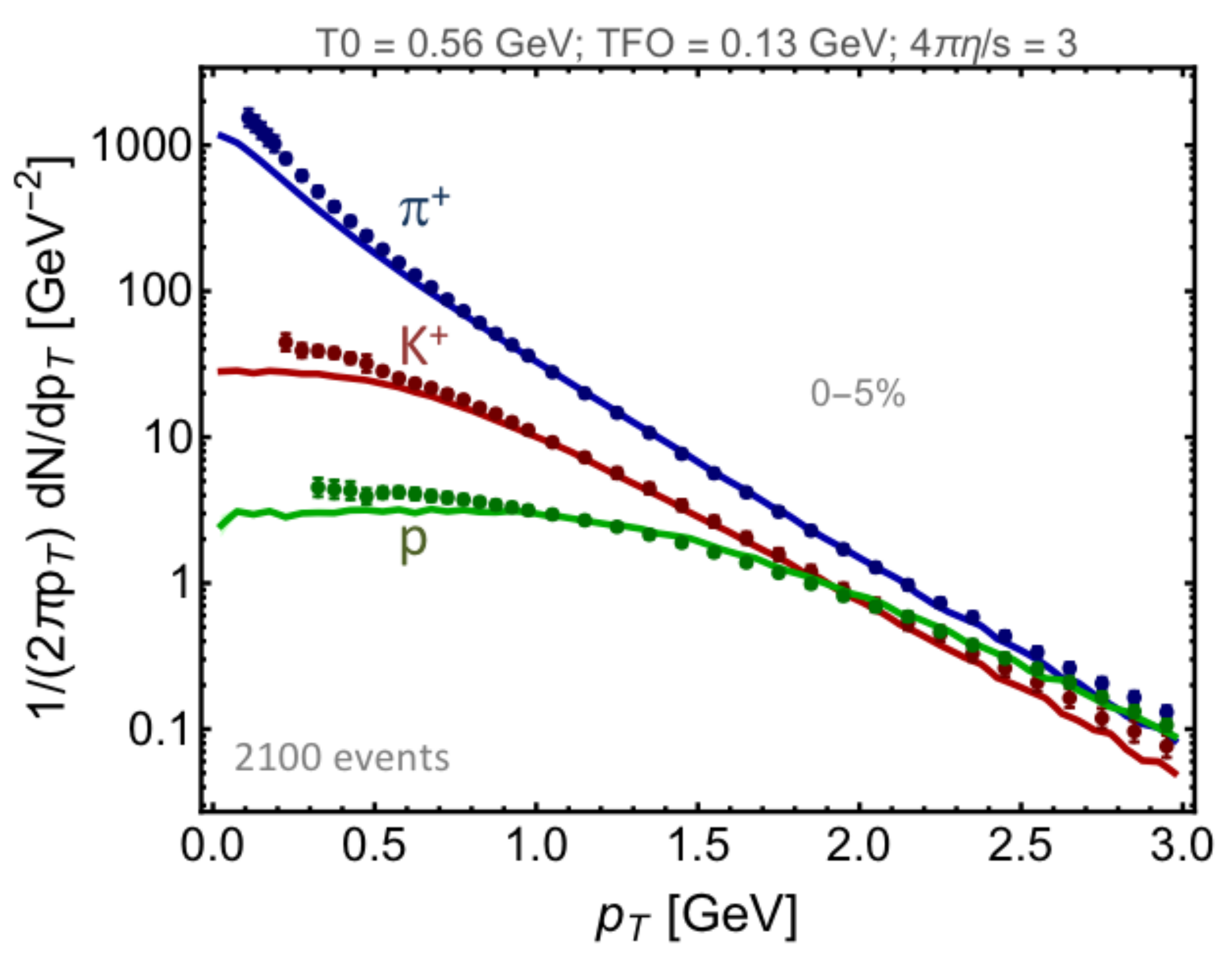}
\includegraphics[width=0.48\linewidth]{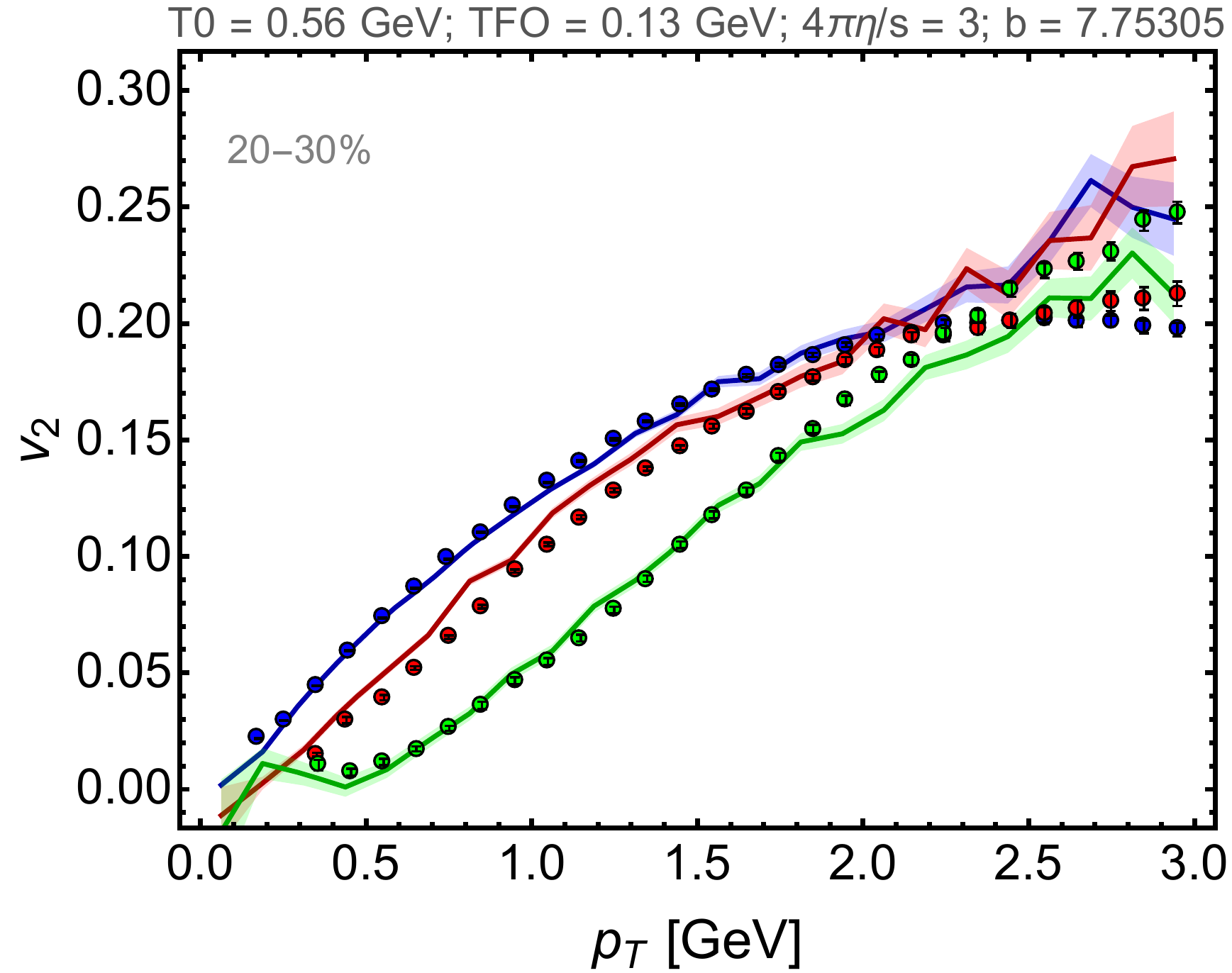}
}
\caption{(Color online) In the left panel, we show comparisons of the pion, kaon, and proton spectra as a function of $p_T$ in $0$-$5\%$ centrality class  obtained using the standard implementation of the equation of state in aHydro with experimental data}. In the right panel, we show the elliptic flow coefficient as a function of $p_T$ in $20$-$30\%$ centrality class. The experimental data is taken from the ALICE collaboration ~\cite{Abelev:2013vea,Abelev:2014pua}.  Figure used with permission from Ref.~\cite{Strickland:2016ezq}.  The initial time for the hydrodynamic simulation was taken to be $\tau_0 = 0.25$ fm/c.
\label{fig:spectras}
\end{figure}

In Fig.~\ref{fig:spectras}, we show some comparisons for standard aHydro with ALICE data published in \cite{Strickland:2016ezq,Nopoush:2016qas}. In panel (a), the spectra of pions, kaons, and protons are shown as a function of the spectra in 0-5\% centrality class. As discussed before, in Sec.~\ref{subsec:aHydros}., we see that the standard aHydro underestimate the spectra at low $p_T$.  In panel (b), on the other hand, we see a quite good agreement between theory and experimental data. However, the agreement of the standard aHydro elliptic flow with data should be interpreted with caution since the spectra are not well-reproduced in this scheme. We showed these comparisons to highlight how some observables are sensitive to the way that EoS is implemented.  

\begin{figure}[t!]
\includegraphics[width=0.99\linewidth]{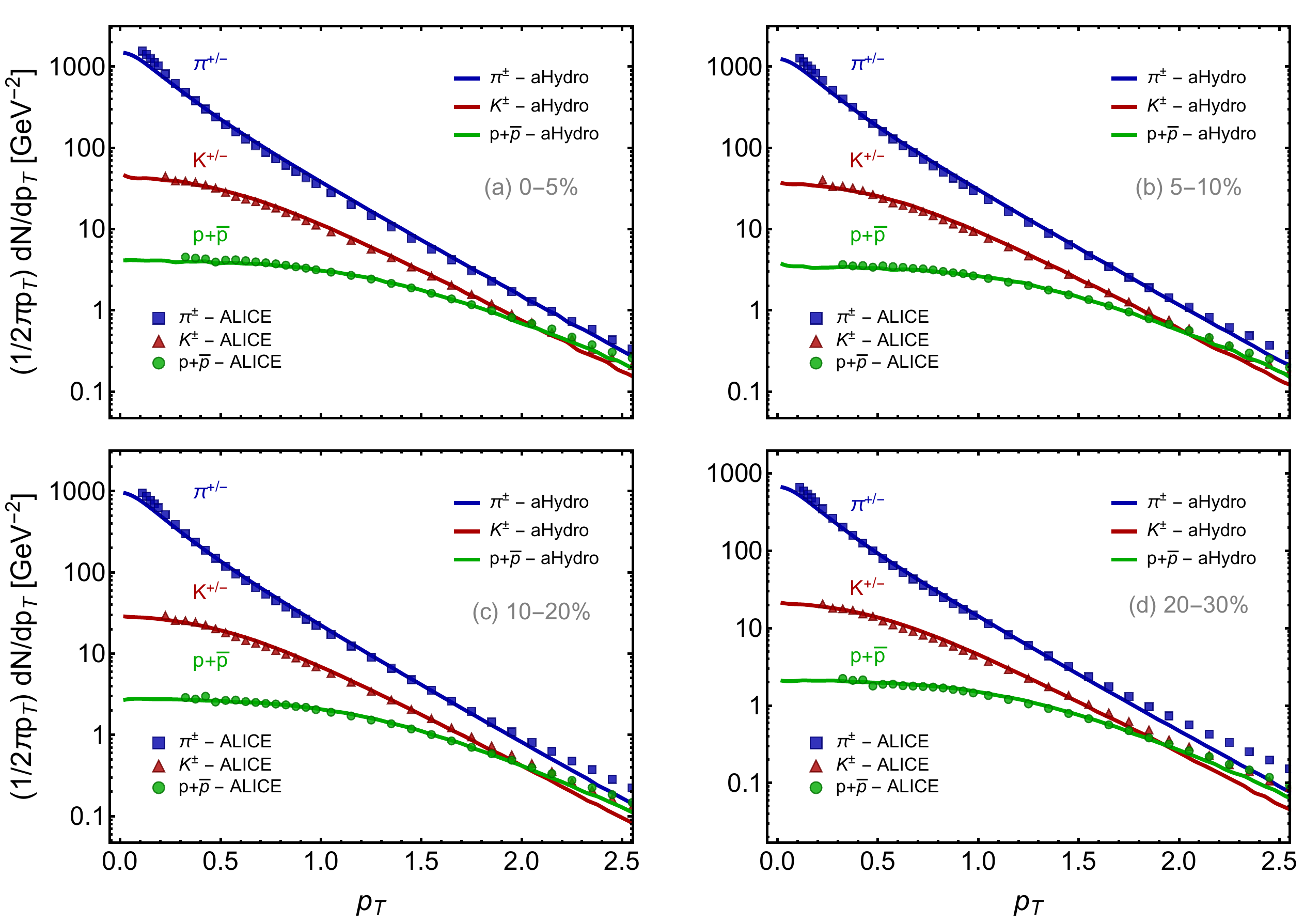}
\caption{(Color online) The spectra of $\pi^\pm$, $K^\pm$, and $p+\bar{p}$ as a function of $p_T$ in four centrality classes is shown with the data taken from the ALICE collaboration \cite{Abelev:2013vea}. 
For this figure we used an admixture of wounded nucleon and binary scattering profiles with a central temperature of $T_0= 600$ MeV at $\tau_0=0.25$ fm/c, $\eta/s = 0.159$, and \mbox{$T_{\rm FO} = 130$ MeV}. For full details of the simulation we refer the reader to Ref.~\cite{Alqahtani:2017tnq}.  Figure used with permission from Ref.~\cite{Alqahtani:2017tnq}.}
\label{fig:spectra}
\end{figure}

\subsection{Quasiparticle anisotropic hydrodynamics}
\label{sec:qppheno}

Next, we turn to 3+1d aHydroQP predictions.  In Fig.~\ref{fig:spectra}, we show the pions, kaons, and protons spectra as a function of the transverse momentum $p_T$ in four centrality classes 0-5\%, 5-10\%, 10-20\%, and 20-30\%. As can be seen from these comparisons, the aHydroQP model shows a good agreement with the experimental data over the entire shown $p_T$ range with some discrepancies at high $p_T$ in relatively higher centrality classes as shown in panel (d). By integrating the spectra over $p_T$, one can get the average transverse momentum $\langle p_T \rangle$ and the multiplicity $dN/d\eta$. In Fig.~\ref{fig:ptavg}, we show the multiplicity as a function of pseudorapidity and the average transverse momentum as a function of centrality in the left and right panels respectively. In panel (a), we show the multiplicity in five centrality classes as shown in the figure. As we can see from this figure that aHydroQP describes the multiplicity very well compared with experimental data. In the right panel, we show $\langle p_T \rangle$ as a function of centrality classes up to quite large centrality classes $\sim 60\%$ where aHydroQP agrees with the data quite well.

Note that the freeze-out temperature we found via fits, $T_{\rm FO} = 130$ MeV, is much lower than what is typically used. As we will present below, even so, we able to describe both the spectra (and hence relative abundances) of pions, kaons, and protons without having to invoke partial chemical equilibrium.  Our extracted temperature is a more consistent way to fit spectra that typical ``thermal fits'' since it folds into the calculation the effects of kinetic non-equilibrium which are discarded by thermal fits which assume perfect kinetic equilibration, tunable fugacities, and blast wave profiles. In addition, we observed that the standard hadron resonance gas EoS based on all particles from the SHARE data table doesn't seem to quantitatively describe the lattice EoS until temperatures less than 135 MeV. To us, this suggests that above such temperatures modeling the system as a non-interacting hadron gas is too simplistic.

\begin{figure*}[t!]
\centerline{
\includegraphics[width=0.51\linewidth]{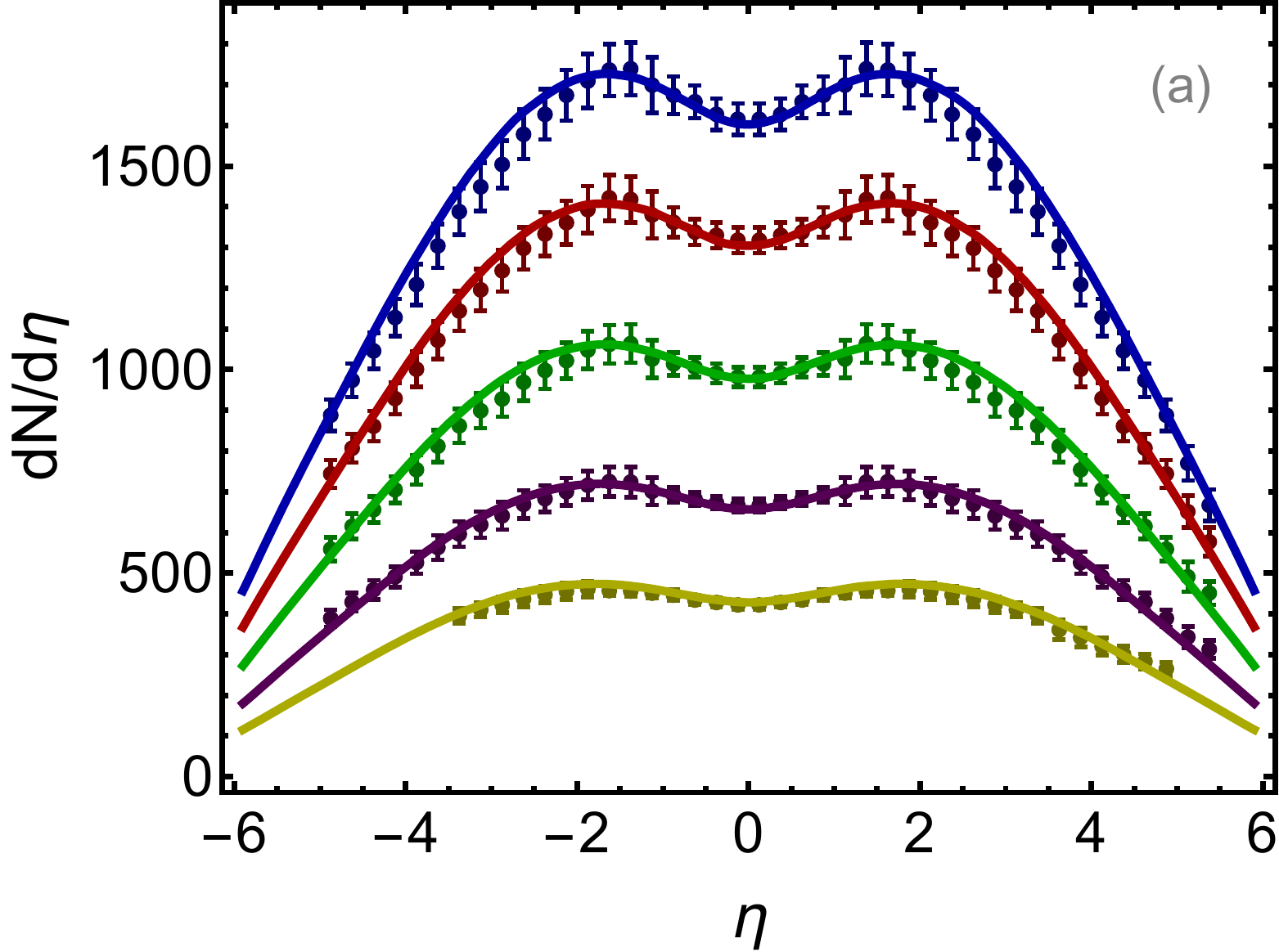}
\includegraphics[width=.49\linewidth]{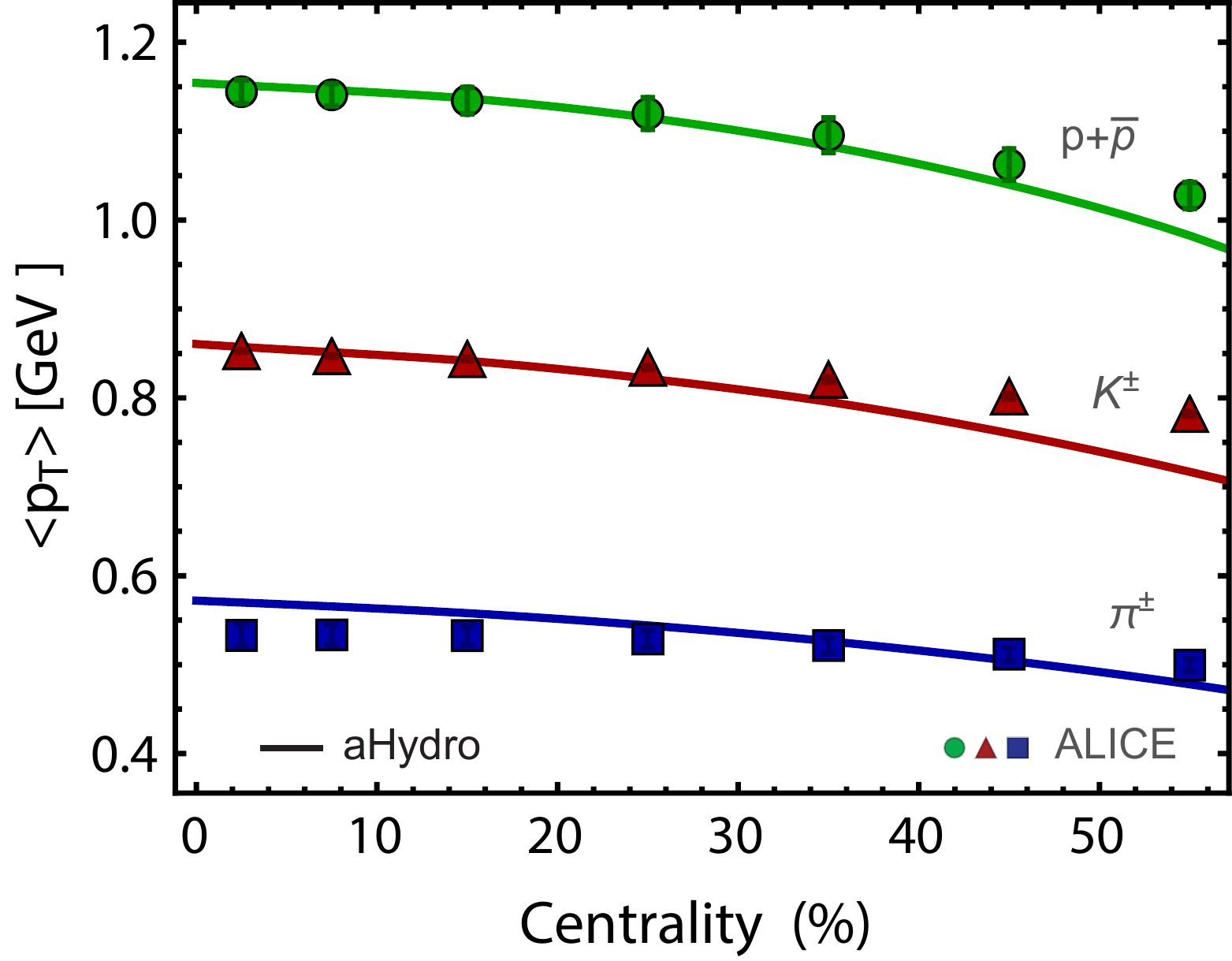}
}
\caption{(Color online) In panel (a), the charged-hadron multiplicity $dN/d\eta$ as a function of pseudorapidity $\eta$  is shown for five centrality classes where  data  are from the ALICE collaboration Refs.~\cite{Abbas:2013bpa,Adam:2015kda}. In panel (b), we show $\langle p_T \rangle $ of pions, kaons, and protons as a function of centrality where data are from the ALICE collaboration Ref.~\cite{Abelev:2013vea}. Figure used with permission from Ref.~\cite{Alqahtani:2017tnq}. }
\label{fig:ptavg}
\end{figure*}

\begin{figure}[t!]
\includegraphics[width=0.99\linewidth]{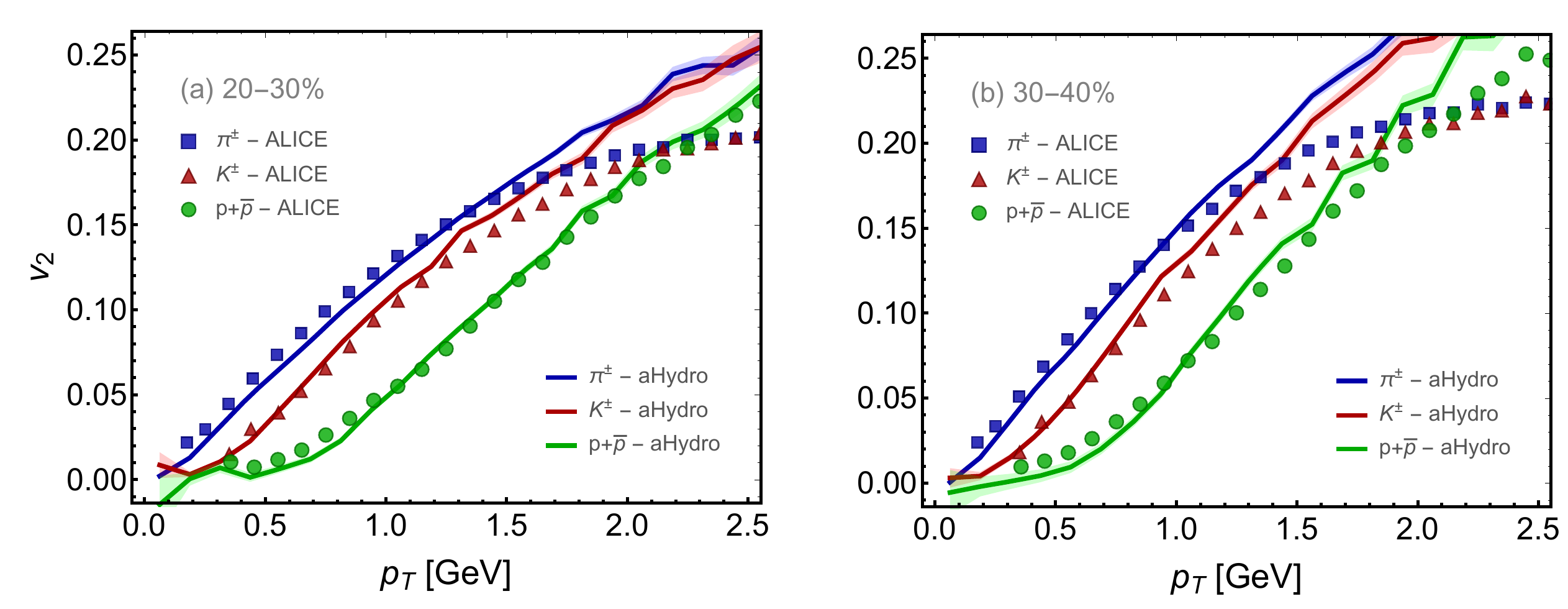}
\caption{ (Color online) The identified elliptic flow coefficient as a function of  $p_T$ is shown for $\pi^\pm$, $K^\pm$, and $p+\bar{p}$ in 20-30\% and 30-40\% centrality classes. The experimental data shown  are from the ALICE collaboration~\cite{Abelev:2014pua}. Figure used with permission from Ref.~\cite{Alqahtani:2017tnq}.}
\label{fig:v2}
\end{figure}

Moving forward with our comparisons, we show, in Fig.~\ref{fig:v2}, the identified elliptic flow as a function of $p_T$ in two different centrality classes 20-30\% and 30-40\%. As can be seen from this figure, aHydroQP was able to describe the data quite well up to $p_T \sim 2$ GeV. At higher $p_T$, our model predictions fail to match the data and they keep increasing. Finally, we show comparisons of HBT radii ratios $R_{\rm out}/R_{\rm side} $, $ R_{\rm out}/R_{\rm long}  $, and $R_{\rm side}/R_{\rm long} $ as a function of the average transverse momentum. In the top row of Fig.~\ref{fig:HBTratios}, we show the HBT ratios for 0-5\% centrality class while in the bottom row of Fig.~\ref{fig:HBTratios} we show the HBT ratios for 20-30\% centrality class. In both centrality classes we see a good agreement between the aHydroQP predictions and the experimental data.  In Ref.~\cite{Alqahtani:2017tnq} we present ratios of HBT radii in more centrality classes along with the separate results for $R_{\rm side}$, $R_{\rm long}$, and $R_{\rm out}$ compared to experimental data. In all cases, the HBT radii were extracted by computing the pair correlation function and making fits to extract the radii using the Therminator 2 codebase.

Next, we would like to talk about the bulk viscosity predicted using aHydroQP model when compared with the data. As can be seen from Fig.~\ref{fig:zeta}, the peak value is $\sim 0.05$ where in some other models~\cite{Ryu:2015vwa}, the peak value of the bulk viscosity used there,  $\zeta/s \sim 0.5$ which is quite large compared with aHydroQP prediction. This difference between models should be investigated more to constrain the bulk viscosity of the QGP.  For additional model-data comparisons, we refer the reader to the original references~\cite{Alqahtani:2017jwl,Alqahtani:2017tnq}.

\begin{figure}[t!]
\centerline{
\hspace{-1.5mm}
\includegraphics[width=1\linewidth]{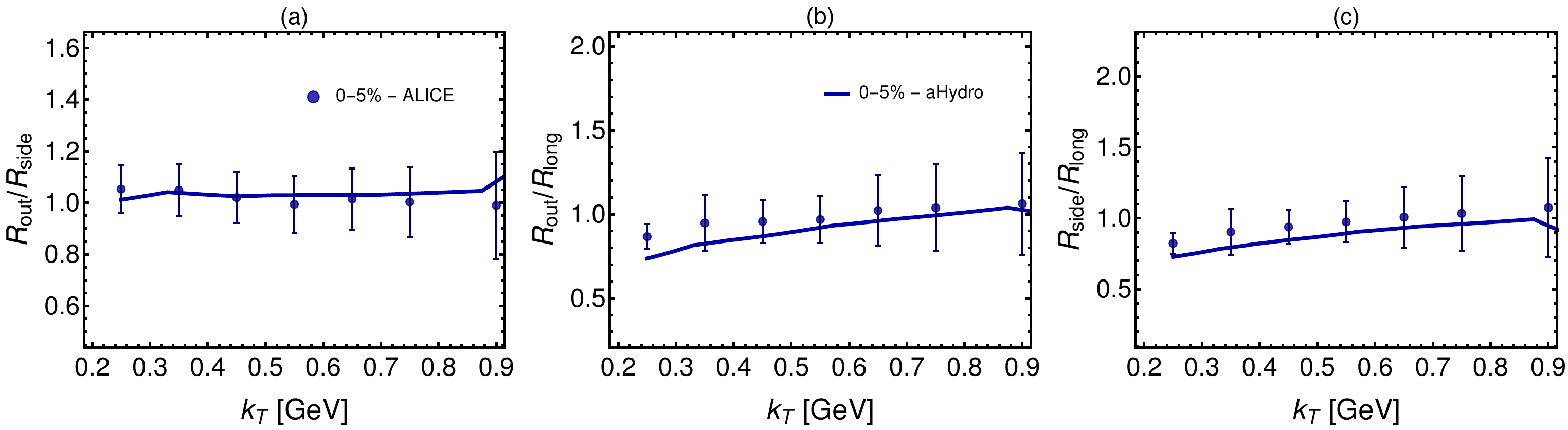}}
\centerline{
\hspace{-1.5mm}
\includegraphics[width=1\linewidth]{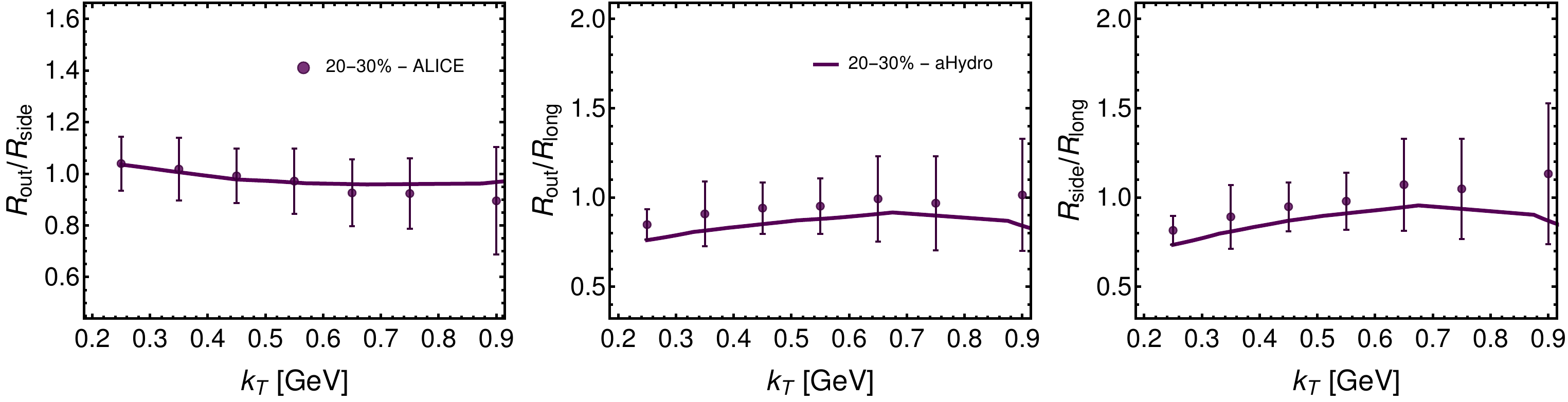}}
\caption{(Color online) The HBT radii ratios are shown as a function of $(k_T)$ for $\pi^+ \pi^+ $ in 0-5\% and 20-30\% centrality classes in the top row and bottom row respectively. The solid lines are the aHydroQP predictions and the experimental data are from the ALICE collaboration~\cite{Graczykowski:2014hoa}. Figure used with permission from Ref.~\cite{Alqahtani:2017tnq}.}
\label{fig:HBTratios}
\end{figure}


\section{Conclusions and outlook}
\label{sec:conclusions}

In this review we have attempted to provide a comprehensive overview of the progress made in anisotropic hydrodynamics since its inception in 2010.  We demonstrated that at leading-order it is possible to find an efficient and accurate description of systems that are momentum-space anisotropic by deforming the argument of the distribution function such that positivity of the one-particle distribution function is guaranteed at all times in the evolution and in all regions of phase space.  After some intermediate developments, the modern aHydro framework now includes full 3+1d dynamics which takes into account three diagonal anisotropy parameters and self-consistently implements a non-conformal equation of state using a quasiparticle model.  This allows for modeling of both shear and bulk viscous corrections in extreme conditions.  Comparisons of soft hadron production with experimental data showed that quasiparticle aHydro is able to describe the data quite well in its initial application \cite{Alqahtani:2017jwl,Alqahtani:2017tnq}.  Along the way we presented updates to the original formalism which have improved and extended the applicability of the aHydro approach.  We presented methods for testing different hydrodynamic frameworks which rely on exact solution to the Boltzmann equation for 0+1d and 1+1d conformal and non-conformal systems.  In all cases tested, researchers have found that aHydro provides the most accurate reproduction of the exact solution and that agreement can be further improved in a systematic manner using the vaHydro method.  We also demonstrated that in the simple test cases currently known, using the anisotropic matching principle results in second-order accuracy at leading-order in the aHydro expansion.

Looking to the future, there are many open questions which are currently being researched or need to be studied
\begin{itemize}
\item All phenomenological aHydro studies to date have used the relaxation-time approximation for the collisional kernel.  There need to be investigations of the impact of the collisional kernel itself on the dynamics. 
\item To date, phenomenological aHydro comparisons with data have used only smooth Glauber-like initial conditions.  Future work will focus on the efficient and stable implementation of fluctuating initial conditions.
\item In order to be complete, the 3+1d aHydro code and associated frameworks need to be able to efficiently take into account off-diagonal anisotropies.
\item The frameworks and associated codes need to be extended to finite chemical potential and initial conditions appropriate for lower-energy collisions need to be prepared.
\item Interfacing aHydro to different hydrodynamic afterburners like URQMD in order to more properly investigate the impact of late-time hadronic kinetic transport.
\item Phenomenological calculations of various signatures of the QGP, such as electromagnetic emissions, heavy-quarkonium suppression, jet quenching, etc. making use of the 3+1d aHydro background.  Some work along these lines has already been done, see e.g. Refs.~\cite{Strickland:2011aa,Krouppa:2015yoa,Bhattacharya:2015ada,Ryblewski:2015hea,Kasmaei:2016apv,Krouppa:2016jcl,Krouppa:2017lsw,Nopoush:2017zbu,Krouppa:2017jlg}, however, in most cases these studies were limited to conformal aHydro with a single anisotropy parameter.
\item Continued exploration of the non-relativistic limit of aHydro and its application to cold atoms, see e.g. Ref.~\cite{Bluhm:2015bzi}.
\end{itemize}
There is much work left to do, but based on works to date it is clear that anisotropic hydrodynamics can further improve our understanding of out-of-equilibrium relativistic systems. Looking to the future, one very promising area where anisotropic hydrodynamics might be expected to provide important improvements over standard viscous hydrodynamics is in the study of small systems, such as $pp$ and $pA$.  Since deviations from isotropy are also large in these systems and the lifetimes are significantly shorter, one expects large viscous corrections at freeze-out.

\section*{Acknowledgments}

We thank J. Noronha-Hostler and J. Noronha for providing us the data of their model used in Fig.~\ref{fig:zeta}. M.~Alqahtani was supported by a PhD fellowship from the Imam Abdulrahman Bin Faisal University, Saudi Arabia.  M.~Nopoush and M.~Strickland were supported by the U.S. Department of Energy, Office of Science, Office of Nuclear Physics under Award No. DE-SC0013470.

\section*{References}

\bibliography{ahydro-review}

\end{document}